\newif\ifRED

\newif\ifFIG
\FIGtrue    

\documentclass{elsarticle}
\makeatletter 

\usepackage[utf8]{inputenc}
\usepackage{amsmath}
\usepackage{amssymb}
\usepackage{color}

\usepackage{geometry}
\usepackage{marginnote}
\usepackage{caption}

\newcommand{\BS}{\boldsymbol}
\newcommand{\RED}{\textcolor{red}}
\newcommand{\BLUE}{\textcolor{blue}}

\title{THEORIE DER ELECTROPHORESE\\
Het Relaxatie-Effect\\J.Th.G. Overbeek's PhD Thesis translation}
\author{Evert Klaseboer, Amitesh S. Jayaraman, Derek Y.C. Chan}
\date{December 2018}

\begin{document}

\begin{abstract}
In this thesis, a theoretical treatment of the relation between electrophoretic velocity and the potential of the double layer of colloidal particles is presented\footnote{\BLUE{In the original thesis by Overbeek written almost entirely in Dutch, this English summary appeared at the end. We have relocated it to the Abstract of this translation. We also replaced the acronym "E.V." by "electrophoretic velocity" everywhere.\\
Comments of the translators are rendered in blue.}}.

The derivation of an electrophoresis equation, expressing the total sum of hydrodynamic and electric forces accompanying electrophoresis was aimed at. In this equation the relaxation effect, due to the distortion of the double layer, should be fully accounted for.

After an introduction the principles underlying the derivation of the electrophoretic velocity of a spherical particle are treated in the second chapter. The problem is separated in two parts; in chapter II A (electric section) the formation of the electric field and the distribution of the ions surrounding the sphere under the influence of electric forces, diffusion and streaming of the liquid is described. Particular attention is paid to the conditions, to which the distribution of the ions at the interface of sphere and liquid is subjected. In chapter II B (hydrodynamic section) is shown how, by application of the usual equations of motion of a viscous liquid, the electrophoretic velocity may be calculated, when the structure of the electric field is known. 

In chapter III a critical survey of the results of other investigators in this field is given.

In chapter IV the calculations are further developed. The hydrodynamic problem can be solved without the introduction of approximations. This leads to equation (48) in which, in addition to the driving force and the viscous resistance the electrophoretic friction, the relaxation effect and the mutual interaction of these effects are accounted for. 

The integration of the differential equations for the distribution of ions in the double layer can not be carried out without the introduction of approximations. The solution may be found by development of a series of increasing powers of $\frac{e\zeta}{kT}$  ($e$ = the charge of the electron, $\zeta =$ potential of the double layer, $k =$ constant of BOLTZMANN, $T =$ absolute temperature). Of this series the first and second term are calculated. With these data concerning the distribution of ions, the electrophoresis equation (89) can be derived from equation (48), which gives the relation between electrophoretic velocity, $\zeta$- potential, radius of the sphere and dimensions of the double layer. In this derivation the sphere is assumed to consists of an insulating substance.

In chapter V the consequences of equation (89) are examined. The results for small values of the $\zeta$- potential can be considered to be certain. Application of (89) to larger values of the $\zeta$- potential, although not justified by the derivation, leads to acceptable results. For values of the $\zeta$- potential not exceeding 25 mV the electrophoresis equation of HENRY requires only small corrections (a few percents). For larger values of $\zeta$ the relaxation effect may effect a considerable, usually diminishing influence upon the electrophoretic velocity. As consequences of the relaxation effect may be considered 
\\$1^{st}$. the fact that only in very exceptional cases electrophoretic velocities surpassing 5 $\frac{\mu \text{cm}}{V \text{sec}}$ are found;
\\$2^{nd}$. the influence of the valency of ions of both signs;
\\$3^{rd}$. the presence of maxima and minima in the electrophoretic velocity- concentration curve at a monotonous course of the $\zeta$-concentration curve. 

In chapter VI the electrophoresis of conducting particles is shortly examined. 

In chapter VII the conductivity of concentrated sols and the influence which may be effected by the electrophoretic forces and the relaxation effect is dealt with. 

Eventually, in chapter VIII an explanation, based on the relaxation effect, is given of the rather high value of the dielectric constant of concentrated hydrophobic sols.
\\
\\
Utrecht, April 1941.

\end{abstract}

\maketitle

\newpage

\section*{\BLUE{Foreword}}

\BLUE{The theory of electrophoresis is one of the foundational topics that underpinned the development of colloid and surface science and ranks with the famous Derjaguin-Landau-Verwey-Overbeek (DLVO) theory of colloidal stability.}

\vspace{5 mm} 
\BLUE{J. Th. G. Overbeek (“Theo” to all who knew him) was the first to develop a complete theoretical analysis of the electrophoretic motion of a charged spherical particle under the influence of an external electric field. This provided the theoretical framework for a widely used experimental method to characterize the state of charge and particle size of small colloidal particles. The solution of this problem required mastery of fluid mechanics, colloidal electrostatics, statistical thermodynamics and transport theory in additional to solid applied mathematics.}

\vspace{5 mm} 
\BLUE{Theo carried out this study as his doctoral thesis under H.R. Kruyt at Utrecht University. The thesis, in Dutch, was later published as a monograph.}

\vspace{5 mm} 
\BLUE{Given the important pedagogic value and historical status of this work, we felt that it deserved to enjoy a wide readership.}

\vspace{5 mm} 
\BLUE{We are grateful for the assistance rendered by Utrecht University and the generosity of the Overbeek family who own the copyright to this work, for granting permission for us to undertake a translation of this work.} 

\vspace{5 mm} 
\BLUE{In supporting our endeavor, Dr. Titia Cohen-Overbeek wrote:}\\

\textit{\BLUE{"What an extremely nice mail and explanation of your effort regarding the thesis of my father J. Th. G. Overbeek. I have shared this information with my three sisters and of course we give you permission to use an English translation of the thesis of our father. It is very dear to learn that the didactic qualities of my father nearly 80 years later are still so relevant. He wrote this thesis during the war and had to choose a theoretical subject because he had no access to a laboratory. Otherwise it had probably become a very different thesis.}}

\textit{\BLUE{My husband, Dr. Adam Cohen added that ...}}

\textit{\BLUE{'That was his strongest side - he could think like no other and understand very complex scientific problems and then explain it clearly. Obviously he became therefore always extremely uncomfortable when the issues in question were irrational, he then couldn't understand it anymore.  I remember once in England I asked him something about a device that we had to measure the blood flow through the arm, which he had never seen before but while driving in his car he figured out the underlying theory and the solution, I believe even the appropriate equations. '}}

\textit{\BLUE{I wish you success with the publication of the article and would like to thank you for your effort to translate the thesis in English."}}\\

\BLUE{The published copy of the thesis came in our possession by a circuitous route. It was first acquired by Dr Robin Arnold, an Australian scientist at the Commonwealth Scientific and Industrial Research Organization when he was doing post-doctoral work in the Netherlands. He gave his copy to Professor Tom Healy of the University of Melbourne who in turn gave it to one of us, Derek Chan who then partnered with Evert Klaseboer because of his unique combination of scientific and linguistic skills required to undertake the project.}

\vspace{5 mm} 
\BLUE{We view this translation as a continuing effort. Initially, we focused on completing the translation of the theory of the electrophoresis of a spherical colloidal particle. To make this accessible, we use modern notation and SI units.  We feel it is important to make the connection between the Overbeek analysis and the subsequent landmark treatment by O'Brien and White some 40 years later. Details of this connection are published in our paper in the Journal of Colloid and Interface Science (2019).}
\noindent
\BLUE{Time permitting we will return to attend to other topics covered in the thesis.}

\begin{center}
\BLUE{Amitesh Jayaraman (Singapore) \\
Evert Klaseboer (Singapore)\\
Derek Y. C. Chan (Melbourne)\\
May 2019}
\end{center}

\newpage
\section*{Preface}

Dissertation to obtain the degree of doctor of mathematics and physics at the State University of Utrecht, on the authority of the Rector Magnificus Dr. H. R. Kruyt, professor in the Faculty of Mathematics and Physics, according to the decision of the Senate of the University to defend against the objections from the Faculty of Mathematics and Physics on Monday 19 May 1941, in the afternoon at 4 o'clock by\begin{center}
Jan Theodoor Gerard Overbeek\\ born in Groningen.\end{center}

\begin{flushright} To my parents\\ To my parents in law \end{flushright}

After the completion of this thesis, I express my gratitude to you, professors, ex-professors and lecturers of the Faculty of Mathematics and Physics at Utrecht for the many things that I have been able to learn from you. 

Professor Errera, Professor Rutgers, Dr. Sack, the time I spend in Brussels and Ghent have been of great importance for my scientific formation. I still remember with joy the many discussions, which I had with you. 

I express my gratitude that I could spend many years with you as assistant, worked with you in the fullest meaning of the word, Professor Kruyt, my thesis supervisor. You have given me a lot, above all the never fading optimism, with which you look at the very complex problems resulting from colloid chemistry. I appreciate the friendship which you also showed to me outside the laboratory. 

I regret that I can only offer you the (although extensive) theoretical part of my study, and had to abandon the originally envisaged combination of mutually complimentary experiments and theory due to the war. 

Professor Bijvoet, I am very grateful for our many discussions and your sharp, yet never disheartening criticism, which has had great influence on the way in which this thesis is written.

Dr.Moesveld, I would like to thank you for the nice way, in which we discussed the technical side of the research.

A great thanks to my sister for the dedication and accuracy with which she checked the many calculations.

The part that you, Annie, played in my work may not seem great to an outsider, but without your continuous support and interest this thesis would not have been finished. 

Finally I thank the people of t'Hoff Laboratory for their warm collaboration\footnote{\BLUE{For more information see "De Collo\"idchemie in Nederland, in Utrecht in het bijzonder" by J.Th.G. Overbeek, in "Werken aan scheikunde", Delft University Press 1993, available online at https://chg.kncv.nl/l/library/...}}.

\newpage

\begin{center}CONTENTS\end{center}
\begin{itemize}
	\item[I -] INTRODUCTION
    \item[II -] FOUNDATIONS OF THE CALCULATION
    \begin{itemize}
        \item [A] The ion distribution
        \item [B] The fluid motion and the forces on the sphere
    \end{itemize}
    \item[III -] CALCULATIONS BY OTHER RESEARCHERS
    \item[IV - ] EXECUTION OF THE CALCULATION
    \begin{itemize}
        \item [A] Calculation of the fluid motion
        \item [B] Calculation of the ion distribution and the electric field
        \item [C] The electrophoresis formula
    \end{itemize}
    \item[V - ] THE ELECTROPHORETIC VELOCITY OF NON-CONDUCTING PARTICLES
    \item[VI -] THE ELECTROPHORETIC VELOCITY OF CONDUCTING PARTICLES
    \item[VII -] CONDUCTIVITY AND TRANSPORT NUMBER
    \item[VIII -] DIELECTRIC CONSTANT
    \item[ ] GLOSSARY OF SYMBOLS
\end{itemize}

\newpage

\section{INTRODUCTION}

Among the electrokinetic phenomena, electrophoresis has always taken a special place. In 1906, Burton\footnote{E.F.BURTON, Phil. Mag. (6), \textbf{11}, 425; \textbf{12},472 (1906).} found that there is a relationship between the electrophoretic velocity and the stability of gold sols; he therewith confirmed the results from Hardy\footnote{W.D.HARDY, Z.physik. Chem. \textbf{33}, 385 (1900).} obtained in 1900 on denatured proteins. From these measurements it follows that the stability of a (hydrophobic) colloid is smaller when the absolute value of the electrophoretic velocity is smaller. Later measurements by Powis\footnote{F.POWIS, Z.physik. Chem \textbf{89}, 186 (1915).} lead to the conclusion, that a sol is only stable when its electrophoretic velocity is above a certain critical value. 

The by Von Helmholtz\footnote{H. VON HELMHOLTZ, Ann. Physik. \textbf{7}, 337 (1879).} deduced and later by Von Smoluchowski\footnote{M. VON SMOLUCHOWSKI, Z. physik. Chem. \textbf{92}, 129 (1918). \BLUE{This appears to be the wrong article, it deals with a mathematical theory for the coagulation of colloidal solutions instead. The correct article should be: M. von Smoluchowski, (1903). "Contribution \`a la th\'eorie de l'endosmose \'electrique et de quelques ph\'enom\`enes corr\'elatifs". Bull. Int. Acad. Sci. Cracovie. 184.}} improved formula\footnote{\BLUE{Due to the use of Gaussian units in the original thesis and SI units by us, a difference of a factor of $4\pi$ often appears in the formulas.}}
\begin{align} \nonumber \tag{1}
      \frac{U}{E} &= \frac{\epsilon_0 \epsilon_r \zeta}{\eta}
\end{align}
gives a simple relationship between the electrophoretic velocity $U$ and the electrokinetic or $\zeta$-potential\footnote{\BLUE{For an explanation of the used symbols, see the Glossary at the end of this document.}}. The $\zeta$-potential is assumed to be the potential jump in the movable part of the double layer\footnote{In chapter V we will come back on the relationship between the $\zeta$-potential and the total double layer potential.}, $\epsilon_r$ and $\eta$ are the dielectric constant and the viscosity of the dispersion medium,  $E$ is the electric field value\footnote{\BLUE{$E$ is the electric field value of the dc-field far away from the particle.}}. Based on (1) and the observations of Hardy, Burton and Powis, one can thus say that the electrophoretic velocity, $U$, is proportional to $\zeta$ and that $\zeta$ is a measure for the stability in that sense, that a hydrophobic colloid is only then stable, if the $\zeta$-potential is larger than a certain value, which we will call critical potential. The fact that the electrophoretic velocity provides us information about the electric double layer and about the stability of colloids explains the never ending interest for electrophoresis measurements. Even after it has become clear that the relationship between electrophoretic velocity and the stability is less simple than 
Powis\footnote{See for example H.C. HAMAKER, Hydrophobic colloids; D.B. Centen, Amsterdam, 1937, pp.16-46. \\ H.EILERS and J.KORFF, Chem. Weekblad \textbf{33}, 358, (1936).\\ S.LEVINE, Proc. Roy. Soc. London. A, \textbf{170}, 145, 165 (1939).\\ S.LEVINE and G.P.DUBE, Trans. Far. Soc. \textbf{35}, 1125 (1939); \textbf{36}, 215 (1940).} assumed and that many objections can be raised against Helmholtz - Smoluchowski's equation (1), this interest is still very much alive. 

If we determine the electrophoretic velocity of a colloid as a function of the amount of neutral electrolyte\footnote{Here a neutral electrolyte is considered to be any electrolyte that does not contain potential determining ions. See also A.KELLERMANN and E.LANGE, Kolloid-Z. \textbf{81}, 88 (1937); \textbf{88}, 341 (1939).}, we often find that the electrophoretic velocity increases after adding a little (usually 1-1 valued) electrolyte, reaches a maximum at a concentration of $10^{-4}$ to $10^{-2}$ and decreases with any further addition of electrolyte. According to equation (1) this would mean, that $\zeta$ would exhibit a maximum as a function of the electrolyte concentration, which does not correspond well with the idea that we nowadays have of the double layer, namely that $\zeta$ decreases continuously with an increasing concentration of the neutral electrolyte. 

Also, the fact noted by Von Hevesey, that the electrophoretic velocity of the most various ions and colloids always give roughly the same value, is hard to reconcile with (1). This velocity, which is about $5 \times 10^{-4} \text{cm}^2/(V \text{sec})$, acts as a sort of maximum velocity, which is seldom exceeded. The electrophoretic velocity can become lower, as low as 0, but values higher than 5 or 6 $\times 10^{-4}$ are great exceptions. This would mean, that the $\zeta$-potential can never be higher than about 70 millivolt, which does not correspond to the much higher values of $\zeta$ which were found from streaming potentials\footnote{See for example A.J.RUTGERS, E.VERLENDE and M.MOORKENS, Proc. Kon. Ned. Akad. v. Wetensch. Amsterdam \textbf{41}, 763 (1938).}. 

Moreover, we know from the work of Verwey and De Bruyn\footnote{E.J.W.VERWEY, PhD thesis Utrecht (1934).\\ H.DE BRUYN, PhD thesis Utrecht (1938).} concerning the \textit{AgI}-sol and from N. Bach\footnote{N.BACH, A.RAKOW and N.BALASCHOWA, Acta Physicochimica U.R.S.S. \textbf{3}, 79 (1935); \textbf{7}, 85, 899 (1937).} and co-workers on the \textit{Pt}-sol that there is a tight connection between the double layer potential in the colloid chemical sense and the electro chemical (Nernst)- potential. This Nernst- potential can easily reach values of many hundreds of millivolts and there is at first sight no reason why the $\zeta$-potential could not be higher than 70 millivolt. 

A closer inspection reveals that the electrophoretic velocity formula (1) is rather wrong. We will first have to try to get a better electrophoretic velocity formula, before thinking of explaining the above mentioned difficulties. Especially the fact that in recent years the stability theories have made remarkable progress, where still a close relationship between the $\zeta$-potential (or particle charge) and the stability is assumed, makes it the more urgent to be able to have reliable values of $\zeta$.

It has above all been the electrolyte theory from Debye and H\"uckel\footnote{P.DEBYE and E.H\"UCKEL, Physik.Z. \textbf{24}, 305 (1923).} that shaked the trust in equation (1) of Helmholtz-Smoluchowski. According to Debye and H\"uckel\footnote{P.DEBYE and E.H\"UCKEL, Physik. Z. \textbf{25}, 49 (1924). \\ E.H\"UCKEL, Physik. Z. \textbf{25}, 204 (1924). \BLUE{The title of this article by H\"uckel is "Die Kataphorese der Kugel" (cataphoresis of a sphere), in which the term cataphoresis refers to electrophoresis of cations (positively charged particles) as opposed to anaphoresis for electrophoresis of negatively charged particles (or anions).}} the factor 4 in (1) must be replaced by another constant which is dependent on the shape of the particle. H\"uckel finds, that this constant must be replaced by a factor 6, such that he finds equation (2). 
\begin{align} \nonumber \tag{2}
      \frac{U}{E}= \frac{2}{3}\frac{\epsilon_0 \epsilon_r \zeta}{\eta}  
\end{align}
In 1931, Henry\footnote{D.C.HENRY, Proc. Roy. Soc. London. \textbf{133}, 106 (1931).} managed to bridge equations (1) and (2). He included the effect of the deformation of the applied field due to the difference in conductivity of the particle and the fluid and could then find theoretically that the formula of Helmholtz - Smoluchowski (1) is valid when the thickness of the double layer is small compared to the dimensions of the particle and if the particle does not conduct the electricity. The equation by H\"uckel (2) is valid for a sphere, which has the same conductivity as the liquid. If the thickness of the double layer is much larger than the radius of the sphere, then the H\"uckel formula is valid irrespective of the conductivity. The "thickness of the double layer" is quantified by\footnote{\BLUE{Nowadays, $\kappa$ is called the Debye parameter and $1/\kappa$ the Debye length. The dimensionless quantity determining the behavior of the system is $\kappa a$ (with $a$ the radius of the particle). When $\kappa a$ is large, there is a thin double layer, when it is small, there is a thick one.}} 
\begin{align} \nonumber
      \frac{1}{\kappa}= \sqrt{\frac{\epsilon_0 \epsilon_r kT}{ \Sigma_i n_i^\infty z_i^2 e^2 }}  
\end{align}

In a flat double layer, at a distance of $1/\kappa$ from the wall, the double layer potential has decreased by a factor $1/e$ compared to the value it has on the wall. Table 1 shows that, at the concentrations usually applied
in colloid chemistry, $1/\kappa$ varies from $3 \times 10^{-5}$ cm to $10^{-7}$ cm. Since the sizes of colloidal particles range from $2 \times 10^{-5}$ cm to $10^{-7}$ cm, there are both cases in which $1/\kappa$ is much larger, but also much smaller than a typical dimension of the particle. But in most cases $1/\kappa$ and the radius of the particle, $a$, are of the same order of magnitude. 

\begin {table} \label{Table1}
\caption*{TABLE I Thickness of the double layer ($1/\kappa$) and the value of $\kappa a$ for several concentrations of 1-1 valued electrolyte. } 
\begin{tabular}{| l | l | l | l |l |}
  \hline
  Concentration & 1/$\kappa$ &  & $\kappa a$& \\ & & a=$3\times10^{-5}$cm & $a=10^{-6}$cm & $a=10^{-7}$cm\\
  \hline
  $10^-6$ N & $3.2\times 10^{-5}$  & 1 & 0.0032 & 0.032\\
  $10^-5$ N & $1\times 10^{-5}$  & 3 & 0.1 & 0.01\\ 
  $10^-4$ N & $3.2\times 10^{-6}$  & 10 & 0.32 & 0.032\\ 
  $10^-3$ N & $1\times 10^{-6}$  & 30 & 1 & 0.1\\ 
  $10^-2$ N & $3.2\times 10^{-7}$  & 100 & 3.2 & 0.32\\ 
  $10^-1$ N & $1\times 10^{-7}$  & 300 & 10 & 1\\
  \hline  
\end{tabular}
\end{table}

In calculating the electrophoresis formula, Von Smoluchowski, H\"uckel as well as Henry have all taken into account three forces that act on a particle. These three forces\footnote{\BLUE{Overbeek uses the symbol $k$ for forces, since the Dutch word for force is 'kracht'.}} are \BLUE{(see Fig. \ref{fig:1})}:
\begin{itemize}
    \item $k_1$, the force on the charge of the particle by the applied field,
    \item $k_2$, the pure hydrodynamic frictional force, which the particle experiences when it moves through the fluid, and
    \item $k_3$ the so-called electrophoretic friction\footnote{\BLUE{In the book "Electrophoresis, Theory, Method and Applications", edited by Milan Bier, Volume II, Academic Press, New York, 1967,  Chapter 1, J.Th.G. Overbeek and P.H. Wiersema, "The interpretation of Electrophoretic Mobilities", the force $k_3$ is called "retarding force" or "electrophoretic retardation".}}.
\end{itemize}
The last force originates from the fact that, under the influence of the applied field, the counter ions are moving in opposite direction as the particle. In doing this they drag the fluid around them with them, which in turn drags the particle again and thus reduces the electrophoretic velocity. 
\ifFIG\begin{figure}
    \centering
    \includegraphics[width=0.5\textwidth]{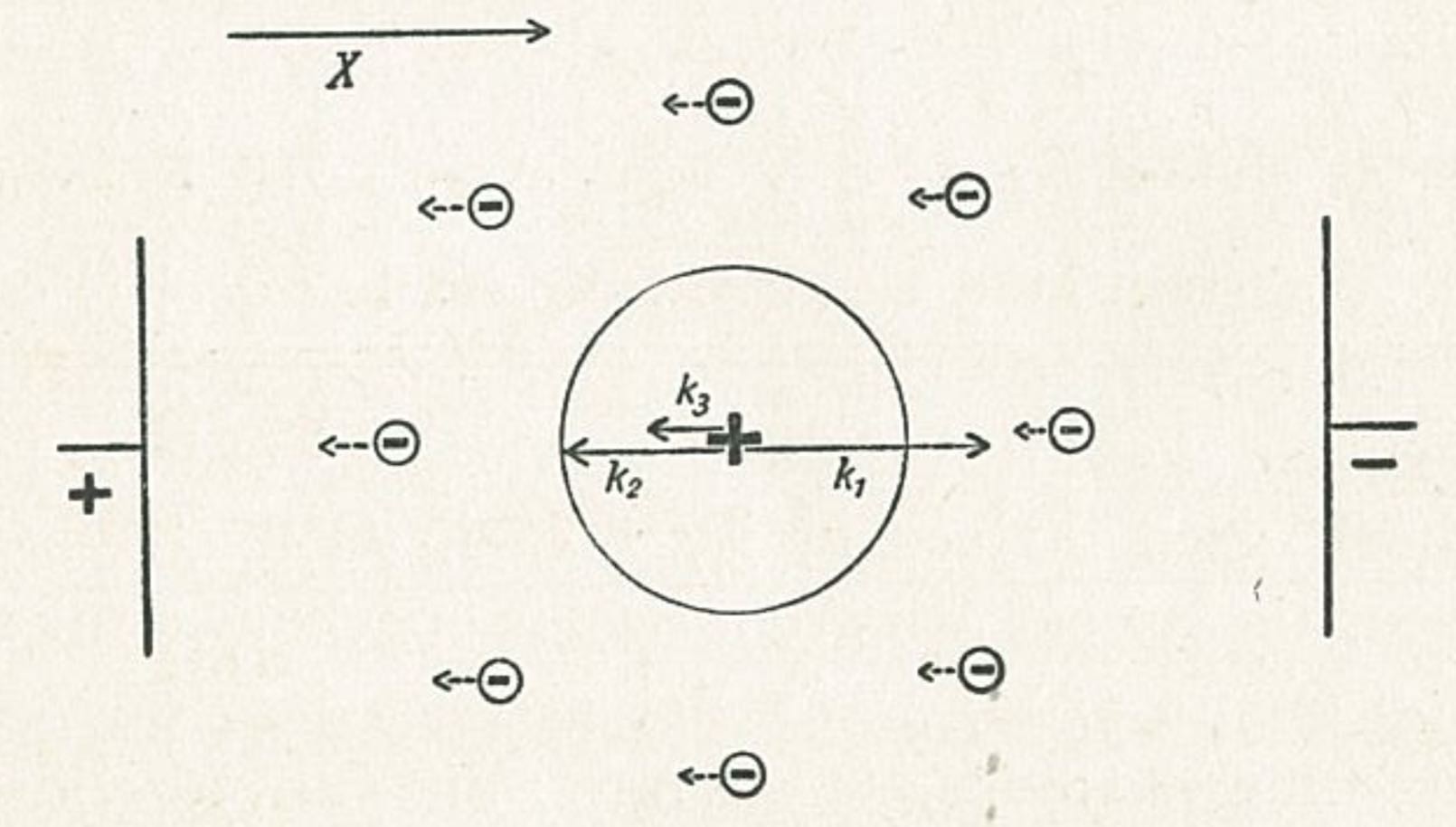}
    \caption{The forces on the sphere. $k_1$ = force excerted by the electric field on the charge of the sphere. $k_2$ = drag (friction) according to Stokes. $k_3$ = electrophoretic drag.}
    \label{fig:1}
\end{figure} \fi

If we limit ourselves at the moment to a spherical particle\footnote{\BLUE{The force $k_2$ then becomes the classical Stokes formula: $k_2=-6\pi\eta a U$.}}, including the possible attached water layer, that has a radius of $a$ and a charge of $+Q$ then, according to H\"uckel\footnote{For a simple derivation of $k_3$ see RUTGERS, Physische Scheikunde, Noordhof, Groningen 1939, page 357 and A.J. RUTGERS and J.T.G. OVERBEEK, Z. physik. Chem. A \textbf{177}, 33 (1936).}
\begin{align} 
\nonumber       k_1 &= + Q E  \\
\nonumber       k_2 &= -6 \pi \eta a U  \\
\nonumber       k_3 &= (4\pi\epsilon_0\epsilon_r \zeta a - Q) E\quad .
\end{align}
If the particle is moving with uniform velocity during electrophoresis, the total force acting on the particle must be equal to zero thus:
\begin{equation}  \nonumber
  k_1 + k_2 + k_3 = 0 
\end{equation}
giving
\begin{equation}  \nonumber
  \frac{U}{E} = \frac{2}{3}\frac{\epsilon_0\epsilon_r \zeta}{\eta}
\end{equation}
The difference in the result of the calculations of Von Smoluchowski, H\"uckel and Henry lies in the difference in the value of the electrophoretic drag $k_3$. It is easy to see, why the formula of H\"uckel (2), which is valid for a thick double layer, needs a correction, when the thickness of the double layer becomes comparable or smaller than the radius of the sphere. $k_3$ is derived by H\"uckel under the assumption that in the whole double layer, the applied field can be set equal to $E$. Since the particle has a different conductivity than the fluid, the field in the neighbourhood of the particle will be distorted. This distortion has no influence on $k_1$; also on $k_3$ it has practically no influence, if $k_3$ is mainly build up from contributions originating from a great distance of the particle, since the distortion is only limited to a distance comparable to the radius of the particle. Is on the other hand the double layer very thin, then $k_3$ will be influenced. A consideration as shown in Fig.\ref{fig:2}, where the electric field lines are shown, can clarify this influence. The field lines bent around an insulator and are being pulled towards a conductor. In front and behind the insulator the field line density is small. The field in the original direction then has become zero there. While thus according to H\"uckel a force $z_i e E$ is exerted on an ion situated in front or behind the particle, which is transmitted to a large extent as friction on the particle, on the contrary, according to Fig.2b, these ions are \textit{not} contributing to the friction. The fact that the ions at the sides in Fig.2b are experiencing a larger force than those of Fig.2a and thus result in a larger friction, can only compensate the first effect partially, such that in total a lower friction results according to the calculations of H\"uckel. In the opposite case, a conducting particle, where the electric field line density in front and behind the particle is on the contrary larger, will have to experience a larger friction and thus a lower electrophoretic velocity. From the analysis of Henry, which can be considered as a summary and extension of older electrophoresis formula, it was determined that the electrophoretic velocity of a sphere can be given by
\ifFIG\begin{figure} 
    \centering
    \includegraphics[width=0.5\textwidth]{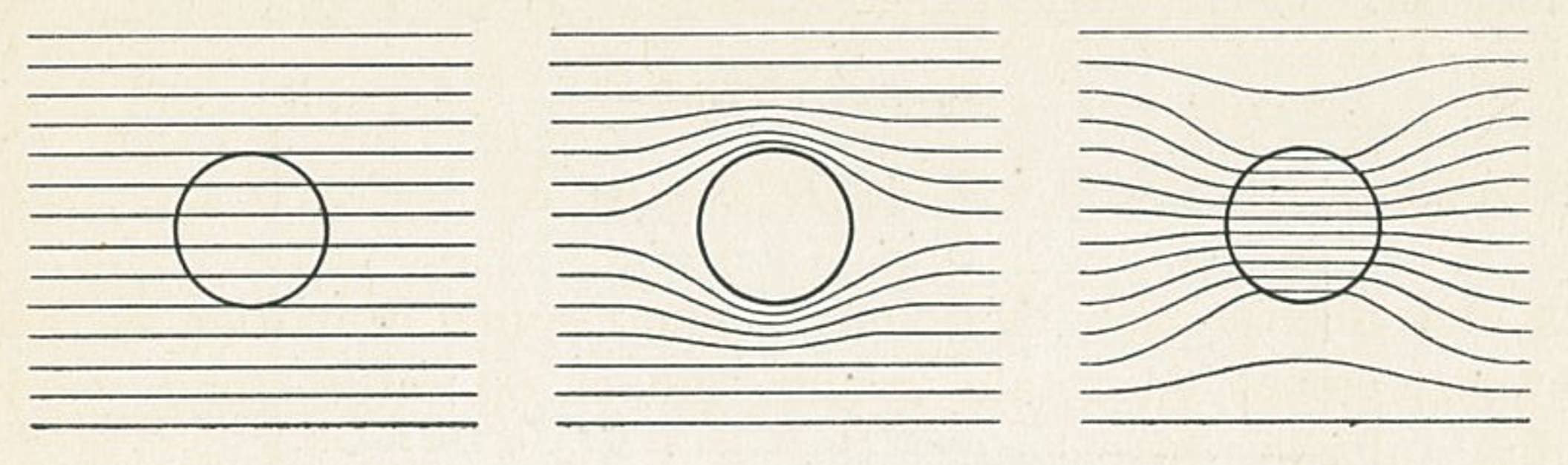}
    \caption{a) H\"uckel \hspace{5mm} b) Insulator. \hspace{5mm} c) Conductor.}
    \label{fig:2}
\end{figure} \fi
\begin{equation}  \nonumber
  \frac{U}{E} = \frac{2}{3}\frac{\epsilon_0 \epsilon_r\zeta}{\eta}f(\kappa a, \mu) 
\end{equation}
in which $\mu$ is the ratio of the conductivity of the sphere and the fluid. Fig.\ref{fig:3} gives a graphical representation of $f(\kappa a, \mu)$ for three values of $\mu$, namely the particle conducts much better $(\mu=\infty)$, the same $(\mu=1)$ and much worse $(\mu=0)$ than the fluid
\ifFIG\begin{figure}
    \centering
    \includegraphics[width=0.45\textwidth]{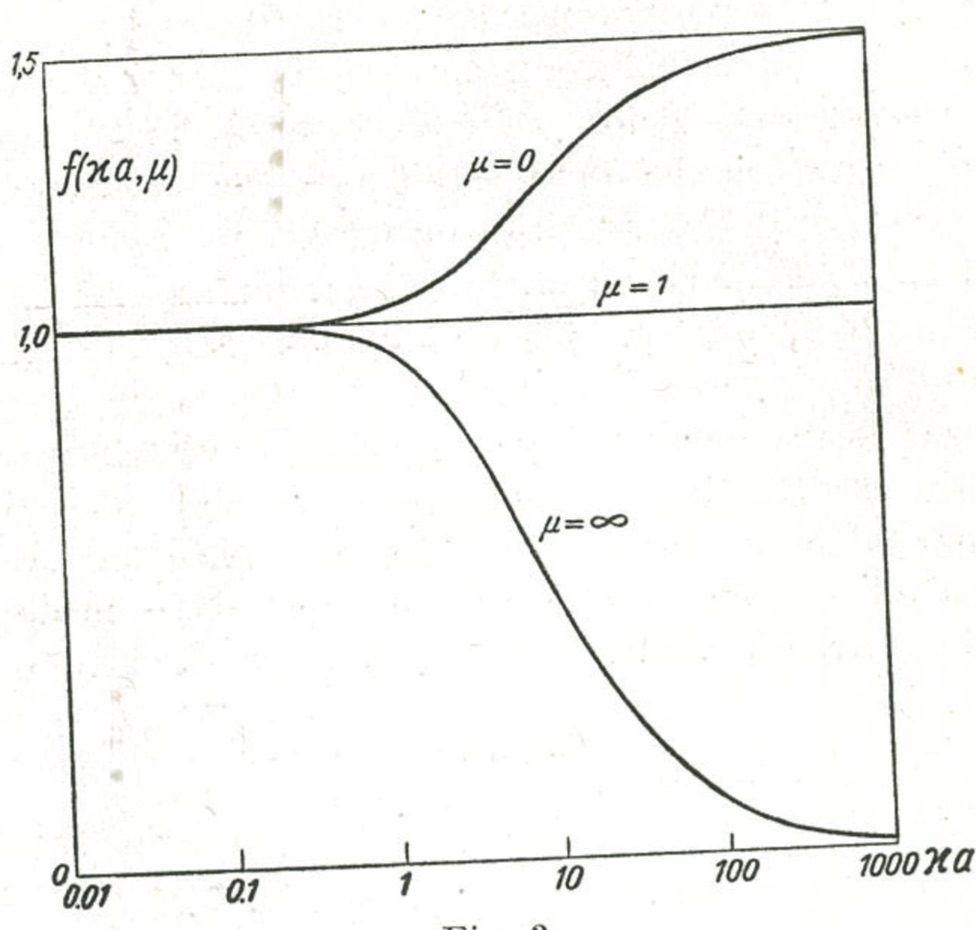}
    \caption{$f(\kappa a,\mu)$ from $\frac{U}{X}=\frac{D\zeta}{6 \pi \eta} f(\kappa a, \mu)$}
    \label{fig:3}
\end{figure} \fi

To get an idea of the influence of the shape of the particle Henry has also calculated the electrophoretic velocity for a cylinder with its axis perpendicular to the field and a cylinder with its axis parallel to the field. For a cylinder parallel to the field the formula of Smoluchowski is valid irrespective of the thickness of the double layer.

 \begin{equation}  \nonumber
   \frac{U}{E} = \frac{\epsilon_0 \epsilon_r \zeta}{\eta}.
\end{equation}
If the cylinder is placed perpendicular to the field and the double layer is \textit{thin}, then for an insulating particle $(\mu=0)$ the formula of Smoluchowski is still valid, while the velocity becomes half as big if $\mu=1$ and 0 if $\mu=\infty$.

However, in the calculations of Von Smoluchowski and H\"uckel as well as those of Henry it is specifically assumed that the original symmetry of the double layer is not disturbed during electrophoresis. From the theory of Debye and H\"uckel for the conductivity of strong electrolytes is it known however, that the ionosphere (double layer) is being deformed and that this deformation creates a friction force on the ion (particle), which is of the same order of magnitude as the electrophoretic friction, which is our $k_3$. The particle (ion) moves through its own double layer and this layer must thus be destroyed behind the particle and being build up again in front of it. In order to do this a small yet finite amount of time is needed, the relaxation time. The double layer will thus always lag \textit{behind} with respect to the particle, and, since the particle and the double layer are of opposite charge sign, will create another drag force as illustrated in Fig.\ref{fig:4}. This force will be called $k_4$\footnote{\BLUE{In the aforementioned book chapter by Overbeek and Wiersema, this force is called "the relaxation effect", it is said there that it is, in most cases, a retarding force.}}.
\ifFIG\begin{figure}
    \centering
    \includegraphics[width=0.5\textwidth]{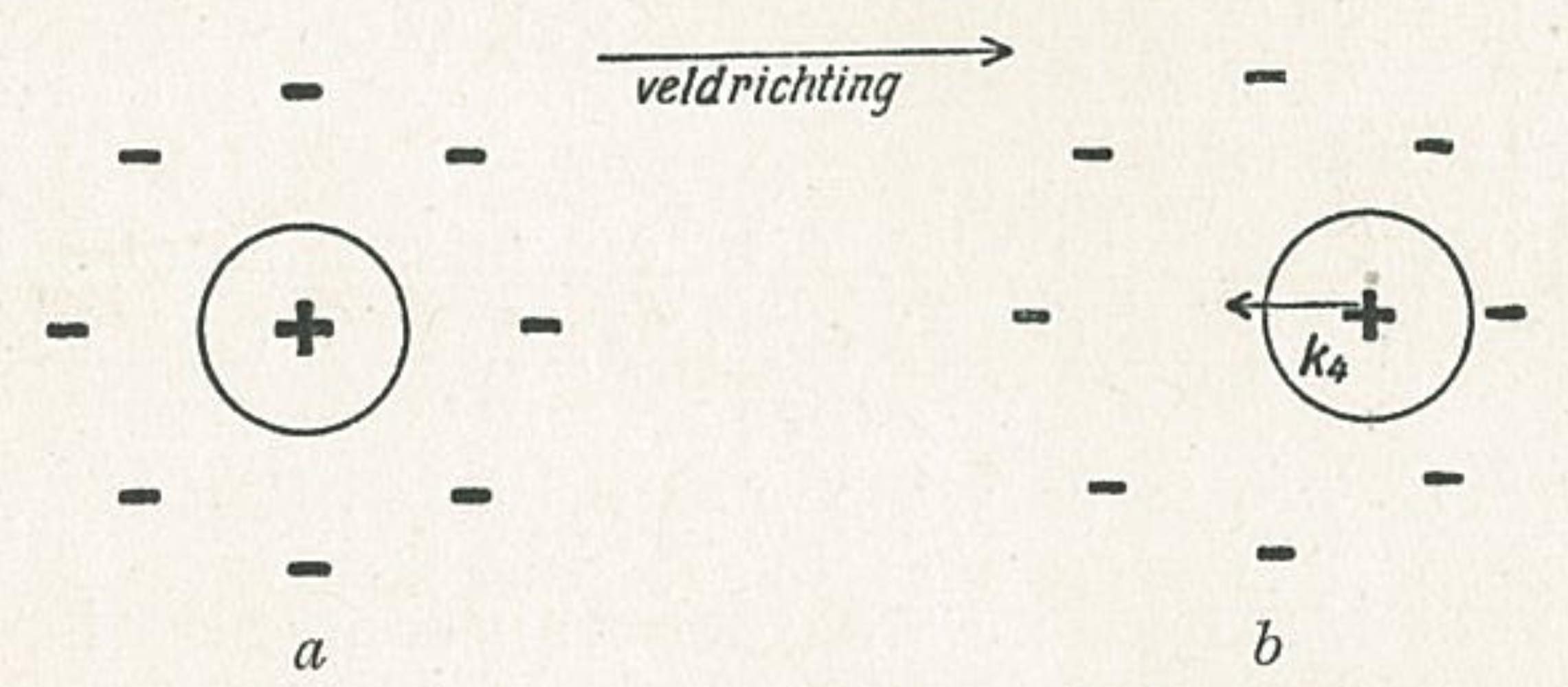}
    \caption{Drag due to the relaxation force $k_4$. Left: at rest. Right: in motion. \BLUE{Direction of the electric field "veldrichting" from left to right.}}
    \label{fig:4}
\end{figure} \fi
One should thus also take into account this relaxation force $(k_4)$ in the electrophoresis equation. The fact that this up to now has not or hardly been done for colloids is more due to the great difficulty to calculate the relaxation effect rather than the conviction that it does not have much influence. 

After all, according to the theory of Debye and H\"uckel and to Onsager\footnote{L.ONSAGER, Physik Z. \textbf{28}, 277 (1927).} who has improved on the examinations of Debye and H\"uckel,
\begin{subequations}
\begin{align}
\nonumber \tag{3}
 k_4 &= -\frac{(z_j e)^3 E \kappa}{ 24\pi\epsilon_0\epsilon_r kT}\; f(q), 
         \quad \text{(Debye-H\"uckel)} \\
         \nonumber \tag{3a}
 k_4 &= -\frac{(z_i e) (z_j e)^2 E \kappa}{24\pi\epsilon_0\epsilon_r kT} \; g(q)
        \quad \text{(Onsager)}
\end{align}
\end{subequations}
Here $z_j e$ is the charge of the ion (particle) under consideration, $z_i e$ the charge of a counter-ion and $f(q)$ and $g(q)$ are functions of the valence and mobility of the ions. For ions with valence one with equal mobility $f(q)=1$ and $g(q)=0.59$. Although these values were calculated for ions with a radius many times smaller than $1/\kappa$, we can still get an indication of the order of magnitude of the relaxation effect in colloids from (3a) and (3b), by calculating the ratio of $k_4$ and $k_1$, where $z_j e$ assumes the value $Q \approx (4\pi\epsilon_0\epsilon_r) a \zeta$. If $z$ is the valence of the counter ions, then according to Onsager\footnote{Equation (3a) would give even larger values for the relaxation effect than (3b).}
\begin{align} \nonumber \tag{4}
    \frac{k_4}{k_1} &\approx \frac{z e a \zeta \kappa}{6kT} 
      \\
      \nonumber     &= 2 \times 10^5 z^2 a \zeta \sqrt{c}\; 
\end{align}
Here $\zeta$ must be expressed in millivolts, $a$ in cm and $c$ in mol/L. If we take for example low values for $z$, $a$, $\zeta$ and $c$ as: $z=1$, $a=10^{-6}$ cm, $\zeta = 25 $ mV., $c=10^{-4} N$ then $k_1/k_4$ is already $1/20$, while for slightly larger values of the concentration, potential and particle radius we can expect corrections of many tenths of percent.

If we thus ever want to conclude anything about the double layer from electrophoretic velocity measurements, it will be a necessity to take into account the relaxation effect in the electrophoresis formula. 

Paine has applied the formulas of Debye and H\"uckel for the relaxation effect directly to colloids. Since in the derivation of those formulas it was assumed that the sizes of the ions (particles) are small compared to the thickness of the double layer, they can only be applied to that case, i.e. $\kappa a$ can at most be 1/10.\footnote{For further discussion on this value, see chapter V, \BLUE{(now in the discussion around the last equation of chapter V.)}}

A number of other researchers, among them Mooney, Bikerman, Komagata and Hermans\footnote{M.MOONEY, J.Phys. Chem \textbf{35}, 331 (1931).\\ J.J.BIKERMAN, Z. physik. Chem., \textbf{A 171}, 209 (1934). \\ S.KOMAGATA, Researches of the electrochemical Laboratory no. 387 (1935). \\ J.J.HERMANS, Phil. Mag., (7) \textbf{26}, 650 (1938).} have studied the relaxation effect, without reaching a definitive formulation, which would enable us to take the relaxation effect into account in general. One can thus state that the considerations of Paine and Komagata only have meaning if the dimensions of the particle are small compared to the thickness of the double layer, while the calculations of Mooney, Bikerman and Hermans are on the contrary applicable to the other extreme case of a very thin double layer (i.e. $\kappa a >25$. Unfortunately, in the great majority of cases for colloids $1/10<\kappa a< 25$, and thus in the range which is of most interest to us we cannot calculate the relaxation effect. 
\ifFIG\begin{figure}
    \centering
    \includegraphics[width=0.6\textwidth]{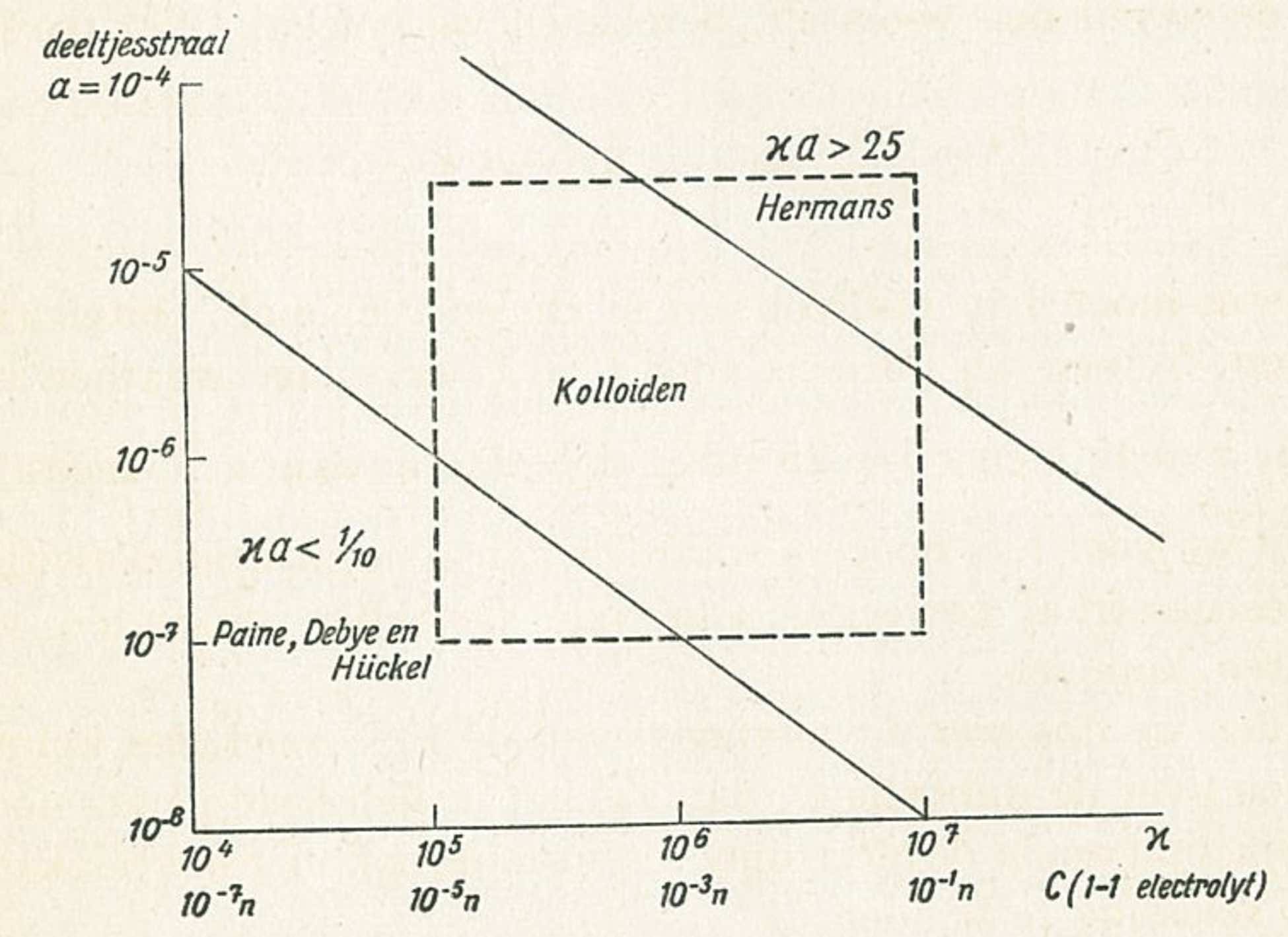}
    \caption{\BLUE{"deeltjesstraal" = particle radius, "Kolloiden" = colloids.}}
    \label{fig:5}
\end{figure} \fi
In Fig.\ref{fig:5} the areas of validity of the approximations of Hermans and Paine (Debye - H\"uckel) are drawn with solid lines, the area of colloids with dashed lines, which, as we can see from a relaxation effect point of view is almost entirely a "terra incognita".

In this thesis a calculation of the relaxation effect will be given, which is valid for arbitrary $\kappa a$, in which as good as possible the various factors that play a role in the deformation of the double layer are accounted for. the calculation is, as far as the potential is concerned, only done in a first approximation, i.e. in the same approximation as the formulas of Paine, Debye and H\"uckel on the one hand and Hermans on the other hand.  

For small $\kappa a$ (thus for small particle radius) and symmetric electrolytes our calculation returns back to the old formulas of Debye and H\"uckel. For non symmetric electrolytes our calculations approach more those of Onsager, except for the fact that by us the Brownian motion is neglected, while he did take into account this effect on the central particle. 

For large $\kappa a$ our result is different from that of Hermans, because we have looked at the boundary conditions in a different, and in our opinion better, way and because we have taken into account the interaction between the electrophoretic drag and the relaxation effect, which is especially important for large $\kappa a$ (see chapter V, \BLUE{ (the discussion around (89aII) and (89bII)).}). 

It is of course true that our calculations are only valid for spherical particles, but in any case from these calculations approximate conclusions can be drawn for other shaped particles. 

We will now give the assumptions and train of thought of the calculations in the next chapter. 

Next in chapter III the work of other researchers will be discussed.

In chapter VI the calculations are then further elaborated upon while chapter V contains the conclusions of the calculations.

In chapters VI, VII and VIII some subject that are closely associated are being discussed briefly and in chapter IX the general summary is presented\footnote{\BLUE{This chapter was written in English and was brought forward in this English translation as the abstract.}}. 

\newpage
\section{FOUNDATIONS OF THE CALCULATION}

To calculate the electrophoretic velocity we are assuming the following:
\begin{enumerate}
  \item The colloidal particle is a rigid sphere, which, optionally with the inclusion of the attached water layer, has a radius $a$. When referring to "the sphere" later on, we will always mean the sphere + the attached water layer. 
  \item A charge $+Q$ is uniformly distributed over the surface of the sphere\footnote{The equations valid for a negatively charged particle can easily be obtained by inverting the signs.}.
  \item The opposite charge $-Q$ can be found in the Gouy atmosphere around the sphere.
  \item In the whole double layer the dielectric constant $\epsilon_r$ and the viscosity $\eta$ have the same value as in the liquid at a large distance from the sphere. 
  \item The influence of colloidal particles on each other is neglected. 
\end{enumerate}

Some objections can be brought in against assumption 4. It would however be premature to assume other values for $\epsilon_r$ in the double layer, when the solution for constant $\epsilon_r$ and $\eta$ is not yet known. 

We consider now a positively charged sphere in a solution of a $z_+$,  $-z_-$ valued electrolyte (concentration $n_+$, respectively $n_-$ ions per cm$^3$). Under the influence of an electric field the sphere moves with a uniform velocity $U$ to the right. Here we thus totally neglect the Brownian motion of the central particle, which is the more justified when the particle is larger. Already when we consider the ion movement, the introduction of the Brownian motion only introduces a small correction to the central ion, such that we can neglect it for the much larger colloidal particles as a first guess. The strength of the electric field at a large distance from the sphere is $E$. We choose a coordinate system that is connected to the sphere, or, which turns out to be the same, we assume that the whole fluid is moving past the sphere to the left, such that the sphere does not move. 

Two problems will have to be solved now. First it must be investigated how the ion distribution around the sphere is being established under the influence of the electric field, the fluid motion and diffusion. Next we must investigate which forces the electric field, the distorted charge cloud and the fluid motion exert on the sphere. From the constraint that in the stationary state the total sum of these forces must be zero, we can deduct the value of the electrophoretic velocity $U$. 

\subsection*{A - THE ION DISTRIBUTION}

In a very short time, after the electric field has been applied, a stationary state establishes, in which the particle is moving with uniform speed towards the cathode. The ions are moving with respect to the particle, but in each volume element the number of positive and negative ions remains constant on average\footnote{Averaged over such a large time, that we can no longer observe the "Schwankungen" caused by thermal motion.}. Per unit of time on average the same amount of ions that will enter such a volume element will leave. We will therefore calculate, how many ions per unit of time, are moving across a surface element $d\omega$ and afterwards set the sum of all ion transport through all surface elements to zero. 

The potential of the double layer at rest is termed $\psi^0$ , the potential at electrophoresis is $(\psi^0 + \delta\psi)$ , where thus in $\delta\psi$ the combined effects of the applied field and the double layer distortion are represented. The electric field is then in vector form given by $-\nabla(\psi^0 + \delta\psi)$ , the force on a positive $z_+$ or negative $z_-$ charged ion is $-z_+ e \nabla (\psi^0 + \delta\psi)$ and $+z_- e \nabla (\psi^0 + \delta\psi)$ respectively. The velocity that an ion achieves under the influence of this force is $-\lambda_+z_+ e \nabla (\psi^0 + \delta\psi)$ and $+\lambda_-z_- e \nabla (\psi^0 + \delta\psi)$  respectively. $1/\lambda_+$ and $1/\lambda_- $ are the friction coefficients of the $+$ and $-$ ions and indicate the force necessary to give an ion a velocity of 1m/s. The number of ions, that passes through a surface element $d\omega$ is determined by the number of ions/m$^3$ $\times$ surface $\times$ velocity component normal to that surface, thus by $-n_+\lambda_+z_+ e \nabla (\psi^0 + \delta\psi)$ and $+n_-\lambda_-z_- e \nabla (\psi^0 + \delta\psi)$ respectively\footnote{The symbol $\nabla \psi^0$ represents a vector, which indicates the direction in which $\psi^0$ increases the largest in which amplitude is proportional to this increase per unit length. Nothing else is stated here than the fact that the positive ions are moving to the location of lowest potential at the velocity which is larger when their charge $(ze)$ in the potential difference $(-\nabla \psi^0)$ is larger and the friction $1/\lambda$ is smaller. The negative ions are moving to locations of higher potential. In general $\nabla (\psi^0 + \delta\psi)$ and the area element $d\omega$ are not perpendicular to each other. This will automatically be taken into account by the assuming the following term $\nabla (\psi^0 + \delta\psi) \cdot d\omega$ as an inner product.}.

The fluid motion will also drag the ions with it, such that each ion obtains the velocity of the fluid element in which it is situated. This velocity will be called $u$. Through the surface element $d\omega$ will be transported $n_+ u \cdot d\omega$ positive and $n_- u \cdot d\omega$ negative ions. 

Finally the ions are moving according to the diffusion tendency (Brownian motion) to places of higher to places of lower concentration. According to Fick's law\footnote{See for example K.JELLINEK, Lehrbuch der physik. Chem. II, 609, Enke, Stuttgart, 1928.}, $-D_+ \nabla n_+ \cdot d\omega$ positive and $-D_- \nabla n_- \cdot d\omega$ negative ions pass through a surface element $d\omega$ per unit time. The diffusion coefficients $D_+$ and $D_-$ are according to a theorem of Einstein\footnote{See for example K.JELLINEK, Lehrbuch der physik. Chem. V, 57, Enke, Stuttgart, 1928.} equal to $D_+ = kT\lambda_+ $ and $D_- = kT\lambda_- $. By summing these contributions, we find for the transport of respectively, positive and negative ions, through a surface element $d\omega$
\begin{subequations}
\begin{align} \nonumber \tag{5}
    \BS t_+ \cdot d\BS \omega &= \Big( -\lambda_+ n_+ z_+ e \; \nabla (\psi^0+\delta\psi) - \lambda_+ kT \nabla n_+ + n_+ \BS u \Big) \cdot d\BS \omega \\
    \nonumber
    \BS t_- \cdot d\BS \omega &= \Big( +\lambda_- n_- z_- e \; \nabla (\psi^0+\delta\psi) - \lambda_- kT \nabla n_- + n_- \BS u \Big) \cdot d\BS \omega
\end{align}
\end{subequations}
We now must, for the various elements $d\omega$ that surround a volume element, calculate the contributions $\BS t_+$ and $\BS t_-$ respectively and impose that $\sum \BS t_+ \cdot d\BS \omega=0$ and $\sum \BS t_- \cdot d\BS \omega=0$. This is satisfied if $\nabla \cdot \BS t = 0$, thus if\footnote{In order to see this, consider a parallelopiped with sides $dx$, $dy$ and $dz$ (Fig. \ref{fig:6}. Through side I, per unit of time an amount $t_x dx dz$ enters the volume with $t_x$ the $x$-component of vector $t$. Through side II per unit time the amount $(t_x)_{x+dx}dy dz = (t_x + \frac{\partial t_x}{\partial x} dx)dy dz$ leaves. The difference is $(t_x)_{x+dx} dy dz - t_x dy dz =\frac{ \partial t_x}{\partial x} dx dy dz$. For the sides $dz dx$ and $dy dx$ analogue formulas are valid, such that in total an amount $(\frac{\partial t_x}{\partial x}+\frac{\partial t_y}{\partial y} + \frac{\partial t_z}{\partial z})dx dy dz$ enters the volume per unit of time. If this is zero then $\nabla \cdot t$=0, when $\nabla \cdot t = \frac{\partial t_x}{\partial x} + \frac{\partial t_y}{\partial y} + \frac{\partial t_z}{\partial z}$}
\begin{subequations}
\begin{align}\nonumber \tag{6}
    \nabla \cdot \Big( -\lambda_+ n_+ z_+ e \; \nabla (\psi^0+\delta\psi) - \lambda_+ kT \nabla n_+ + n_+ \BS u \Big) &= 0 \\
    \nonumber
    \nabla \cdot \Big( +\lambda_- n_- z_- e \; \nabla (\psi^0+\delta\psi) - \lambda_- kT \nabla n_- + n_- \BS u \Big) &= 0
\end{align}
\end{subequations}
These are thus the fundamental equations for the ion movement. They can still be written somewhat differently, by splitting the ion concentrations $n_+$ \& $n_-$ in concentrations, $n_+^0$ \& $n_-^0$, as they are in the unperturbed equilibrium state, and the changes $\delta n_+$ \& $\delta n_-$, which occur during motion
\begin{equation} \nonumber \tag{7}
    n_+ = n_+^0 + \delta n_+ \; \quad \text{and} \quad n_- = n_-^0 + \delta n_- \; 
\end{equation}
(6) can now be written as\footnote{From now onwards we will often combine the equations for both ion species in one equation, in which the upper sign represents the cations and the lower sign the anions.}
\begin{align} \nonumber \tag{6a}
    \nabla \cdot \Big( \mp\lambda_{\pm}  n_{\pm}^0 z_{\pm} e \; \nabla \psi^0 \; &\mp \; \lambda_{\pm} \delta n_{\pm} z_{\pm} e \; \nabla \psi^0   
     \; \mp \; \lambda_{\pm} n_{\pm}^0 z_{\pm} e \; \nabla \delta\psi \; \mp \; \lambda_{\pm} \delta n_{\pm} z_{\pm} e \; \nabla \delta\psi \\  \nonumber
    & - \lambda_{\pm} kT \nabla n_{\pm}^0 - \lambda_{\pm} kT \nabla \delta n_{\pm}
        + n_{\pm}^0 u  + \delta n_{\pm} \BS u  \Big) = 0
\end{align}
Since in the unperturbed state, $\delta\psi$, $u$ as well as $\delta n_\pm$ are all zero, we can write for the double layer at rest
\begin{equation} \nonumber \tag{8}
    \nabla \cdot \Big( \mp n_{\pm}^0 z_{\pm} e \; \nabla \psi^0 - kT \nabla n_{\pm}^0 \Big) = 0
\end{equation}
(8) will be satisfied if\footnote{$n_+^\infty$ and $n_-^\infty$ are the ion concentrations at large distances from the particle.}
\begin{equation} \nonumber \tag{9}
    n_{\pm}^0 = n_{\pm}^\infty \exp(\mp z_{\pm} e \psi^0/kT) 
\end{equation}
From which it follows, that in the unperturbed state the ion concentration is given according to Boltzmann\footnote{Actually it is precisely the possibility to derive the Boltzmann distribution (9) from equation (8) that the \BLUE{(just before Eq.(5))} mentioned theorem of Einstein, $D_\pm = \lambda_{\pm} kT \; $, is proven.}.
Except for the "equilibrium terms", whose sum is zero, (6a) also contains terms, which are proportional to $\delta\psi$, $\delta n_\pm$ or $u$, thus with the field $E$ , and terms, which contain products of $\delta\psi$, $\delta n_\pm$ and $u$. These last terms, which are thus proportional to the square of the electric field, can be neglected, as long as the electrophoretic velocity is proportional to the field. (6a) then transforms into 
\begin{equation} \nonumber \tag{10}
  \nabla \cdot \Big(\mp \lambda_\pm n_\pm^0 z_\pm e \nabla \delta\psi \mp  \lambda_\pm \delta n_\pm z_\pm e \nabla \psi^0 -kT \lambda_\pm \nabla \delta n_\pm + \delta n_{\pm}^0 \BS u \Big) = 0 \;  
\end{equation}
The potential distribution and the ion concentrations should not only satisfy (6) and the relationships derived from it ((9) and (10)), but also satisfy Poisson's law\footnote{See for example A.HAAS, Einf\"uhrung in die theoretische Physik I, \textbf{176} Leipzig 1930.}, which relates the potential and the charge density $\rho$. In our case the charge density is
\begin{equation}  \nonumber
 \rho = e(n_+ z_+ - n_- z_-) . 
\end{equation}
Thus according to Poisson\footnote{$\Delta$ is the Laplace operator, equal to $\frac{d^2}{dx^2} + \frac{d^2}{dy^2} + \frac{d^2}{dz^2}$ in Cartesian coordinates.} 
\begin{equation} \nonumber \tag{11}
       \nabla^2 (\psi^0 + \delta\psi) = -\frac{e}{\epsilon_0\epsilon_r} (n_+ z_+ - n_- z_-) 
\end{equation}
In the equilibrium state
\begin{equation} \nonumber \tag{12}
    \nabla^2 \psi^0 = -\frac{e}{\epsilon_0\epsilon_r} (n_+^0 z_+ - n_-^0 z_-)
\end{equation}
By subtracting (11) and (12) from each other, we find a relationship between $\delta\psi$, $\delta n_+$ and $\delta n_-$  .
\begin{equation} \nonumber \tag{13}
    \nabla^2 \delta\psi = -\frac{e}{\epsilon_0\epsilon_r} (\delta n_+ z_+ - \delta n_- z_-) 
\end{equation}
Equation (12) and the two equations (9) consist of three differential equations, from which the quantities $\psi^0$, $n_+^0$ and $n_-^0$can be determined as function of location, provided there are enough constraints to determine the constants, that appear in the solution of the differential equations. Also, from (13), the two equations (10) and the relationship $\nabla^2 \delta \psi_i = 0$with ($\psi_i^0 + \delta\psi_i$)being the potential inside the sphere, the four quantities $\delta n_+$, $\delta n_-$, $\delta\psi$ and $\delta\psi_i$can be determined\footnote{The value of $u$, which is at the moment also unknown, will be determined in the hydrodynamic considerations of \BLUE{(the last page of chapter II)}.}.

The quantity $\nabla^2 \delta\psi$from (13) is for our considerations very important, since the relaxation effect relies on the fact that in the double layer the additional charge density, which is proportional to $\nabla^2 \delta\psi$is not everywhere equal to zero. 

The boundary conditions for the equations (9) and (12) are:
\begin{enumerate}
    \item at a large distance from the sphere $\psi^0$approaches zero; $n_+^0$ and $n_-^0$approach the values $n_+^\infty$ and $n_-^\infty$there, which are the equilibrium concentrations.
    \item On the boundary of the sphere and the fluid the potential equals $\psi^0=\zeta$, (alternatively some information about the charge on the sphere can be provided here).
\end{enumerate}

The boundary conditions for equations (10), (13) and $\nabla^2 \delta\psi_i = 0$are:
\begin{itemize}
    \item \textbf{1. and 2.} At large distance from the sphere $\delta n_+$ and $\delta n_-$approach zero.
    \item \textbf{3.} at large distance from the sphere the electric field is equal to $E$.
    \item \textbf{4.} The components of the field parallel to the boundary of the sphere-fluid are equal at both sides of the boundary.
    \item \textbf{5.} The components of the field perpendicular to this boundary $E^{liq}$ and $E^{sph}$ are related to the surface charge density of the sphere by\footnote{The additional charge density of the movement does not appear in this boundary condition, since it is not considered as a surface charge but as a volume charge and is already accounted for in the potential distribution outside the sphere.}

\begin{equation}
\nonumber
\epsilon_0\epsilon_r^{liq} E^{liq} = \epsilon_0\epsilon_r^{sph} E^{sph} + \frac{Q}{4\pi a^2}
\end{equation}
     \item \textbf{6.} There cannot be any discontinuities in the field, except at the boundary of the two media. The potential cannot be infinite inside the sphere.
     \item \textbf{7. and 8.} In the stationary state, no ions nor any other electricity bearing objects can accumulate at the boundary of the sphere and the fluid. For an insulator this means that the radial components of the vectors $t_+$ and $t_-$ from equation (5) must both be zero at the boundary of the sphere. On a conducting sphere $(\BS  t_+ z_+ e - \BS t_- z_- e)$must be equal to the electrical current density inside the sphere. Moreover in that case more information must be given regarding the discharging mechanism at the boundary, for example the electricity transfer from the fluid to the sphere happens by a discharge of one of the two sorts of ions. For more details we refer to chapter VI.
\end{itemize}
In total there are now thus 8 boundary conditions, exactly enough to determine the 8 constants, that appear in the solution of the four differential equations of the 2nd order (10), (13) and $\nabla^2 \delta\psi_i = 0$. 

Hereby the whole problem of the ion concentrations is determined, and we can now proceed to investigate the hydrodynamic part of the problem.

\subsection*{B - THE FLUID MOTION AND THE FORCES ON THE SPHERE}

Our treatment of the fluid motion and the by this motion exerted force on the sphere completely connects to the treatments of H\"uckel and Henry for the electrophoresis effect. without much effort these considerations can be made much more general, such that they don't only take into account the originally applied field (H\"uckel), and the distortion of the field by the difference in electric conductivity of the sphere and the fluid (Henry), but moreover also the under A discussed distortion of the field due to the relaxation effect (the interaction between the relaxation effect  and the electrophoretic friction). 

We will assume the following two hydrodynamic equations\footnote{See for example L.HOPF, Handbuch der Physik VII, 91, Springer, Berlin 1927.}. 
\begin{subequations}
\begin{align} \nonumber \tag{14a}
    \eta \nabla \times \nabla \times \BS u  + \nabla p + \rho \nabla(\psi^0 + \delta\psi) = -\rho_m \frac{D \BS u}{D t}&  \\
    \nabla \cdot \BS u = 0 \nonumber \tag{14b} \qquad \qquad \qquad \qquad &
\end{align}
\end{subequations}
Equation (14a) expresses that the sum of all forces, acting on a volume element, is equal to the mass of that volume element $\rho_m$, multiplied by its acceleration ($D \BS u/D t$).

\ifFIG\begin{figure}
    \centering
    \includegraphics[width=0.45\textwidth]{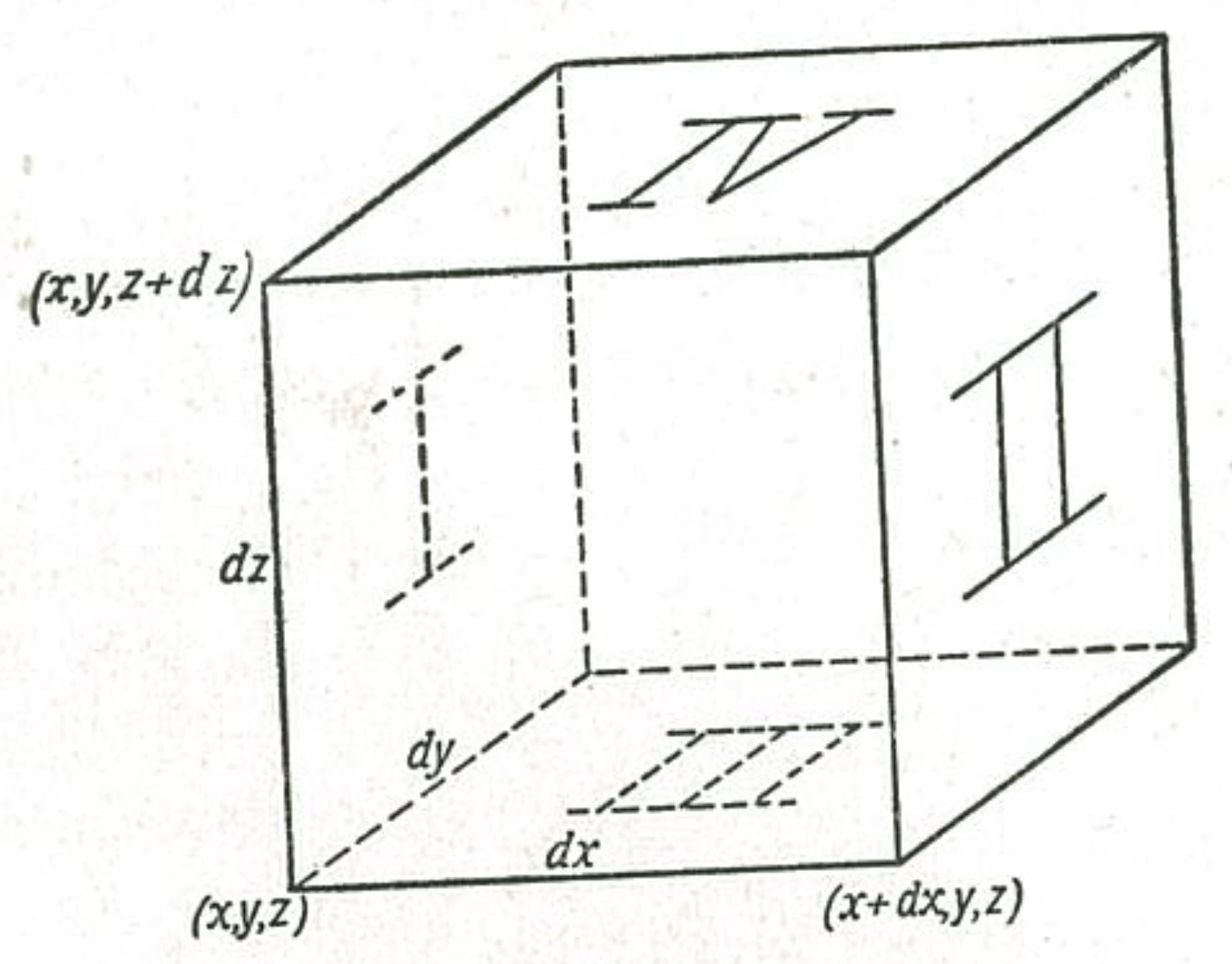}
    \caption{\BLUE{Parallelopiped used to explain the divergence theorem}.}
    \label{fig:6}
\end{figure} \fi
$\eta \nabla \times \nabla \times \BS u$ is the force, that acts on a volume element by the viscous friction of the surrounding fluid\footnote{Without going into the general form of the derivation of this formula, we shall calculate the magnitude of this friction force for a special case. Consider a parallelopiped with corners $dx$, $dy$ and $dz$ in a fluid flowing in the x-direction with a velocity which only depends on the z-coordinate (see Fig.6). The friction force in the x-direction which the fluid under surface III exerts on the liquid in the parallelopiped, is given by the viscosity $\times$ area $\times$ velocity-drop, thus $k_1 = \eta dx dy \frac{\partial u}{\partial z}$. The liquid above plane IV exhibits a force on the liquid below it $k_2=+\eta dx dy \frac{\partial u}{\partial z}_{z+dz}=\eta dx dy \Big(\frac{\partial u}{\partial z} + \frac{\partial^2u}{\partial z^2} dz \Big)$. In the case considered here no other forces are acting on the fluid element, such that the total friction force is directed in the $x$-direction and is equal to $k_1+k_2=\eta dx dy dz \frac{\partial^2 u}{\partial z^2}$. The vector $\nabla \times \nabla \times u $ has in this special case also only a $x$-component of magnitude $\frac{\partial^2 u}{\partial z^2}$, with which thus the formula for the friction force is verified for this special case.} $\nabla p$ is the force which drives the volume element from higher to lower pressure, while $\rho \nabla(\psi^0 + \delta\psi)$is the electric force, applied by the potentials $\psi^0$ and $\delta\psi$on the ions of the volume element. This force can also be considered as a force, which is acting on the volume element itself as can be shown as follows. On the ions, the field is applying a force $\rho \nabla(\psi^0 + \delta\psi)$. The ions achieve after a short period a uniform motion, because they are decelerated by the surrounding liquid. This friction force is then equal to $-\rho \nabla(\psi^0 + \delta\psi)$and, because action=reaction, the fluid experiences a force from the ions of $\rho \nabla(\psi^0 + \delta\psi)$\footnote{H.A.LORENTZ has shown in general that the state of motion of a fluid, caused by forces which act on a certain volume element $V$ is independent of the distribution of these forces over $V$ (except in the immediate vicinity of location of application of those forces). Abhandlungen \"uber theoretische Physik. Teubner Leipzig, Berlin (1906), Volume I, 27.}.

Equation (14b) shows that the amount of fluid, which enters a volume element per unit time, is equal to the amount of fluid exiting. In other words, the fluid is incompressible. 

The right hand side of (14a) can be neglected, since we are dealing with a stationary state and the velocities that occur are sufficiently small: the right hand side, valid for the acceleration of a \textit{fluid} element, can also be written such that it describes the velocity change in a certain point (\textit{volume} element) and then it is written as $-\rho_m (\partial \BS u/\partial t + \BS u \cdot \nabla \BS u)$. The first term $-\rho_m (\partial \BS u/\partial t)$is zero for a stationary state and the second term $ \BS\rho_m  (\BS u \cdot \nabla \BS u)$is proportional to the square of the electric field and can thus be neglected. 

(14a) is a collection of three equations, which are valid for each of the three components of the vector $\eta \nabla \times \nabla \times \BS u  + \nabla p + \rho \nabla(\psi^0 + \delta\psi)= -\rho_m \frac{D \BS u}{Dt}$. These are three differential equations of the second order. In addition to this we will have one more differential equation of order one, namely $\nabla \cdot \BS u=0$. In total there will thus be 7 integration constants while solving (14a and b), such that we also need 7 boundary conditions to determine these. These are: \begin{itemize}
    \item \textbf{1.} At large distance from the sphere the velocity in the $x$-direction is equal to $-U$
    \item \textbf{2. and 3.} At large distance from the sphere the velocity components in the $y$- and $z$-direction are zero.
    \item \textbf{4.,5. and 6.} At the boundary of the sphere and the fluid the fluid velocity is zero.
    \item \textbf{7.} At large distance the hydrostatic pressure $p$ is constant.
\end{itemize}

Together with \BLUE{(the boundary conditions just before chapter II.B)}, these conditions now enable us to solve completely the differential equations (10), (13), $\nabla^2 \delta\psi_i = 0$, (14a) and (14b). 

From the amplitude of the pressure and the fluid velocity in the neighbourhood of the boundary of the sphere and the fluid, the force can be calculated that the fluid is exerting on the sphere. 

The electric force on the sphere is equal to the sum of the forces exerted by the applied field and the distorted double layer on the charge of the sphere
\begin{equation}  \nonumber
   QE + \int \frac{\rho Q}{4\pi \epsilon_0\epsilon_r r^2} dv.   
\end{equation}
By setting the sum of all forces on the sphere to zero, we can find the sought after relation for $U/E$, since the hydrodynamic force is proportional to $U$ and the electric force proportional to $E$.

\newpage
\section{CALCULATIONS BY OTHER RESEARCHERS}

Now that we have established, according to our views, how the electrophoretic velocity of colloids should be calculated, we will give an overview of the methods given by other researchers in this area and the results obtained by them. 

The work of Henry can be considered as a summary of all calculations, which include electrophoretic drag, but exclude the distortion of the double layer.

In this chapter we will in particular look at other researches concerning the relaxation effect. The by various author used methods can all be brought back to the method as described in chapter II, where a number of allowed or not allowed assumptions or extensions (Onsager) are applied. We will therefore in the discussion that follows shine some light on the differences between these methods and our own work. 

Debye and H\"uckel\footnote{F.DEBYE and E.H\"UCKEL, Physik. Z. \textbf{24}, 305 (1923).} calculated the relaxation effect for ions, i.e. particles with a very \textit{small} $\kappa a$.

In the calculation of the influence of the electric field on the ion distribution (the term \\$ \mp \lambda_\pm n_\pm^0 z_\pm e \nabla \delta\phi$  from (10)), they replace the ion concentrations by the equilibrium concentrations, $n_\pm^0$ which is, at least for asymmetrical electrolytes, not allowed. See also chapter VI, \BLUE{(the discussion just after (60))} and this chapter . 

They formulate their boundary conditions differently from what we did in chapter II. It can however be shown, that their formulation and ours lead to the same result for small $\kappa a$.

The boundary conditions of Debye and H\"uckel can however not be applied if the particle radius is not small, while our boundary conditions are valid for any value of $a$.

Debye and H\"uckel  consider the relaxation effect and the electrophoretic drag as two different small corrections and did thus not take into account the interaction between these two effects. The fact that this is allowed for small $\kappa a$ and only for small $\kappa a$, will be further highlighted in chapter V . 

Their calculation is only valid for, as is ours, small $\zeta$ potentials. They can however defend the application on ions (particles) with larger potentials, by pointing at the fact that the principal contribution to the relaxation effect is delivered by the most outer part of the ionic atmosphere\footnote{\BLUE{In the aforementioned book chapter by Overbeek and Wiersema the ionic atmosphere is defined as the region in the vicinity of a particle, where the ions are distributed unequally, resulting in a net charge opposite in sign to that of the particle.}} (double layer) where indeed the potential is small. 

They find for the velocity of a cation the formula
\begin{equation} \nonumber
     \frac{U}{E}= \lambda_+ z_+ e \Big( 1 - \kappa a - \frac{(z_+/\lambda_+  + z_-/\lambda_-)}{(z_+ + z_-) e}  \frac{\lambda_+ z_+^2 e^2}{24\pi \epsilon_0\epsilon_r kT} \kappa \Big)
\end{equation}
or, if the charge of the central ion $z_+ \varepsilon$ is expressed in terms of $\zeta$
\begin{equation}  \nonumber
  \zeta = \frac{z_+ e}{4\pi\epsilon_0\epsilon_r a}(1-\kappa a) 
\end{equation}
and the friction coefficient $1/\lambda_+$ is set equal to $6 \pi \eta a$
\begin{equation}  \nonumber
     \frac{U}{E }=\frac{2\epsilon_0\epsilon_r \zeta}{3 \eta} \Big[1 - \frac{(z_+ /\lambda_+ + z_- /\lambda_-)}{(z_+ + z_-) e} \frac{2\epsilon_0\epsilon_rkT}{3 \pi \eta e} \Big(\frac{e \zeta}{kT} \Big)^2 \frac{\kappa a}{6} \Big] 
\end{equation}
Onsager finds an electrophoresis formula of the same shape as the Debye and H\"uckel one, but with a different coefficient for the relaxation effect (last term in between brackets). 

As for the boundary conditions, magnitude of $\kappa a$ and $\zeta$, his calculation is entirely consistent with that of Debye and H\"uckel. 

He takes the influence of the electric field on the ion distribution (the term $\mp n_\pm z_\pm e\lambda_\pm \nabla \delta\phi$ from (10)) better into account than Debye and H\"uckel. 

Moreover, he takes into account the Brownian motion of the central particle, while Debye and H\"uckel (as we did) have prescribed a linear path for this particle. 

Both of these differences express themselves in the electrophoretic formula, which shows itself clearest when a formula of Debye and H\"uckel with our formula, where the term $\mp n_\pm z_\pm e\lambda_\pm \nabla \delta\phi$ 
is represented correctly, and one from Onsager, where moreover the Brownian motion of the central particle is taken into account. 

According to Onsager\footnote{L.ONSAGER, Physik. Z., \textbf{28}, 277 (1927).} the electrophoretic velocity of a positive ion is
\begin{equation}  \nonumber
\frac{U}{E}=\frac{2\epsilon_0\epsilon_r \zeta}{3 \eta} \Big[1 - \frac{z_- }{\lambda_+z_+ e} \frac{2q}{1+\sqrt{q}}\frac{2\epsilon_0\epsilon_rkT}{3 \eta e} \Big(\frac{e \zeta}{kT} \Big)^2 \frac{\kappa a}{6} \Big].
\end{equation}
where
\begin{equation} \nonumber
 q=\frac{z_+ /\lambda_- + z_- /\lambda_+}{(z_+ + z_-) (1/\lambda_+ + 1/\lambda_-)} \qquad 
.
\end{equation}

Our own formula for very small $\kappa a$ (89a) is
\begin{equation}   \nonumber
\frac{U}{E}=\frac{2\epsilon_0\epsilon_r \zeta}{3 \eta} \Big[1 - \frac{z_- }{\lambda_+ z_+ e}\frac{z_+ /\lambda_- + z_- /\lambda_+}{(z_+ + z_-)/ \lambda_+} \frac{2\epsilon_0\epsilon_rkT}{3 \eta e} \Big(\frac{e \zeta}{kT} \Big)^2 \frac{\kappa a}{6} \Big].
\end{equation}
For symmetrical electrolytes ($z_+ = z_-)$ this formula is identical to that of Debye and H\"uckel, for non symmetrical electrolytes these formulas are different and our formula is closer to that of Onsager. 

Hermans\footnote{J.J.HERMANS, Phil. Mag. (7) \textbf{26}, 650 (1938).} gives a calculation of the electrophoretic velocity of a non conducting sphere for large $\kappa a$. 

He neglects the term $\mp \delta n_\pm z_\pm e\lambda_\pm \nabla \delta\phi$, which is allowed for small $\kappa a$, but not for large $\kappa a$.

Instead of our boundary conditions 7 and 8 concerning the ion transport through the boundary of the sphere and the fluid, Hermans demands that at this boundary the additional charge density should be zero. He motivates this by pointing out that at the boundary of the particle the radial components of the fluid velocity and the electric field (insulating particle!) become zero and that consequently no charge can accumulate there. He did not realize, that this accumulation can actually take place before the stationary state has been reached, and is kept this way afterwards, precisely since the boundary conditions 7 and 8 are satisfied. 

Furthermore he does not take into account the interaction between the electrophoretic drag and the relaxation effect.

In chapter V \BLUE{(discussion after Equation (89bII))} it will be shown that especially due to this last neglect, the relaxation correction of Hermans is a factor on the order of $\kappa a$ too large. 

His electrophoresis formula is
\begin{equation}   \nonumber
\frac{U}{E}=\frac{\epsilon_0\epsilon_r \zeta}{\eta} \Big[1 - (z_- - z_+) \frac{1}{2}\Big(\frac{e \zeta}{kT}\Big)- \frac{z_+ /\lambda_+ + z_-/ \lambda_-}{(z_+ + z_-) e} \frac{2\epsilon_0\epsilon_rkT}{3 \eta e} \frac{5}{12}\Big(\frac{e \zeta}{kT} \Big)^2  \Big].
\end{equation}
while our formula for large $\kappa a$ approaches to (89b)\footnote{\BLUE{This formula does not appear to be in the thesis, the closest one is (89bII), however it does not contain a term with $(z_+^2-\frac{2z_+z_-}{5}+z_-^2)$ (in fact none of the formulas contains such a term...).}}
\begin{align}   \nonumber
\frac{U}{E} =  \frac{\epsilon_0\epsilon_r \zeta}{\eta} \Big[1 &- \frac{3}{\kappa a} -  (z_- - z_+) \frac{3}{4 \kappa a}\Big(\frac{e \zeta}{kT}\Big) \\ \nonumber
& - \frac{z_+ /\lambda_+ + z_- /\lambda_-}{(z_+ + z_-) e} \frac{2\epsilon_0\epsilon_rkT}{3 \eta e} \frac{3}{4 \kappa a}\Big(\frac{e \zeta}{kT} \Big)^2
 - \Big(z_+^2 - \frac{2 z_+ z_-}{5} + z_-^2 \Big) \frac{5}{24 \kappa a} \Big(\frac{e \zeta}{kT} \Big)^2 \Big].
\end{align}
The calculation of Komagata\footnote{S.KOMAGATA, Res. of the Electrotechnical Laboratory Tokyo no. \textbf{387} (1935).} completely complements the method followed in this thesis, which among other things can be observed from the fact that he takes into account the interaction of the relaxation and the electrophoretic drag explicitly. Against the execution of his calculations a number of serious objections can be made, which makes his work virtually worthless. The objections being:\begin{enumerate}
    \item The way, in which he accounts for the influence of the fluid flow, is only justified for small $\kappa a$
    \item In the boundary conditions for the boundary of the sphere and the fluid he only takes into account the dielectric constant (our conditions 4, 5 and 6), but not the conductivity and the ion transport (conditions 7 and 8). 
    \item Finally he calculates the force on the particle as the sum of $k_1$, $k_2$ and $k_3$ and he forgets $k_4$, the force which the distorted double layer exerts on the particle. While he thus takes into account the influence of the relaxation effect on the electrophoretic drag, the relaxation force itself is neglected. How bad this all is, is clear when one applies his final formula for the case of small $\kappa a$, where crazy large \textit{accelerations} of the electrophoresis are being found.
\end{enumerate}

Mooney\footnote{M.MOONEY, J. Phys. Chem. \textbf{35}, 331 (1931).} has formulated the problem of electrophoretic velocity very well and has also attempted to solve it in general. He encountered insurmountably mathematical problems and had to be satisfied with a formula, which is valid for large $\kappa a$, while there is no limitation on the value of $\zeta$. Unfortunately he does not indicate, how his formula is deduced, such that it is difficult to judge its value.  An error must have crept in his formulation, since not all terms of his final formula have the same dimension. He finds for the electrophoretic velocity of an insulating sphere at large $\kappa a$ 
\begin{equation} \nonumber
\frac{U}{E}=\frac{\epsilon_0\epsilon_r \zeta}{\eta} \Big[ 1 - f(\zeta) \frac{1}{\kappa a} \Big] .
\end{equation}
Bikerman\footnote{J.J.BIKERMAN, Z. physik. Chem. \textbf{A 171}, 209 (1934).} assumes, to take into account the deformation of the double layer, that in general the double layer has a larger conductivity than the fluid. He considers a long cylinder shaped insulating particle with the axis of the cylinder parallel with the field direction. The presence of the additional surface conductivity causes in the double layer a counter force, which counteracts the original electric field. The particle is then subjected to a weaker field, such that the electrophoretic velocity is being reduced. His consideration does indeed take into account the influence of the electric field on the ion distribution $\mp (n_\pm z_\pm e/\lambda_\pm) \nabla \delta\psi$, but ignores the influence of diffusion and liquid flow entirely. 

The value he calculates for the electrophoretic velocity is
\begin{equation}  \nonumber
 \frac{U}{E}=\frac{\epsilon_0\epsilon_r \zeta}{\eta} \frac{(L-r)}{(L-r) + 2 \chi_s/\chi_f} \qquad \qquad .
\end{equation}

Here $L$ is the length and $r$ is the radius of the cylinder, $\chi_s$ is the surface conductivity, and $\chi_f$ is the conductivity of the fluid. Since the ratio of $\chi_s/\chi_f$ is getting larger, when the ion concentration of the fluid is getting smaller, he can explain a maximum in the $U-c$ curve, without having a maximum in the $\zeta -c$ curve at the same time.

Besides the objections, which can be brought against the neglect of the fluid flow and the diffusion, there is another big objection to the theory of Bikerman, which is that it is only valid for very long particles with a thin double layer and abandons us with the more common case of spherical or near spherical shaped particles. 

Finally in Table II we give once more a global overview of the conditions, where according to the authors, their formulas have validity.

\begin {table} [!ht] \label{Table2}
\caption*{TABLE II } \label{tab:title} 
\begin{tabular}{| l | l | l | l | l | l |}
  \hline
    & Relaxation & $\kappa a \hspace{2cm} $ & Shape & Conductivity& $\zeta$ \hspace{2cm} \\ 
  \hline
  Helmholtz & no  & large & any & 0 & any\\
  Smoluchowski & no  & large & any & 0 & any\\ 
  H\"uckel & no  & small & sph./cyl. & - & any\\ 
  Henry & no  & any & sph./cyl. & $0-\infty$& small\\ 
  Debye/H\"uckel & yes  & small & sphere & - & small\\ 
  Onsager & yes  & small & sphere & - & small\\ 
  Paine & yes  & small & sphere & - & small \\
  Mooney & yes  & large & sphere & 0 & any \\
  Bikerman & yes  & large & cylinder & 0 & any \\
  Komagata & yes  & any & sphere & 0 & small \\
  Hermans & yes  & large & sphere & 0 & small \\
  Current thesis & yes  & any & sphere & $0-\infty$ & small \\
  \hline  
\end{tabular}
\end{table}

\newpage
\section{EXECUTION OF THE CALCULATION}

In this chapter we will invert the order of the treatment of the ion-distribution and the fluid motion, because we will need some of the results of the hydrodynamic calculation to calculate the ion-distribution\footnote{The reader, who is not interested in the full derivation of the electrophoresis formula can skip chapter IV without any objection. }.
We will use the polar coordinates, r, $\theta$ and $\varphi$, where the center of the sphere is the origin of the coordinate system with the direction $\theta=0$ coinciding with the direction of the electric field and the flow at large distance from the sphere, see Fig.\ref{fig:7}.

\ifFIG\begin{figure}[!ht]
    \centering
    \includegraphics[width=0.45\textwidth]{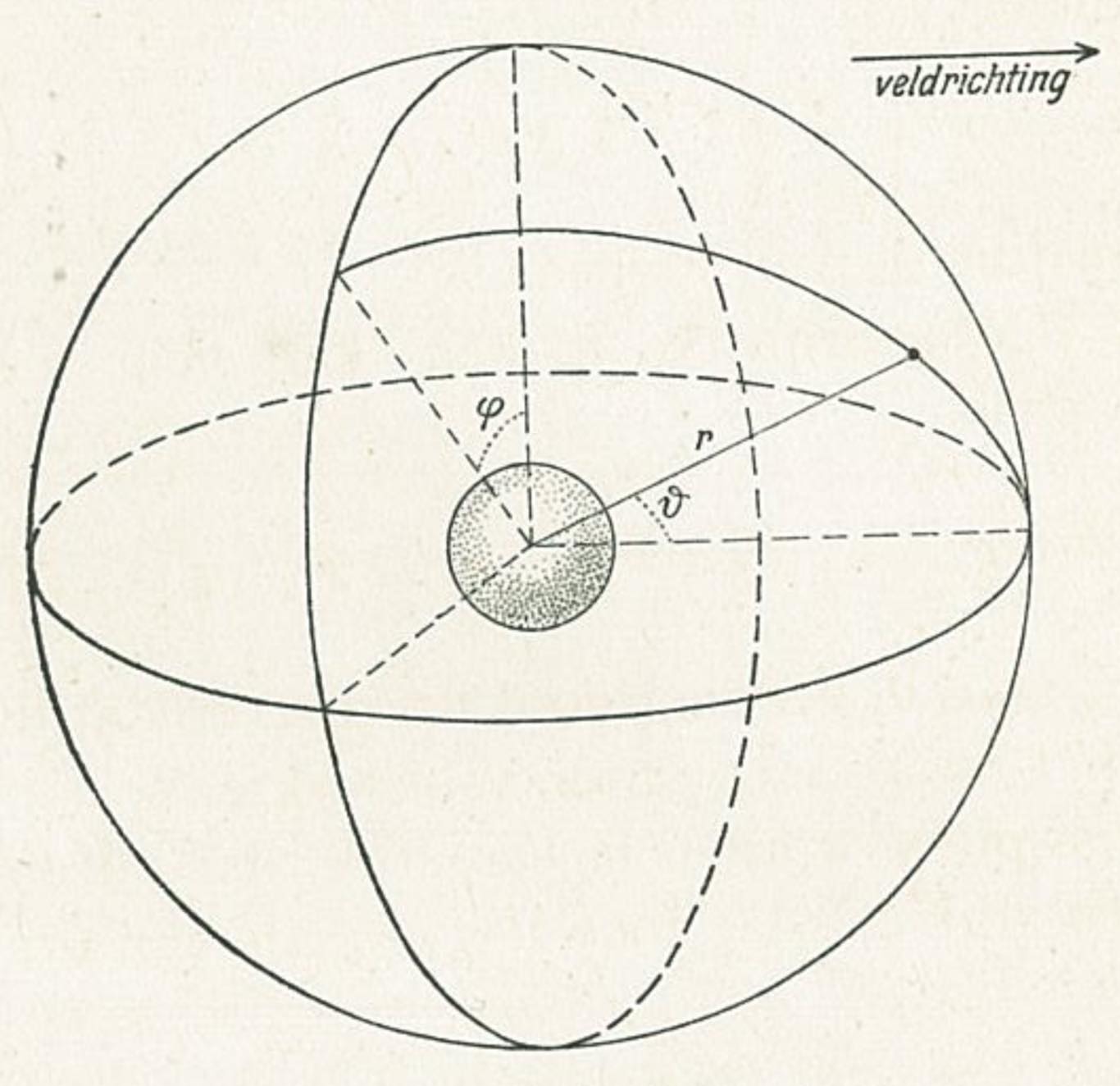}
    \caption{Spherical coordinates. \BLUE{"veldrichting" = direction of the field.}}
    \label{fig:7}
\end{figure} \fi

\subsection*{A - CALCULATION OF THE FLUID MOTION}

We assume (14a) and (14b) from chapter II are valid, where we have reasoned that $\rho_m D \BS u/Dt$ is equal to zero\footnote{\BLUE{Equations (14a) and (14b) appear twice in Overbeeks' thesis, once with the right hand side of (14a) non-zero (with inertia) and once with a zero right hand side (without inertial terms).}}
\begin{subequations} \nonumber
\begin{align} \nonumber \tag{14a}
\eta \nabla \times \nabla \times \BS u  + \nabla p + \rho \nabla(\psi^0 + \delta\psi)&= 0 \\  \nonumber  \tag{14b}
\nabla \cdot \BS u &= 0 
\end{align}
\end{subequations}
For the charge density $\rho$ we can write according to Poisson
\begin{equation} \nonumber \tag{15}
    \rho = - \epsilon_0\epsilon_r\nabla^2(\psi^0 + \delta\psi) = \rho^0 + \delta\rho
\end{equation}
Here $\psi^0$ and $\rho^0$ are functions of $r$ alone. For $\delta\psi$, which assumes at infinity the form:

\begin{equation} \nonumber \tag{16}
\delta\psi_\infty = - E \; r \cos\theta 
\end{equation}
we can assume

\begin{equation} \nonumber \tag{17}
\delta\psi = - E \; R(r) \cos\theta 
\end{equation}
where $R(r)$ is a function of $r$ alone.
Using (15), we can write for $\rho \nabla (\psi^0 + \delta\psi)$ 
\begin{align} \nonumber \tag{18}
\rho\nabla (\psi^0 + \delta\psi) = \rho^0 \nabla \psi^0 + \rho^0 \nabla \delta\psi + \delta\rho \nabla \psi^0 + \delta\rho \nabla \delta\psi
\end{align}
Here $\delta\rho \nabla \delta\psi$ can be neglected, since this term is proportional to the second power of the electric field $E$. 

Now we can split $p$ in a piece that only depends on $r$, \BLUE{$(p^0)$}, and a piece that depends on both $r$ and $\theta$\footnote{\BLUE{The radial part is now the "osmotic" pressure, $p^0$ the remainder has a $\cos\theta$ term due to the asymmetry of the front and back as far as the pressure is concerned.}}, by assuming:

\begin{equation} \nonumber \tag{19}
p = \delta p(r,\theta) + p^0(r) = \delta p- \int_\infty^r \rho^0 \Big(\frac{d\psi^0}{dx} \Big) dx
\end{equation}
Then (14a) transforms into

\begin{align} \nonumber \tag{20}
\eta \nabla \times \nabla \times \BS u  + \nabla \delta p + \delta\rho \nabla\psi^0 + \rho^0 \nabla \delta\psi= 0
\end{align}

From (20), $\BS u$ can be eliminated by applying the divergence operator on it. Since $\nabla \cdot \nabla \times =0$ and $ \nabla \cdot \nabla = \nabla^2 $, we then find
\begin{align} \nonumber 
\nabla^2 \delta p &= - \nabla \cdot (\delta\rho \nabla \psi^0 + \rho^0 \nabla \delta\psi) \\
\nonumber \tag{21}
 &= - \delta\rho \nabla^2 \psi^0 - \rho^0 \nabla^2 \delta\psi - (\nabla \delta\rho \cdot\nabla \psi^0) - (\nabla \rho^0 \cdot\nabla \delta\psi)
\end{align}
Take into consideration that the whole problem is symmetric in the $\theta=0$ direction, and thus\footnote{\BLUE{Note that the expression $d^2f/dr^2+2/r df/dr - 2f/r^2$ with $f$ a function of $r$ appears often later on. This 'operator' was called the $L$-operator by Ohshima, Healy and White, J. Chem. Soc. Faraday Trans. 2, 1983, 1613-1628 (their Eq.30) as $L(f)=\frac{d}{dr} \frac{1}{r^2} \frac{d}{dr} (r^2 f) = \frac{d^2f}{dr^2} + \frac{2}{r} \frac{df}{dr} - \frac{2}{r^2} f$.}}
\begin{align}  \nonumber
    \nabla^2 \delta\psi \equiv \frac{1}{r^2}\frac{\partial}{\partial r}\Big(r^2\frac{\partial \delta\psi}{\partial r}\Big) + \frac{1}{r^2 \sin\theta}\frac{\partial}{\partial \theta }\Big(\sin\theta \frac{\partial \delta\psi}{\partial \theta}\Big) 
    \nonumber \tag{22}
    = - E\cos\theta \Big( \frac{d^2R}{dr^2}  + \frac{2}{r}\frac{dR}{dr} - \frac{2R}{r^2}\Big) 
\end{align}
$(\nabla \delta\rho \cdot \nabla \psi^0)=\frac{\partial \delta\rho}{\partial r}\cdot \frac{d\psi^0}{dr}$ and $(\nabla \rho^0 \cdot \nabla \delta\psi)=\frac{d \rho^0}{d r}\cdot \frac{\partial\delta\psi}{\partial r}$ and substitute for $\delta\rho$ and $\rho^0$ the values from (15) then 
\begin{align} \nonumber
\nabla^2 \delta p = -\epsilon_0\epsilon_r E \cos\theta & \Big[\Big( 2\frac{d^2R}{dr^2} + \frac{4}{r}\frac{dR}{dr}-\frac{4R}{r^2}\Big) \nabla^2\psi^0 + \\ 
\nonumber \tag{23}
& \frac{d}{dr} \Big(\frac{d^2R}{dr^2} + \frac{2}{r}\frac{dR}{dr} - \frac{2R}{r^2}\Big)\frac{d\psi^0}{dr} + \frac{dR}{dr}\frac{d}{dr} \nabla^2\psi^0 \Big]
\end{align}
The solution of this differential equation can be written in the form

\begin{equation} \nonumber \tag{24}
\delta p = - \epsilon_0\epsilon_r E \cos \theta P_1(r)  ,
\end{equation}
where $P_1(r)$ is a function of $r$ only. Then
\begin{align}  \nonumber
    \nabla^2 \delta p \equiv \frac{1}{r^2}\frac{\partial}{\partial r}\Big(r^2\frac{\partial \delta p}{\partial r}\Big) + \frac{1}{r^2 \sin\theta}\frac{\partial}{\partial \theta }\Big(\sin\theta \frac{\partial \delta p}{\partial \theta}\Big) \nonumber
    = -\epsilon_0\epsilon_r E\cos\theta \Big( \frac{d^2P_1}{dr^2}  + \frac{2}{r}\frac{dP_1}{dr} - \frac{2P_1}{r^2}\Big) 
\end{align}
And thus 
\begin{align} \nonumber \tag{25}
\frac{d^2P_1}{dr^2} + \frac{2}{r}\frac{dP_1}{dr} & - \frac{2P_1}{r^2} =2 \Big( \frac{d^2R}{dr^2} + \frac{2}{r}\frac{dR}{dr}-\frac{2R}{r^2}\Big) \nabla^2\psi^0 + \\  \nonumber
& \frac{d}{dr} \Big(\frac{d^2R}{dr^2} + \frac{2}{r}\frac{dR}{dr} - \frac{2R}{r^2}\Big)\frac{d\psi^0}{dr} + \frac{dR}{dr}\frac{d}{dr} \nabla^2\psi^0 \equiv f_1(r)
\end{align}
The solution of this differential equation is given by
\begin{equation} \nonumber \tag{26}
P_1(r) = c_1r + \frac{c_2}{r^2} +r \int_\infty^r \frac{1}{x^4} \int_\infty^x  y^3 f_1(y) dy dx.
\end{equation}
For large $r$, $\delta p$ and thus also $P_1$ must approach 0 and thus $c_1$ must be zero. Some discussion on the value of $c_2$ will be  given \BLUE{before (42a)}. To satisfy (21), $c_2$ can assume any value. However, if it also has to satisfy (20) (and this is what it is all about), then $c_2$ will be partly determined by $\BS u$.
The integral
\begin{equation} \nonumber \tag{27}
\chi(r) \equiv r \int_\infty^r \frac{1}{x^4} \int_\infty^x  y^3 f_1(y) dy dx
\end{equation}
can be simplified by using $f_1(r)$ from (25), to perform the integrations noting that
\begin{equation} \nonumber
\nabla^2 \psi^0 =  \frac{1}{r^2}\frac{d}{dr} \Big(r^2 \frac{d\psi^0}{dr} \Big) 
  .
\end{equation}
We then get
\begin{subequations}
\begin{align}
\nonumber 
\chi=r\int_\infty^r \frac{1}{x^4}\int_\infty^x y^3 \Big[\frac{dR}{dy}\frac{d^3 \psi^0}{dy^3} +\Big( 2 \frac{d^2R}{dy^2} +\frac{6}{y}\frac{dR}{dy}-\frac{4R}{y^2}\Big) \frac{d^2 \psi^0}{dy^2}\\
\nonumber
+\Big(\frac{d^3R}{dy^3} + \frac{6}{y}\frac{d^2R}{dy^2} +\frac{2}{y^2}\frac{dR}{dy} - \frac{4R}{y^3}\Big) \frac{d\psi^0}{dy} \Big]dydx\\
\nonumber
=r\int_\infty^r \frac{1}{x^4} \int_\infty^x \frac{d}{dy}\Big[ y^3 \frac{dR}{dy} \frac{d^2 \psi^0}{dy^2}+y^3 \frac{d^2R}{dy^2} \frac{d\psi^0}{dy} +3y^2 \frac{dR}{dy}\frac{d\psi^0}{dy}-4yR \frac{d\psi^0}{dy}\Big] dy dx\\
\nonumber
=r\int_\infty^r\Big[ \frac{1}{x}\frac{dR}{dx}\frac{d^2\psi^0}{dx^2}+ \frac{1}{x}\frac{d^2R}{dx^2} \frac{d\psi^0}{dx} +\frac{3}{x^2} \frac{dR}{dx} \frac{d\psi^0}{dx}-\frac{4}{x^3}R \frac{d\psi^0}{dx}\Big] dx\\
\nonumber
=r\int_\infty^r \frac{d}{dx}\Big[ \frac{1}{x} \frac{dR}{dx}\frac{d\psi^0}{dx}\Big] dx + r \int_\infty^r \Big[ \frac{4}{x^2} \frac{dR}{dx} \frac{d\psi^0}{dx}-\frac{4}{x^3} R \frac{d\psi^0}{dx}\Big] dx = 
\end{align}
\end{subequations}
\begin{equation} \nonumber \tag{28}
    \chi = \frac{dR}{dr}\frac{d\psi^0}{dr} + 4r \int_\infty^r \Big(\frac{1}{x^2}\frac{dR}{dx}- \frac{R}{x^3} \Big)\frac{d\psi^0}{dx} dx
\end{equation}
The full solution for the pressure $p$, can be found by combining (19), (24), (26), (27) and (28) to give
\begin{align} \nonumber
p = \epsilon_0\epsilon_r & \int_\infty^r \nabla^2 \psi^0 \frac{d\psi^0}{dx} dx   
 \tag{29}
  - \epsilon_0\epsilon_r E \cos \theta\Big[\frac{c_2}{r^2} + \frac{dR}{dr} \frac{d\psi^0}{dr}+4r \int_\infty^r \Big(\frac{1}{x^2}\frac{dR}{dx}-\frac{R}{x^3} \Big)\frac{d\psi^0}{dx}dx\Big]
\end{align}
To determine the velocity distribution, we apply the curl operator \BLUE{$(\nabla \times )$} on (20), whereby $p$ will be eliminated, since curl(grad) equals zero.
\begin{align} \nonumber \tag{30}
\eta \nabla \times \nabla \times \nabla \times \BS u + \nabla \times (\delta\rho \nabla \psi^0 + \rho^0 \nabla \delta\psi) = 0. 
\end{align}
The expression for $\nabla \times \BS u$ in polar coordinates is\footnote{\BLUE{Overbeek seems to have defined $\nabla \times$ with a sign difference when compared to modern notation. The operator $\nabla \times \nabla \times$ will then have a plus sign again. In the appendix on vector identities (that we did not translate), the same equations can be found, apparently taken from \\
E.MADELUNG, Die mathematischen Hilfsmittel des Physikers, Springer Berlin 1936, page 145.}}
\begin{subequations}
\begin{align} \nonumber
(\nabla \times \BS u)_r &= \frac{1}{r\sin\theta} \Big[ -\frac{\partial}{\partial \theta}(u_\varphi \sin\theta) + \frac{\partial u_\theta}{\partial \varphi} \Big]\\
\nonumber \tag{31}
(\nabla \times \BS u)_\theta &= \frac{1}{r} \Big[- \frac{1}{\sin\theta}\frac{\partial u_r}{\partial \varphi} + \frac{\partial}{\partial r}(r u_\varphi) \Big]\\
\nonumber
(\nabla \times \BS u)_\varphi &= \frac{1}{r} \Big[ - \frac{\partial}{\partial r}(r u_\theta) + \frac{\partial u_r}{\partial \theta} \Big]
\end{align}
\end{subequations}
Due to symmetry considerations, $u_\varphi$ must be zero, thus $\nabla \times \BS u$ has no $r$ or $\theta$ component. 

We now define $\BS w$, a vector that only has a $\varphi$ component,:
\begin{align} \nonumber \tag{32}
(\BS w)_\varphi \equiv w = (\nabla \times \BS u)_\varphi = \frac{1}{r}\Big[- \frac{\partial}{\partial r} (ru_\theta) + \frac{\partial u_r}{\partial \theta} \Big].
\end{align}
so that $\nabla \times \BS w$ has a $r$ and a $\theta$ component, but no $\varphi$ component, and $\nabla \times \nabla \times \BS w$ has again only a $\varphi$ component, namely
\begin{align} \nonumber
(\nabla \times \nabla \times \BS w)_\varphi = -\frac{1}{r} \Big \{\frac{\partial^2(r w)}{\partial r^2} + \frac{1}{r} \frac{\partial}{\partial \theta}\Big[\frac{1}{\sin \theta} \frac{\partial}{\partial \theta} (w \sin \theta) \Big] \Big \}.
\end{align}
Also $\nabla \times (\delta\rho \nabla \psi^0 + \rho^0 \nabla \delta\psi)$ is a vector with only a $\varphi$ component, which is:
\begin{equation} \nonumber
    \big(\nabla \times (\delta\rho \nabla \psi^0 + \rho^0 \nabla \delta\psi)\big)_\varphi = \frac{1}{r}\Big( \frac{\partial \delta\rho}{\partial \theta}\frac{d\psi^0}{dr} - \frac{d\rho^0}{dr} \frac{\partial \delta\psi}{\partial \theta} \Big)
\end{equation}
Thus equation (30) now transforms into:
\begin{equation} \nonumber \tag{33}
    \frac{\eta}{r} \Big\{ \frac{\partial^2(rw)}{\partial r^2}+ \frac{1}{r} \frac{\partial}{\partial \theta} \Big[ \frac{1}{\sin \theta} \frac{\partial} {\partial \theta} (w \sin \theta) \Big] \Big\}=\frac{1}{r} \Big( \frac{\partial \delta\rho}{\partial \theta}\frac{d\psi^0}{dr} - \frac{d\rho^0}{dr} \frac{\partial \delta\psi}{\partial \theta} \Big)
\end{equation}
or after substitution of $\rho^0$, $\delta\rho$ and $\delta\psi$ according to (15), (17) and (22):

\begin{align} \nonumber
\frac{\partial^2(rw)}{\partial r^2}  + \frac{1}{r} \frac{\partial}{\partial \theta} \Big[ \frac{1}{\sin \theta} \frac{\partial} {\partial \theta} (w \sin \theta) \Big] =  \tag{33'}
 \frac{\epsilon_0\epsilon_r E \sin \theta}{\eta} \Big[ R \frac{d}{dr} \nabla^2 \psi^0 - \Big(\frac{d^2R}{dr^2} +\frac{2}{r} \frac{dR}{dr} - \frac{2R}{r^2} \Big) \frac{d\psi^0}{dr} \Big].
\end{align}
The solution of (33’) has the form
\begin{equation} \nonumber \tag{34}
    w = \frac{\epsilon_0\epsilon_r E \sin\theta}{\eta} W
\end{equation}
where $W$ is a function of $r$ alone. If $w$ is substituted into (33’) according to (34), then it turns out that $W$ must satisfy the following differential equation
\begin{equation} \nonumber \tag{35}
    \frac{d^2W}{dr^2}+ \frac{2}{r} \frac{dW}{dr} - \frac{2W}{r^2} = \frac{R}{r} \frac{d}{dr} \nabla^2 \psi^0 - \frac{1}{r}\Big(\frac{d^2R}{dr^2} +\frac{2}{r} \frac{dR}{dr} - \frac{2R}{r^2} \Big) \frac{d\psi^0}{dr} \equiv f_2(r),
\end{equation}
of which the solution is given by
\begin{equation} \nonumber \tag{36}
W = c_3 r + \frac{c_4}{r^2}+r\int_\infty^r \frac{1}{x^4} \int_\infty^x y^3 f_2(y) dy dx. 
\end{equation}
The integration constants $c_3$ and $c_4$ are, for now, undefined. We define $\xi(r)$ as:
\begin{equation} \nonumber \tag{37}
    \xi(r) \equiv r\int_\infty^r \frac{1}{x^4} \int_\infty^x y^3 \Big\{\frac{R}{y}\frac{d}{dy}\nabla^2 \psi^0 - \frac{1}{y}  \Big(\frac{d^2R}{dy^2}+\frac{2}{y}\frac{dR}{dy}-\frac{2R}{y^2} \Big)\frac{d\psi^0}{dy}\Big\} dy dx
\end{equation}
In our calculations this function fulfills the same role as the function with the same name in Henry. By using the relationship $\nabla^2 \psi^0 = \frac{1}{r^2} \frac{d}{dr} \Big( r^2 \frac{d\psi^0}{dr}\Big)$ and perform the integrations\footnote{\BLUE{This can easiest be done by partially integrating $\int_\infty^x y^3\Big\{\frac{R}{y} \frac{d}{dy} \nabla^2 \psi^0\Big\} dy$, by which many terms will cancel out with the terms in round brackets. }}, we can bring $\xi$ in the following form\footnote{\BLUE{The function $\xi$ from Henry can be recovered by setting $R=r+\lambda a^3/r^2$ and using $\nabla^2 \psi^0 = \frac{1}{r^2} \frac{d}{dr} \Big( r^2 \frac{d\psi^0}{dr}\Big)$. Henry's formula is $\xi=\frac{d\psi^0}{dr}+ \lambda a^3 r\int_\infty^r \frac{1}{x^4} \nabla^2 \psi^0 dx$ (page 112 of Henry). In fact, (40b) is identical to Henry's Eq.11, but with $c_3=c_4=0$ (which Overbeek also finds later).}}:
\begin{equation}  \nonumber \tag{37'}
\xi = \frac{R}{r}\frac{d\psi^0}{dr} - 2r \int_\infty^r \Big(\frac{1}{x^2}\frac{dR}{dx}-\frac{R}{x^3} \Big)\frac{d\psi^0}{dx}dx.
\quad
\end{equation} 
With the full solution for $w$:

\begin{equation} \nonumber \tag{38}
w = \frac{\epsilon_0\epsilon_r E \sin \theta}{\eta}\Big( c_3 r + \frac{c_4}{r^2} + \xi\Big)
\end{equation}
which determines thus $\nabla \times \BS u$  from (32) and (38) and $\nabla \cdot \BS u$ from (14b), while due to symmetry considerations it follows that $u_\varphi = 0$. 
Now it remains to solve $u_r$ and $u_\theta$ from
\begin{equation}  \nonumber 
\nabla \cdot \BS u \equiv \frac{1}{r^2} \frac{\partial}{\partial r} (r^2 u_r) + \frac{1}{r \sin \theta} \frac{\partial}{\partial \theta} (u_\theta \sin \theta)=0.
\end{equation}
\begin{align}  \nonumber
(\nabla \times \BS u)_\varphi & \equiv \frac{1}{r} \Big\{-\frac{\partial} {\partial r} (r u_\theta) + \frac{\partial u_r}{\partial \theta}\Big\} \\
\nonumber \tag{39}
& =\frac{\epsilon_0\epsilon_r E \sin \theta }{\eta} \Big( c_3 r +\frac{c_4}{r^2} + \xi \Big)
\end{align}
The solution of this equation can be written as
\begin{equation}
\nonumber
u_r = R_1(r) \; \cos \theta \qquad \text{and} \qquad u_\theta = R_2(r) \; \sin \theta
\end{equation}
where $R_1(r)$ and $R_2(r)$ are pure functions of $r$. Substituting this into (39), it turns out that $R_1(r)$ and $R_2(r)$ must satisfy the following differential equations\footnote{\BLUE{From (39) and the fact that $\nabla \times u$ must die out at infinity if follows that $c_3=0$.}}

\begin{subequations}
\begin{align}
\nonumber \tag{40a}
  r R_2 = -\frac{1}{2} \frac{d}{dr}(r^2 R_1) \\ \nonumber
 -\frac{2}{r}\frac{dR_2}{dr}-\frac{2R_2}{r^2}-\frac{2R_1}{r^2} & =\frac{d^2R_1}{dr^2} +\frac{4}{r} \frac{dR_1}{dr} \\ 
 \nonumber \tag{40b}
 & =\frac{2\epsilon_0\epsilon_r E}{\eta} \frac{1}{r} \Big( c_3 r + \frac{c_4}{r^2} +\xi\Big) \equiv f_3(r). \\ \nonumber
 &
\end{align}
\end{subequations}
From (40b) it follows that\footnote{\BLUE{In order to avoid problems with improper integrals, we can easily see that by taking testfunction $r^n$ in (40b) that the solution for the coefficients $c_3$ and $c_4$ in (41a) is correct. The lower limit of the integrals with $R_1$ and $R_2$ can actually be any number, for example choosing $a$ instead of $\infty$ will finally result in the same answer, since the differences will be absorbed in the terms with $c_5$ and $c_6$.}}
\begin{subequations} \nonumber
\begin{align} \nonumber
    R_1(r) &= \frac{c_5}{r^3} + c_6 + \int_\infty^r \frac{1}{x^4} \int_\infty^x y^4 f_3(y) \; dy \; dx \\ \nonumber
    & = \frac{c_5}{r^3} + c_6 + \int_\infty^r \frac{1}{x^4} \int_\infty^x \frac{2\epsilon_0\epsilon_r E}{\eta} y^3 \Big(c_3 y +\frac{c_4}{y^2} +\xi \Big) \; dy \; dx
    \\ \nonumber
    &
\end{align}
\end{subequations}
Substituting this value of $R_1$ into (40a) we find for $R_2$:
\begin{align} \nonumber
    R_2(r) = \frac{c_5}{2r^3} -c_6 & -\int_\infty^r \frac{1}{x^4} \int_\infty^x \frac{2\epsilon_0\epsilon_r E }{\eta} y^3 \Big(c_3 y +\frac{c_4}{y^2} +\xi \Big) \; dy \; dx \\ \nonumber
    & -\frac{1}{2r^3 } \int_\infty^r \frac{2\epsilon_0\epsilon_r E }{\eta} x^3 \Big(c_3 x +\frac{c_4}{x^2} +\xi \Big) dx
\end{align}
After carrying out the integrals we can now write $u_r$ and $u_\theta$ as
\begin{subequations} 
\begin{align} \nonumber \tag{41a}
u_r &= \cos \theta \left\{ \frac{c_5}{r^3} +c_6 + \frac{\epsilon_0\epsilon_r E }{\eta} \Big(\frac{c_3 r^2}{5} - \frac{c_4}{r} - \frac{2}{3r^3} \int_\infty^r x^3 \xi dx +\frac{2}{3} \int_\infty^r \xi dx\Big) \right\} \\  \nonumber \tag{41b}
u_\theta &= \sin \theta \left\{ \frac{c_5}{2r^3} -c_6 -\frac{\epsilon_0\epsilon_r E }{\eta} \Big(\frac{2c_3 r^2}{5} - \frac{c_4}{2r} + \frac{1}{3r^3} \int_\infty^r x^3 \xi dx +\frac{2}{3} \int_\infty^r \xi dx \Big)\right\}
\end{align}
\end{subequations}
Aside from the constants $c_2$ to $c_6$ in equations (29) and (41), the pressure and velocity distribution are now fully determined. We now come back to the statement that $c_2$ is partly determined by $\BS u$, $p$ and $\BS u$ must satisfy (14a). The term $\nabla ( -\epsilon_0\epsilon_r E \cos\theta c_2/r^2)$ which arises when expanding $\nabla p $ \BLUE{(from (24) and (26))} must thus be compensated by a term of the same nature from $\eta \nabla \times \nabla \times \BS u$, since $\rho \nabla(\psi^0 + \delta\psi)$ does not contain a term proportional to $\nabla (\cos\theta/ r^2)$. The vector $\eta \nabla \times \nabla \times \BS u$ has  $r$- and $\theta$- components, thus with the help of (31), (32) and (38) we find\footnote{\BLUE{Note again the minus-sign difference in the definition of the curl-operator when compared to modern notation.}}: 

\begin{align}\nonumber
\eta (\nabla \times \nabla \times \BS u)_r = \eta (\nabla \times \BS w)_r &=- 2\epsilon_0\epsilon_r E \cos \theta \Big( c_3 + \frac{c_4}{r^3} + \frac{\xi}{r}\Big) \\ \nonumber 
&\\
\nonumber
\eta (\nabla \times \nabla \times \BS u)_\theta  = \eta (\nabla \times \BS w)_\theta &= \epsilon_0\epsilon_r E \sin \theta \Big[ 2 c_3 - \frac{c_4}{r^3} + \frac{1}{r} \frac{d}{dr} (r \xi)\Big]\\ \nonumber
& .
\end{align}
$\nabla (-\epsilon_0\epsilon_r E c_2 \cos \theta/r^2)$ is again a vector with an $r-$ and $\theta-$ component as well
\begin{align} \nonumber
    \nabla_r \Big( - \epsilon_0\epsilon_r E \cos \theta \frac{c_2}{r^2}\Big) &= 2\epsilon_0\epsilon_r E \cos\theta \frac{c_2}{r^3} \qquad
    \\\nonumber
    \nabla_\theta \Big( - \epsilon_0\epsilon_r E \cos \theta \frac{c_2}{r^2}\Big) &= \epsilon_0\epsilon_r E \sin\theta \frac{c_2}{r^3} \qquad
\end{align}
Apparently for each component, the sum of the $c_2$ and $c_4$ terms must be equal to zero, which can be satisfied if:
\begin{equation} \nonumber \tag{42a}
    c_2 = c_4
\end{equation}
The boundary conditions for the fluid velocity and the potential are:
\begin{itemize}
    \item For $r\to \infty$,  $u_r=-U \cos\theta$ and $u_\theta=U \sin\theta$, $\psi^0=0$ 
    \item For $r=a$, \quad $u_r=0$ \quad \quad \quad and $u_\theta=0$,\quad \quad \quad $\psi^0=\zeta$ .
\end{itemize}
From the conditions for $r\to \infty$, when applied to (41), we can learn that
\begin{equation} \nonumber \tag{42b}
    c_3 = 0 \qquad ; \qquad c_6 = -U
\end{equation}
With the conditions for $r=a$ we find from (41)\footnote{\BLUE{When the force is calculated later, it turns out that a direct consequence is that actually $c_4=0$ (and thus $c_2=0$ as well with (42a)).}}
\begin{equation} \nonumber \tag{42c}
    c_4 = a \int_\infty^a \xi \; dx - \frac{3 \eta a U}{2\epsilon_0\epsilon_r E} 
\end{equation}
and
\begin{align} \nonumber
     c_5  = -\frac{Ua^3}{2} + \frac{2\epsilon_0\epsilon_r E}{3 \eta}\int_\infty^a x^3\xi dx + \frac{\epsilon_0\epsilon_r E a^3}{3 \eta}\int_\infty^a \xi \; dx  
\end{align}
With these values for the constants $c_2$ to $c_6$ the full expression for the pressure and velocity will be
\begin{subequations} 
\begin{align} \tag{43} \nonumber
p &= \epsilon_0 \epsilon_r \int_\infty^r \Delta \psi^0 \frac{d\psi^0}{dx} dx + \cos\theta \Big\{ \frac{3 \eta a U}{2r^2} - \epsilon_0 \epsilon_r E \chi - \frac{\epsilon_0 \epsilon_r Ea}{ r^2} \int_\infty^a \xi \; dx\Big\}\\
 \nonumber
u_r &= \cos\theta \Big\{ \Big( -1 + \frac{3a}{2r} - \frac{a^3}{2r^3} \Big)U + \frac{2\epsilon_0\epsilon_r E}{3\eta} \Big( \int_\infty^r \xi \; dx + \frac{1}{r^3} \int_r^a x^3 \xi \; dx \Big) \\ \nonumber
& \hspace{60 mm}-\frac{\epsilon_0\epsilon_r E}{\eta} \Big( \frac{a}{r} - \frac{a^3}{3r^3} \Big) \int_\infty^a \xi \; dx\Big\}\\ \nonumber
u_\theta &= \sin\theta \Big\{ \Big( 1 - \frac{3a}{4r} - \frac{a^3}{4r^3} \Big)U - \frac{2\epsilon_0\epsilon_r E}{3\eta} \Big( \int_\infty^r \xi \; dx - \frac{1}{2r^3} \int_r^a x^3 \xi \; dx \Big) \\ \nonumber
& \hspace{60 mm} +\frac{\epsilon_0\epsilon_r E}{\eta} \Big( \frac{a}{2r} + \frac{a^3}{6r^3} \Big) \int_\infty^a \xi \; dx\Big\}\\ \nonumber
u_\varphi &= 0.
\end{align}
\end{subequations}
One can imagine the fluid flow being build up from a pure Stokesian flow (proportional to $U$), a second flow, directly caused by the in the fluid acting forces of electrical origin, that does not satisfy the boundary conditions on the boundary of the fluid and the sphere (terms with $\chi$, $\int_\infty^r \xi dx$ and $\frac{1}{r^3} \int_r^a x^3 \xi dx $), and a third flow (proportional to $\int_\infty^a \xi dx  $), that is not exposed to external forces, but combined with the second flow is satisfying the boundary conditions at $r=a$.

\ifFIG\begin{figure}
    \centering
    \includegraphics[width=0.5\textwidth]{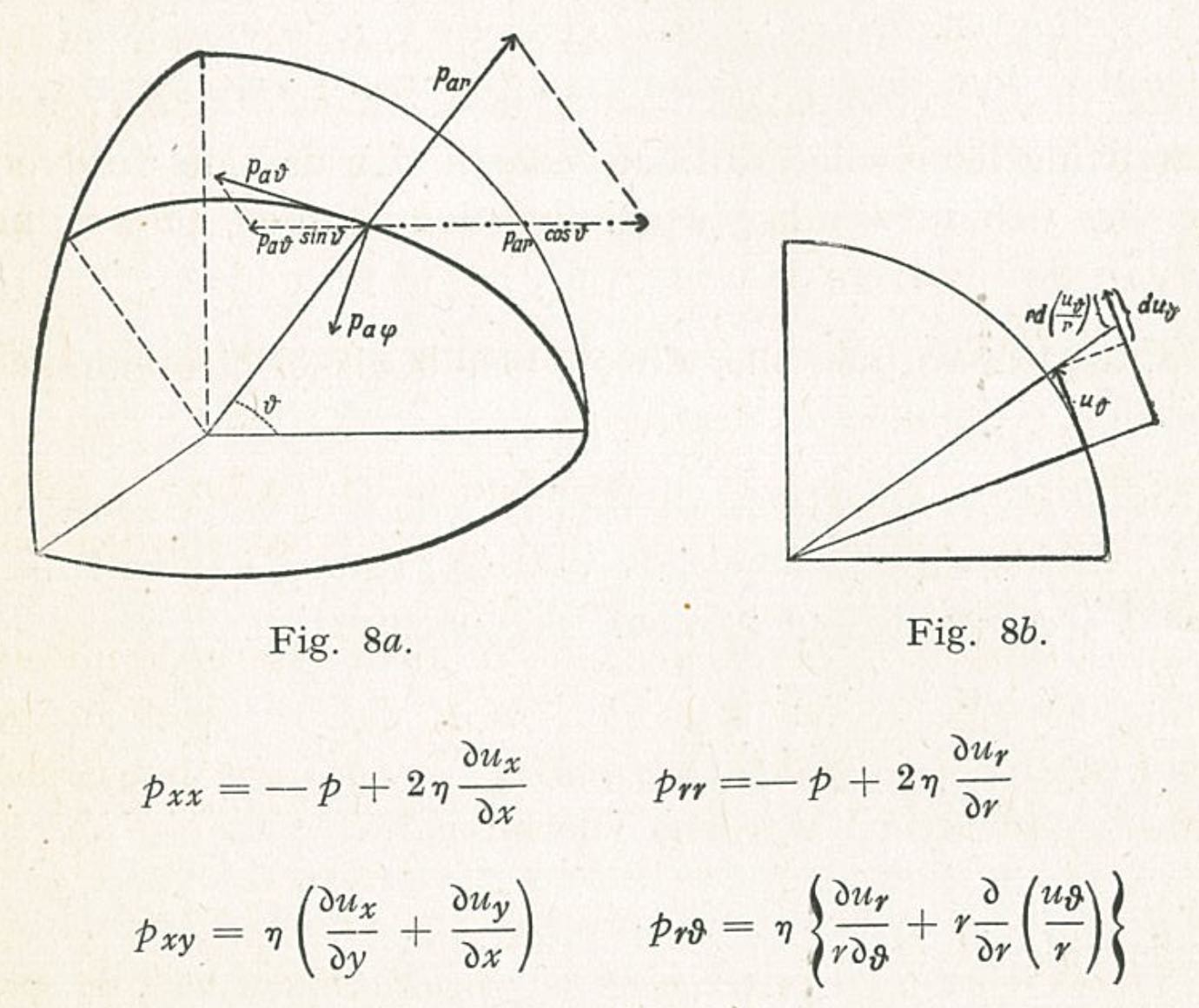}
    \caption{Only that part of the change of $u_\theta$ with $r$ must be taken into account, which leads to torsion, thus the change of the angular velocity $u_\theta/r$ (and not the linear velocity $u_\theta$); see Fig.8b.}
    \label{fig:8}
\end{figure}\fi
To calculate the force, that the fluid defined by (43) is exerting on the sphere, we must integrate $p_{ax}$, the pressure\footnote{\BLUE{This is actually the traction (the force per unit area) and not the pressure (Overbeek used the term "pressure" here).}} on a point on the surface of the sphere in the direction of $\theta=0$, over the whole surface of the sphere.  Here we have\footnote{This value for $p_{ax}$ can be deduced from the stress tensor in Cartesian coordinates (DEBYE and H\"UCKEL, Phys. Z., \textbf{24}, 315 (1932). With the help of Fig.\ref{fig:8}a, it can be seen that $p_{ax}$ must have the indicated form, if one writes the components of the stress in spherical coordinates. In the change of $u_\theta$ with $r$ only that part must be counted, which results in torsion, thus the change in angular velocity $u_\theta/r$ but not that of the linear velocity $u_\theta$, see Fig.\ref{fig:8}b. 
\BLUE{The stress tensor is $\sigma_{ij} = -p \delta_{ij} + \eta[\partial u_i/\partial x_j + \partial u_j/\partial x_i]$. In the current case $u_i=\frac{U_j}{U}\Big\{ \frac{x_i x_j}{r^2}(R_1+R_2) -\delta_{ij} R_2\Big\}$. A further simplification of (44) can be obtained by realizing that $u_\theta=0$ and $\partial u_r/\partial \theta = 0$ at $r=a$. }}
\begin{subequations}
\begin{align} \nonumber  
p_{ax} & \equiv p_{ar} \cos \theta - p_{a\theta} \sin \theta \\ \nonumber \tag{44}
   & = \Big( -p + 2 \eta \frac{\partial u_r}{\partial r} \Big)_{r=a} \cos \theta
-\eta \Big( \frac{\partial u_\theta}{\partial r} - \frac{u_\theta} {r} + \frac{1}{r} \frac{\partial u_r}{\partial \theta}\Big)_{r=a} \sin \theta.
\end{align}
\end{subequations}
The force on the sphere (which thus corresponds to the forces $k_2+k_3$ from Chapter I, \BLUE{just before (2)} is now given as
\begin{equation}
\nonumber
k_2+k_3=\int_0^\pi p_{ax} \; 2 \pi a^2 \; \sin \theta \;d \theta.
\end{equation}
By using (43) and (44) we find after a simple but lengthy calculation
\begin{align}
\nonumber
k_2+k_3 &= -6\pi \eta a U + \frac{4\pi\epsilon_0\epsilon_r E a^2}{3} \chi(a) + \frac{8 \pi \epsilon_0 \epsilon_r E a^2}{3}\xi(a)+ 4\pi\epsilon_0\epsilon_r E a \int_\infty^a \xi \; dx
\\ \nonumber 
 &
\end{align}
By substituting $\chi$ and $\xi$ according to Eqs. (28) and (37') this will become:
\begin{equation} \nonumber \tag{45}
    k_2+k_3=-6\pi \eta a U + \frac{4 \pi \epsilon_0\epsilon_r E a}{3} \Big(\frac{d\psi^0}{dr} \Big)_a \Big( 2R + a \frac{dR}{dr} \Big)_a +  4\pi\epsilon_0\epsilon_r E a \int_\infty^a \xi \; dx.
\end{equation}
When we compare (45) with the value that we obtained \BLUE{in Chapter I} for $k_2$ and $k_3$, then we see that in the right hand side of (45) the following terms in order of appearance are
\begin{enumerate}
    \item the Stokesian drag (as before)
    \item the friction due to that part of the flow that is caused directly by electrical forces, and
    \item the friction due to that part of the flow that is free of external forces and is required to satisfy the boundary conditions at $r=a$.
\end{enumerate}
Points 2. and 3. combined represent the \BLUE{in Chapter I} defined electrophoretic drag force $k_3$.

The total \emph{electrical} force on the sphere is the sum of the forces $k_1$ and $k_4$, which are exerted on the sphere by the applied field, $E$  and the additional charge density, $\delta\rho$ respectively. 

\begin{align} \nonumber
 k_1+k_4 & = QE - \int_0^\pi \int_a^\infty \frac{Q \; \delta\rho}{4 \pi\epsilon_0\epsilon_r r^2} \cos \theta\; 2\pi r \; \sin \theta\; r d\theta \; dr\\ \nonumber
 &
\end{align}
Here, according to (15) \BLUE{(and (22))}
\begin{equation} \nonumber
    \delta\rho  = - \epsilon_0\epsilon_r \nabla^2 \delta\psi = \epsilon_0\epsilon_r E \cos \theta\Big(\frac{d^2R}{dr^2} + \frac{2}{r} \frac{dR}{dr} - \frac{2R}{r^2} \Big)
\end{equation}
From a simple integration we can learn that \footnote{\BLUE{Here we have used  that $(dR/dr +2R/r)$ at $r=\infty$ becomes $3$ since $R$ behaves as $r$ there. Then this cancels out exactly with the $QE$ term on the right hand side of two equations ago.}}
\begin{equation} \nonumber
\tag{46}
    k_1+k_4=\frac{Q E}{3}\Big(\frac{dR}{dr} + \frac{2R}{a} \Big)_{r=a}
\end{equation}
To express the total charge of the sphere, $Q$ in terms of $\psi^0$ and $a$, we use the fact that this charge is equal but opposite to the total charge on the double layer. Thus
\begin{equation} \nonumber \tag{47}
    Q =-\int_a^\infty \rho^0 4 \pi r^2dr = 4 \pi \epsilon_0\epsilon_r \int_a^\infty \frac{1}{r^2} \frac{d}{dr}\Big( r^2 \frac{d\psi^0}{dr} \Big) r^2 dr = -4 \pi \epsilon_0\epsilon_r a^2 \Big(\frac{d\psi^0}{dr} \Big)_{r=a}
\end{equation}
The sum of all forces on the sphere must be zero, thus $k_1+k_4+k_2+k_3=0$. From Eqs. (45), (46) and (47) it then follows that\footnote{The cancellation of $k_1+k_4$ against the second term of $k_2+k_3$ (Eq. 45) is no coincidence, but has essential meaning, since this part of $k_2+k_3$ is exactly caused by forces that act on the charge in the fluid and this charge is equal and opposite to the charge on the sphere.}
\begin{equation} \nonumber
    -6 \pi \eta a U + 4 \pi \epsilon_0\epsilon_r E a \int_\infty^a \xi \; dx = 0
\end{equation}
%
or\footnote{One can easily indicate, how the final formula would have looked like, if the interaction between the relaxation and the electrophoresis was not taken into account. In that case $k_2+k_3$ would be calculated assuming that $R=r+\lambda a^3/(2r^2)$, where $\lambda=(1-\mu)/(2+\mu)$. Then $k_2+k_3=-6\pi \eta a U +4\pi \epsilon_0 \epsilon_r Ea^2 \Big(\frac{d \psi^0}{dr} \Big)_a+4\pi \epsilon_0 \epsilon_r Ea \int_\infty^a \xi dr$,  
where $\xi$ is calculated as if $R=r+\lambda a^3/(2r^2)$. The force $k_1+k_4 $ remains unchanged 
$ k_1+k_4=-4 \pi \epsilon_0 \epsilon_r
Ea^2 \Big(\frac{d \psi^0}{dr} \Big)_a
+4\pi \epsilon_0 \epsilon_r Ea^2
\Big(\frac{d \psi^0}{dr} \Big)_a
\Big\{1-\frac{1}{3}
\frac{dR}{dr}-\frac{2R}{3a}\Big\}_{r=a}$ 
\\ where $R=r+\lambda a^3/(2r^2)+f(r)$. By setting $k_1+k_2+k_3+k_4$ to zero, we find after some manipulation\\
$U=\frac{\epsilon_0 \epsilon_r E} {3 \eta}\Big\{\int_\infty^a \xi dr -  \int_\infty^a \Big\{ \frac{f(r)}{r} \frac{d\psi^0}{dr} -2r \int_\infty^r \Big(\frac{1}{r^2} \frac{df(r)}{dr}-\frac{f(r)}{r^3} \Big) \frac{d\psi^0}{dr} dr\Big \} dr - \frac{a}{3} \Big( \frac{d\psi^0}{dr}\Big)_a \Big(\frac{df(r)}{dr}+\frac{2}{a} f(r) \Big)_{r=a}\Big\}$ \\
\\
where $\xi$ has the same value as in (48a). How big the difference between this formula and (48a) is, is difficult to estimate, however it is easy to see, that they are not identical. See also chapter V .\\ 
\BLUE{Translator's note: It appears there is an error in the formula above because to be consistent with the work of Henry, it should have been $R=r+\lambda a^3/r^2$, without the factor "2" in the denominator.}
} 
\BLUE{(using (37'))}
%
\begin{align} \nonumber \tag{48a}
    U &=\frac{2\epsilon_0\epsilon_r E}{3\eta} \int_{\infty}^a \xi \; dx \\
    \nonumber \tag{48b}
    U &=\frac{2\epsilon_0\epsilon_r E}{3\eta} \int_{\infty}^a \Big \{ \frac{R}{x} \frac{d\psi^0}{dx} -2x \int_{\infty}^x\Big( \frac{1}{y^2} \frac{dR}{dy}-\frac{R}{y^3} \Big)\frac{d\psi^0}{dy} dy \Big \}dx
\end{align}
This formula is thus very generally valid for the electrophoresis velocity of a sphere. We have not assumed any special assumptions concerning the functional form of $\psi^0$ , nor that of $R$. The formula of Henry (and consequently those of H\"uckel and Smoluchowksi)\footnote{\BLUE{The formula of H\"uckel (2) can easily be recovered by setting $R=r$, then $\xi=\frac{d\psi^o}{dr}$ and $\frac{U}{E}=\frac{2\varepsilon_0 \varepsilon_r} {3\eta} \psi^0_{r=a}=\frac{2\varepsilon_0 \varepsilon_r }{3\eta} \zeta$}} can be derived as special cases from (48b), by applying the appropriate form of the function $R$. 

By using (48a), the expressions for $u_r$ and $u_\theta$ from (43) can be written even shorter as\footnote{\BLUE{This is closely related to the $h$-function of Ohshima et al. (H.Ohshima, T.W.Healy and L.R.White J.Chem. Soc., Faraday Trans. 2, 1983, 79, 1613-1628), in which the velocity vector is given as:\\
$\BS{u} = -2\frac{h}{r}E \cos \theta  \BS{e}_r +\frac{1}{r} \frac{d}{dr} [rh] E \sin \theta \BS{e}_\theta$
with $h=-\frac{\epsilon_0\epsilon_r}{3\eta} \big\{ r \int_a^r \xi dx -\frac{1}{r^2} \int_a^r x^3 \xi dx\big\}.$ \\
From (43), with the help of the first equation after (47): $U=\frac{2\epsilon_0 \epsilon_r E}{3\eta} \int_\infty^a \xi dr$,  we can write for the pressure: 
$p=\epsilon_0 \epsilon_r [\int_\infty^r \nabla^2 \psi^0 \frac{d\psi^0}{dx} dx -\cos \theta  E \chi]$, with $\chi$ defined in (28). From this it follows that outside the double layer, the pressure is zero, thus a "viscous potential flow" or zero pressure Stokes flow exists there. The term with the integral represents the osmotic pressure and is independent of $E$. \\
For completeness sake the pressure and vorticity have also been added in (49).\\
The functions $\chi$ and $\xi$ are related to each other as $\chi=-2\xi +\Big(2\frac{R}{r} +\frac{dR}{dr}\Big) \frac{d\psi^0}{dr}$\\
$\xi$ is related to $h$ as $\xi = -\frac{\eta}{\varepsilon_0 \varepsilon_r} \Big[\frac{d^2h}{dr^2} +\frac{2}{r}\frac{dh}{dr} - 2 \frac{h}{r^2}\Big]$}}: 
\begin{align} \nonumber \tag{49a}
    u_r &= \; \; \frac{2\epsilon_0\epsilon_r E \cos \theta}{3 \eta} \Big[ \int_a^r \xi \; dx -\frac{1}{r^3} \int_a^r x^3\xi \; dx \Big] \BLUE{\; \;=\frac{2\epsilon_0\epsilon_r E \cos \theta}{3 \eta}  \int_a^r \Big( 1 -\frac{x^3}{r^3} \Big) \xi \; dx } \\ \nonumber \tag{49b}
    u_\theta &= -\frac{2\epsilon_0\epsilon_r E \sin \theta}{3 \eta} \Big[ \int_a^r \xi \; dx +\frac{1}{2r^3} \int_a^r x^3\xi \; dx \Big] \BLUE{=\frac{2\epsilon_0\epsilon_r E \sin \theta}{3 \eta}  \int_a^r \Big(- 1 -\frac{x^3}{2r^3} \Big) \xi \; dx }
   \\
   \nonumber
   \BLUE{ \tag{49c}
    p \;} & \BLUE{ =\epsilon_0 \epsilon_r \int_\infty^r \nabla^2 \psi^0 \frac{d\psi^0}{dx} dx -\epsilon_0 \epsilon_r E \chi  \cos \theta}\\
    \nonumber \BLUE{ \tag{49d}
    w \;} & \BLUE{= (\nabla \times u)_\varphi = \frac{\epsilon_0 \epsilon_r E}{\eta} \xi \; \sin \theta}
\end{align}
\BLUE{with $\chi$ in the pressure from (28), which can be simplified using (37') as: $\chi=-2\xi +\big(\frac{2R}{r}+\frac{dR}{dr}\big) \frac{d\psi^0}{dr}$.}

\subsection*{B - CALCULATION OF THE ION DISTRIBUTION AND THE ELECTRIC FIELD}

In determining $R$ from the ion distribution we did not reach the same generality that we achieved in writing down the electrophoresis formula (48). We had to apply approximations here, such that the obtained results were only valid exactly for small double layer potentials. 

By eliminating $n_+^0$ and $n_-^0$ from (12) and (9) we can find for $\psi^0$ the following differential equation 
\begin{equation} \nonumber \tag{50}
 \nabla^2 \psi^0 \equiv \frac{1}{r^2} \frac{d}{dr}\Big(r^2 \frac{d\psi^0}{dr} \Big)= - \frac{e}{\epsilon_0\epsilon_r}\Bigg ( z_+ n_+^\infty \; \text{e}^{-\frac{z_+e \psi^0}{kT}}
-z_- n_-^\infty \; \text{e}^{+\frac{z_-e \psi^0}{kT}}\Bigg)
\end{equation}

We now introduce new variables $x$ and $y$ for $r$ and $\Psi$, which will be used in the remainder of this work
\begin{align} \nonumber \tag{51}
    x &\equiv \kappa r = \Big[ \frac{(n_+^\infty z_+^2 + n_-^\infty z_-^2) e^2}{\epsilon_0\epsilon_r kT} \Big]^{1/2} r
\\ \nonumber \tag{52}
    y &\equiv \frac{e\psi^0}{kT}
\end{align}
At 25$^{\circ}$ C
\begin{equation}
\nonumber
 y= 0.0389 \psi^0 \approx \frac{1}{25} \psi^0 
\end{equation}
with $\psi^0$ expressed in millivolts\footnote{\BLUE{See also "Electrophoresis, Theory, Method and Applications", edited by Milan Bier, Volume II, Academic Press, New York, 1967,  Chapter 1, J.Th.G. Overbeek and P.H. Wiersema, "The interpretation of Electrophoretic Mobilities", the discussion after their Eq.28.}}. 

For the value of $\kappa$ one can refer to table 1, while it is easy to remember that $\kappa=10^{+5}$ cm$^{-1}$ when the concentration of a 1-1 electrolyte is $10^{-5} N$.

Remember further that $z_- n_-^\infty = z_+ n_+^\infty$ , then (50) transforms with the new variables into
\begin{equation}
\nonumber
 \frac{d^2 y}{d x^2} + \frac{2}{x}\frac{dy}{d x} = \frac{\text{exp}(z_-y) - \text{exp}(-z_+y)}{z_+ + z_-}
\end{equation}
Developing the exponential power on the right hand side leads to
\begin{equation}   \nonumber \tag{53}
\frac{d^2 y}{d x^2} + \frac{2}{x}\frac{dy}{d x} = y + (z_- - z_+) \frac{y^2}{2} + (z_+^2 - z_+ z_- + z_-^2) \frac{y^3}{6} +....
\end{equation}
In the approach of Debye and H\"uckel only the first term of this sum is taken into account. Thus the contributions of order $y^2$ are neglected if we deal with a non-symmetric electrolyte and the terms of order $y^3$ are being neglected if we deal with a symmetric electrolyte. 

Since it will turn out that the corrections of the relaxation effect are also of order $y^2$ and $y^3$ respectively, it is necessary in certain parts of the calculation to take into account the 2nd and 3rd order approximations of (53)\footnote{For the potential distribution around ions, thus for small $\kappa a$, this problem has been extensively studied by\\
T.H.GRONWALL, V.LA MER and K.SANDVED, Physik Z. \textbf{29}, 358 (1928) and\\ T.H.GRONWALL, V.LA MER and L.GREIFF, J.Phys. Chem. \textbf{35}, 2245 (1931).}. In a first approximation the solution of (53) is
\begin{equation}  \nonumber \tag{54}
Y_1 = \frac{A \text{e}^{-x}}{x} + \frac{A_2 \text{e}^{+x}}{x}.
\end{equation}
The constant $A_2$ has the value zero, since otherwise $y$ would become infinitely large when $x$ approaches infinity. For the solution in the second order we can now set:
\begin{equation} \nonumber
  y = \frac{A \text{e}^{-x}}{x} + Y_2.
\end{equation}
By substituting this into (53), we can find an equation from which $Y_2$ can be determined.
\begin{equation} \nonumber
\frac{d^2 Y_2}{d x^2} + \frac{2}{x}\frac{dY_2}{d x} - Y_2 =  (z_- - z_+) \frac{A^2 \text{e}^{-2x}}{2x^2} + \text{higher order terms of } A \text{e}^{-x} 
\end{equation}
The general solution of this differential equation has the same shape as (54). The particular integral can be written as
\begin{align}  \nonumber
    Y_2 &=  \frac{1}{2} \int_\infty^x (z_- -z_+) \frac{A^2 \text{e}^{-2t}}{2t^2} \Big[\frac{\text{e}^x}{x}\frac{\text{e}^{-t}}{t} - \frac{\text{e}^{-x}}{x}\frac{\text{e}^t}{t} \Big] t^2 \; dt  \\ \tag{55a}
        &= - (z_- -z_+) A^2 \text{e}^{-2x} \Big[\frac{\text{e}^x}{4x}\int_\infty^x \frac{\text{e}^{-t}dt}{t} - \frac{\text{e}^{3x}}{4x}\int_\infty^{3x} \frac{\text{e}^{-t}dt}{t} \Big]
\end{align}
Here we have used the property that
\begin{equation}
\nonumber
\int_\infty^x \frac{e^{-\alpha t }dt}{t} = \int_\infty^{\alpha x} \frac{e^{-t} dt}{t}
\end{equation}
which is easy to prove by differentiation. 

The value of the function $\int_\infty^{x} \frac{e^{-t} dt}{t}$ can be looked up in tables\footnote{JAHNKE and EMDE, Funktionentafeln, Teubner, Leipzig, Berlin (1933) 78.} for various values of $x$. These values can also be calculated by using one out of two following series\footnote{G.BRUNEL Bd 2, part 1, Enzyklop\"adie der mathematischen Wissenschaften page 174.}, for which the first one is convergent for any $x$, but not practical for large $x$ while the second one is very suitable for large $x$

For small $x$\footnote{\BLUE{The value 0.577... is Euler's constant.}}:
\begin{equation}
\nonumber
\int_\infty^{x} \frac{e^{-t} dt}{t} = +0.577 .... + \text{ln} \; x - \frac{x}{1} + \frac{x^2}{2. 2!} - \frac{x^3}{3 .3!} + \frac{x^4}{4 .4!} - ....
\end{equation}

For large $x$:
\begin{equation}
\nonumber
\int_\infty^{x} \frac{e^{-t} dt}{t} = e^{-x}\Big(-\frac{1}{x} + \frac{1}{x^2} - \frac{2!}{x^3} + \frac{3!}{x^4}- .....\Big).
\end{equation}
The series for large $x$ is a half converging series. If this series is truncated at the term $\frac{(n-1)!}{x^n} e^{-x}$ then the term $n! \int_\infty^x \frac{e^{-t}dt}{t^{n+1}}$ will be neglected. 
The absolute value of this is certainly smaller than $\Big| n! e^{-x} \int_\infty^x \frac{dt}{t^{n+1}}\Big| = \Big| \frac{(n-1)! e^{-x}}{x^n}\Big|$ such that the error made is always smaller than the last computed term. 
Since the function $\int_\infty^{x} \frac{e^{-t} dt}{t}$ appears repeatedly in the next calculation, we will introduce a special symbol for it, namely
\begin{equation}  \nonumber \tag{56}
  E(x) \equiv\int_\infty^{x} \frac{e^{-t} dt}{t} 
\end{equation}
If the approximate values for $E(x)$ are substituted in (55a) then $Y_2$ takes on the form
\begin{align} \nonumber
   \text{for small $x$}\qquad Y_2 &\approx (z_--z_+) \frac{A^2 \text{e}^{-2x}}{x^2}\frac{x \ln 3}{4}  \\ \nonumber \\ \nonumber 
   \text{for large $x$}\qquad Y_2 &\approx (z_--z_+) \frac{A^2 \text{e}^{-2x}}{x^2} \Big( \frac{1}{6} - \frac{2}{9x} ....\Big).
\end{align}

Near the particle $(x=\kappa a)$, $\frac{Ae^{-x}}{x}$ is roughly equal to $y_{r=a}$ and lies in between 0 and 4 for $\zeta$ potentials between 0 and 100 mV. With increasing $x$, $\frac{Ae^{-x}}{x}$ quickly diminishes, such that at large distances from the particle, $Y_2$ is always much smaller than $Y_1$; also near the particle $Y_2\ll Y_1$ if $x$ is small (thus if $\kappa a \ll 1$), but if $\kappa a \gg 1$, the term $Y_2$ cannot be neglected with respect to $Y_1$. As it will turn out, taking into account higher order approximations of the solution of (53) is important especially for large $\kappa a$.

In the second order approximation, the solution of (53) is thus
\begin{equation}
\nonumber
y=\frac{A \; \text{e}^{-x}}{x} - (z_- - z_+) \frac{A^2 \; \text{e}^{-2x}}{x^2}\Big\{ \frac{x \;\text{e}^{x}}{4} E(x) - \frac{x \;\text{e}^{3x}}{4} E(3x)\Big\}
\end{equation}
In a fully analog manner the solution can be found in the third order approximation

\begin{align} \nonumber 
y &=\frac{Ae^{-x}}{x} - (z_- - z_+) \frac{A^2 e^{-2x}}{x^2}\Big\{\frac{xe^x}{4} E(x)- \frac{xe^{3x}}{4} E(3x)\Big\}\\
\nonumber 
&-(z_--z_+)^2 \frac{A^3 e^{3x}}{x^3} \Big\{ \frac{x^2e^{4x}}{8} \int_\infty^x \frac{e^{-3t}E(t)}{t} dt - \frac{x^2 e^{4x}}{8} \int_\infty^x \frac{e^{-t} E(3t)}{t} dt\\
\nonumber 
&- \frac{x^2 e^{2x}}{8} \int_\infty^x \frac{e^{-t}E(t)}{t} dt + \frac{x^2 e^{2x}}{8} \int_\infty^x \frac{e^t E(3t)}{t}dt\Big\}\\
\nonumber \tag{55b}
&-(z^2_+ - z_+ z_- + z^2_-) \frac{A^3 e^{-3x}}{x^3}\Big\{ \frac{x^2 e^{4x}}{3} E(4x) - \frac{x^2 e^{2x}}{6} E(2x)\Big\}.
\end{align}
The value of the constant $A$ must be chosen such, that $y$ will take on the value $e \zeta/kT$ , if on the right hand side of (55b) $\kappa a$ is substituted for $x$.

The series for $y$, of which the first terms are written down here, converges more quickly, as $x$ becomes smaller, but also for large $x$, it is still better than the series of equation (53). We are aware, that taking into account these extra few terms in general does not make any significant improvement over the first approximation of Debye and H\"uckel, but in our case it is necessary to indeed include these terms, since they can be of the same order of magnitude as the relaxation terms. See also \BLUE{last page before chapter V}.

To get from equations (10) and (13) the ion distribution in the by electrophoresis distorted double layer, we rewrite (10) in the following form
\begin{subequations}
\begin{align}
\nonumber    
0=\mp n^0_\pm z_\pm e \lambda_\pm \nabla^2 (\delta \psi) \mp \Big ( \nabla [n^0_\pm z_\pm e \lambda_\pm] \cdot \nabla (\delta \psi)\Big) \mp \delta n_\pm z_\pm e \lambda_\pm \nabla^2 \psi^0
\\
\nonumber
\tag{10'} \\
\nonumber
\mp\Big( \nabla [\delta n_\pm z_\pm e \lambda_\pm] \cdot \nabla \psi^0 \Big) -kT\lambda_\pm\nabla^2 \delta n_\pm + n^0_\pm \nabla \cdot u + (u \cdot \nabla n^0_\pm).
\end{align}
\end{subequations}
After introducing the new variables $x$ (from (51)) and $y$ (from (52)) and using the relationships (9) for $\nu_+$ and $\nu_-$, (17) for $\Phi$, (50) for $\Delta \Psi$, (52) for $\Psi$ and (14b) for $u$, (10') and (13) will transform into (57a, b and c).
\begin{subequations}
\begin{align}
\tag{57a and  b}
&0=\kappa n^0_\pm z_\pm e E \lambda_\pm \cos \theta \Big\{ \mp \frac{\nabla^2 \delta \psi}{\kappa E \cos \theta} e^{\mp s \pm y} - z_\pm e^{\mp z \pm y} \frac{dy}{dx} \frac{d(xR)}{dx}
\nonumber
\\
&\mp \frac{z_\pm e \delta n_\pm}{\varepsilon_0 \varepsilon_r \kappa  X \cos \theta} \Big(e^{z_- y} - e^{-z_+ y} \Big) \mp (z_+ + z_-) \frac{z_\pm e}{\varepsilon_0 \varepsilon_r \kappa E \cos \theta} \frac{\partial \delta n_\pm}{dx} \frac{dy}{dx}
\nonumber \\
&-\frac{z_+ + z_-} {z_\pm} \frac{z_\pm e}{\varepsilon_0 \varepsilon_r E \kappa \cos \theta} \frac{\nabla^2 \delta n_\pm}{\kappa^2} \mp \frac{u_r}{E \cos \theta}  \frac{1}{\lambda_\pm e} e^{\mp z \pm y} \frac{dy}{dx}
\nonumber \Big\}.
\end{align}
\end{subequations}
\begin{equation}
\nonumber 
\frac{\nabla^2 \delta \psi}{-\kappa E \cos\theta}= \frac{ z_+ e}{\varepsilon_0 \varepsilon_r E \kappa \cos\theta} \delta n_+ - \frac{ z_- e}{\varepsilon_0 \varepsilon_r E \kappa \cos \theta} \delta n_-
\tag{57c} 
\end{equation}

It is now convenient to introduce new variables as
\begin{subequations}
\begin{align}
\nonumber
\tag{58} S = -\frac{\nabla^2 \delta \psi}{\kappa E \cos \theta}  \hspace{8 mm}
&\Sigma_+ = -\frac{ z_+ e}{\epsilon_0 \epsilon_r \kappa E \cos\theta} \delta n_+\\ \nonumber
\hspace{8 mm} I = R\kappa=-\frac{\kappa \delta \psi}{E \cos\theta} \hspace{10 mm}
&\Sigma_- = \frac{z_- e}{\epsilon_0 \epsilon_r \kappa E \cos\theta} \delta n_-
\end{align}
\end{subequations}
Concerning $S$ and $I$ we know according to (22) and (17) that these are functions of $r$ alone, thus also of $x$ alone. We assume, that this is also the case for $\Sigma_+$ and $\Sigma_-$. With the definitions (58) we deduce then from (57a, b and c) and from (22) the following set of equations:
\begin{subequations}
\begin{align}
\nonumber 
\tag{59a and b)}
&\frac{d^2\Sigma_\pm}{dx^2} + \frac{2}{x} \frac{d \Sigma_\pm}{dx} - \frac{2 \Sigma_\pm}{x^2} \pm  z_\pm\frac{d \Sigma_\pm}{dx} \frac{dy}{dx} \pm \frac{z_\pm}{z_+ + z_-} \Sigma_\pm (e^{z_-y}- e^{-z_+y})\\
\nonumber
&\mp \frac{z_\pm}{z_+ + z_-} S e^{\mp z \pm y} + \frac{z_\pm^2}{z_++z_-} e^{
\mp z \pm y} \frac{dI}{dx} \frac{dy}{dx}\pm \frac{u_r}{\ifRED\else E \fi \cos \theta} \frac{z_\pm \ifRED\else /\lambda_\pm\fi}{(z_++z_-)\ifRED\else e\fi} e^{\mp z \pm y}\frac{dy}{dx}=0.
\end{align}
\end{subequations}
\begin{equation}
\nonumber
\tag{59c}  \Sigma_+ - \Sigma_- = S    
\end{equation}
\begin{equation}
\nonumber
\tag{59d} \hspace{10 mm} S= \frac{d^2I}{dx^2} + \frac{2}{x}\frac{dI}{dx} - \frac{2I}{x^2}.
\end{equation}

The function $\frac{u_r}{E \cos \theta}$ , which appears in (59a and b) can according to (49a) and (37') be expressed in terms of $R$, $\psi^0 $ and $r$ and thus in terms of $I$, $y$ and $x$. 
\begin{equation}
\nonumber
 \tag{59e} \frac{u_r}{\ifRED\else E\fi \cos \theta} = -\frac{\ifRED\else 4 \pi \varepsilon_0 \varepsilon_r \fi kT}{6 \pi \eta \ifRED\else e\fi} \int_{\kappa a}^x d\Big( \frac{1}{x^3}\Big) \int_{\kappa a}^x x^3 \Big\{\frac{I}{x} \frac{dy}{dx} - 2x \int_\infty^x\Big( \frac{1}{x^2} \frac{dI}{dx} - \frac{I}{x^3} \Big)\frac{dy}{dx} dx \Big\} dx.
\end{equation}
Unfortunately we were unable to find the complete solution of (59a). We therefore had to expand the functions $\Sigma_+$, $\Sigma_-$, $S$ and $I$ in series of increasing powers of $y$, where the coefficients are quite complex functions of $x$. These series were truncated after a few terms only, such that they are only valid for small $\zeta$. In principle however, it is possible to include more terms in these series according to the above described procedure, but it will entail an enormous amount of calculations. We have only included terms in the series as far as was needed to get the first term that influences the relaxation effect on the electrophoresis. The approach that we followed is more or less the following. 

By subtracting equations (59a) and (59b) from each other, using (55b) and (59c) and expanding $e^{\mp z \pm y} $ we find for the $S$ the following differential equation
\begin{subequations}
\begin{align}
\nonumber
 \tag{59f} \frac{d^2S}{dx^2} + \frac{2}{x} \frac{dS}{dx} - \frac{2S}{x^2}-S(1+(z_-- z_+) y ......) \\
 \nonumber
 -\Big(z_+ \frac{d \Sigma_+}{dx} + z_- \frac{d \Sigma_-}{dx} \Big) \Big(A e^{-x} \Big(\frac{1}{x}+ \frac{1}{x^2}\Big)+ ....  \Big)\\
 \nonumber
 +(z_+ \Sigma_+ + z_-\Sigma_-) \Big( y+ \frac{(z_--z_+) y^2}{2} .... \Big)\\
 \nonumber
 +\Big\{ (z_-- z_+)+(z_+^2 -z_+ z_- + z_-^2)y + ....\Big\} \frac{dI}{dx} \Big\{Ae^{-x} \Big(\frac{1}{x}+\frac{1}{x^2} \Big)\\
 \nonumber -(z_--z_+)A^2 \frac{e^{-2x}}{4}\Big( \big(
 \frac{1}{x}+ 
\frac{1}{x^2} \big) e^x E(x) + \big( \frac{1}{x} - \frac{1}{x^2}\big) e^{3x}E(3x) \Big) \Big\}\\
\nonumber
-\frac{u_r}{\ifRED\else E\fi \cos\theta} \frac{z_+ \ifRED\else /\lambda_+\fi + z_-\ifRED\else /\lambda_-\fi + (z_-^2\ifRED\else /\lambda_-\fi - z_+^2\ifRED\else /\lambda_+\fi)y....}{(z_++z_-)\ifRED\else e\fi}\Big(A e^{-x}\Big(\frac{1}{x}+\frac{1}{x^2} \Big).... \Big)=0.
 \end{align}
\end{subequations}
Remember now that the series for $I$ starts with a term proportional to $y^0$, while the series for $S$, $\Sigma_+$, $\Sigma_-$ and $\frac{u_r}{E \cos \theta}$ start with a term proportional to $y^1$, then we can set up the following scheme to get a solution for the set of equations (59a to f). In succession the solution is being determined of\\
$\boldsymbol{A.}$ $I$ from (59d) in the 0th approximation (i.e. with the neglect of $y$ leads to (61')\\
$\boldsymbol{B.}$ $S$ from (59f) in the 1st approximation (i.e. with the neglect of $y^2$ leads to (63)\\
$\boldsymbol{C.}$ $I$ from (59d) in the 1st approximation (i.e. with the neglect of $y^2$ leads to (65)\\
$\boldsymbol{D.}$  $\Sigma_\pm$ from (59a and b) in the 1st approximation (i.e. with the neglect of $y^2$ leads to (66)\\
$\boldsymbol{E.}$ $u$ from (59e) in the 1st approximation (i.e. with the neglect of $y^2$ leads to (67)\\
$\boldsymbol{F.}$ $S$ from (59f) in the 2nd approximation (i.e. with the neglect of $y^3$ leads to (70)\\
$\boldsymbol{G.}$ $I$ from (59d) in the 2nd approximation (i.e. with the neglect of $y^3$ leads to (71)\\
$\boldsymbol{H.}$ $\Sigma_\pm$ from (59a and b) in the 2nd approximation (i.e. with the neglect of $y^3$ leads to (73)\\
$\boldsymbol{I.}$ Afterward the solutions of $S$, $I$ and $\Sigma_\pm$ are being substituted in the boundary conditions to determine the integration constants, which leads to (82), (84) and (86). 

Finally $U$ is being determined (in chapter IV C) in 3rd order approximation from (59e) or slightly simpler from (48b).

Before we proceed to do this, we once more write (59f), with the inclusion of all terms of order $y^2$ and lower and neglect all higher powers of $y$. We then find
\begin{subequations}
\begin{align}
\nonumber
 \tag{60}  \frac{d^2S}{dx^2} + \frac{2}{x} \frac{dS}{dx} - \frac{2S}{x^2}-S\Big\{1+(z_-- z_+)\frac{Ae^{-x}}{x} \Big\} \\
 \nonumber
 -\Big(z_+ \frac{d \Sigma_+}{dx} + z_- \frac{d \Sigma_-}{dx} \Big) A e^{-x} \Big(\frac{1}{x}+ \frac{1}{x^2} \Big) +(z_+\Sigma_++z_-\Sigma_-)\frac{Ae^{-x}}{x}\\
 \nonumber
 +\Big\{ (z_-- z_+)+(z_+^2 -z_+ z_- + z_-^2)\frac{Ae^{-x}}{x}\Big\} \frac{dI}{dx} Ae^{-x} \Big(\frac{1}{x}+\frac{1}{x^2} \Big)\\
 \nonumber -(z_--z_+) \frac{A^2e^{-2x}}{4} \Big\{ \big(
 \frac{1}{x}+ 
\frac{1}{x^2} \big) e^x E(x) + \big( \frac{1}{x} - \frac{1}{x^2}\big) e^{3x}E(3x)  \Big\}\frac{dI}{dx}\\
\nonumber
-\frac{u_r}{\ifRED\else E\fi \cos\theta} \frac{z_+ \ifRED\else /\lambda_+\fi + z_-\ifRED\else /\lambda_-\fi}{(z_++z_-)\ifRED\else e\fi}A e^{-x}\Big(\frac{1}{x}+\frac{1}{x^2} \Big)=0.
\end{align}
\end{subequations}
Debye and H\"uckel as well as Hermans have used this equation as a basis to calculate the relaxation effect. They however did not use the entire equation. Debye and H\"uckel worked with 
\begin{equation}
\nonumber
 \frac{d^2S}{dx^2} + \frac{2}{x} \frac{dS}{dx} -\frac{2S}{x^2}-S -\frac{U}{\ifRED\else E\fi} \frac{z_+\ifRED\else /\lambda_+\fi + z_- \ifRED\else /\lambda_-\fi}{(z_++z_-)\ifRED\else e\fi} Ae^{-x} \Big( \frac{1}{x} + \frac{1}{x^2}\Big)=0.
\end{equation}
while Hermans wrote
\begin{equation}
\begin{aligned}
\nonumber
 \frac{d^2S}{dx^2} +\frac{2}{x} \frac{dS}{dx} -\frac{2S}{x^2}-S +(z_--z_+) \frac{dI}{dx} Ae^{-x} \Big(\frac{1}{x}\Big)
 -\frac{u_r}{\ifRED\else E\fi \cos\theta} \frac{z_+ \ifRED\else /\lambda_+\fi + z_-\ifRED\else /\lambda_-\fi}{(z_++z_-) \ifRED\else e\fi}Ae^{-x} \Big(\frac{1}{x}\Big)=0.
 \nonumber
 \end{aligned}
\end{equation}
The approximations of Debye and H\"uckel may be justified partly, since they limited themselves to very small and very large $\kappa a$. The neglect of the term
\begin{equation}
\nonumber
 (z_- - z_+) \frac{dI}{dx}A e^{-x}\Big( \frac{1}{x} + \frac{1}{x^2}\Big)
\end{equation}
however, by Debye and H\"uckel and the terms $\Sigma_+$ and $\Sigma_-$ by Hermans appears not to be justified in the by them treated extreme cases (see \BLUE{Chapter III and first page of Chapter V etc.}).\\
\\
\textbf{A.} \textit{Zeroth approximation of I.}\\
The solution of $I$ from (59d) has the form
\begin{equation}
\nonumber
  \tag{61}  I=B_1x + \frac{B}{x^2} +x \int^x \frac{dx}{x^4}\int^x x^3 S dx. 
\end{equation}
Since $S$ is of order $y$ (and higher), the integral of (61) only contains terms of $y$ and higher, such that the zeroth approximation for $I$ can be found as (61')
\begin{equation}
\nonumber
  \tag{61'}  I=x + \frac{B}{x^2}.
\end{equation}
The constant $B_1$ is set to 1, since $I$ must approach $x$ if $x$ approaches infinity (see equations (58), (52) and (16)).\\ 
\\
\textbf{B.} \textit{First approximation of S.}\\
To determine $S$ in the first order approximation, we rewrite (59f) while neglecting all terms of order $y^2$ and higher. Te terms with $u_r$ and $\Sigma_\pm$ disappear entirely, such that (59f) transforms into  
\begin{equation}
\nonumber
 \tag{62} 
 \frac{d^2S}{dx^2}+\frac{2}{x}\frac{dS}{dx}- \frac{2S}{x^2}- S + (z_+ -  z_-) \Big(1 - \frac{2B}{x^3}\Big) A e^{-x}\Big(\frac{1}{x} + \frac{1}{x^2}\Big)=0.
\end{equation}
The solution is\footnote{\BLUE{In the original manuscript something is missing in the first exponential function under the integral sign, where now the correct symbol $\xi$ has been put.}}
\begin{subequations}
\begin{align}
\nonumber
 S= P\Big(\frac{1}{x}+\frac{1}{x^2}\Big)e^{-x} +Q\Big(\frac{1}{x}-\frac{1}{x^2}\Big)e^x -\frac{1}{2}\int^x(z_--z_+)\Big(1-\frac{2B}{\xi^3}\Big)Ae^{-\xi} \times\\
 \nonumber\Big(\frac{1}{\xi}+\frac{1}{\xi^2}\Big)\xi^2\Big\{ \Big( \frac{1}{x} -\frac{1}{x^2}\Big) e^x \Big( \frac{1}{\xi}+\frac{1}{\xi^2}\Big) e^{-\xi} - \Big( \frac{1}{x}+\frac{1}{x^2}\Big)e^{-x}\Big( \frac{1}{\xi}-\frac{1}{\xi^2}\Big) e^{\xi}\Big\} d\xi.
 \end{align}
\end{subequations}
where $P$ and $Q$ are integration constant.

After execution of the integrals this gives
\begin{subequations}
\begin{align}
\nonumber
\tag{63} S= P\Big(\frac{1}{x}+\frac{1}{x^2}\Big)e^{-x} +Q\Big(\frac{1}{x}-\frac{1}{x^2}\Big)e^x\\
+(z_--z_+)Ae^{-x} \Big(\frac{1}{2} +\frac{3}{4x}+\frac{3}{4x^2} \Big) +(z_--z_+)BAe^{-x} \frac{1}{2x^3}.
\nonumber
\end{align}
\end{subequations}
$Q$ must be equal to zero, otherwise $S$ would become infinitely large when $x\rightarrow \infty$.
\\
\\
\textbf{C.} \textit{First approximation of I.}\\
By substituting this function $S$ into (59d), we find the differential equation
\begin{subequations}
\begin{align}
\nonumber
 \tag{64} \frac{d^2I}{dx^2}+\frac{2}{x}\frac{dI}{dx}- \frac{2I}{x^2}=P\Big(\frac{1}{x}+\frac{1}{x^2}\Big)e^{-x}+(z_--z_+)Ae^{-x}\Big(\frac{1}{2} +\frac{3}{4x}+\frac{3}{4x^2} \Big)\\
 \nonumber
 +(z_--z_+)BAe^{-x} \frac{1}{2x^3},
 \end{align}
\end{subequations}
from which $I$ can be solved
\begin{subequations}
\begin{align}
\nonumber
 \tag{65}  I=x +\frac{B}{x^2}+P\Big(\frac{1}{x}+\frac{1}{x^2}\Big)e^{-x}+(z_--z_+)Ae^{-x}\Big(\frac{1}{2} +\frac{7}{4x}+\frac{7}{4x^2} \Big)\\
 \nonumber
 +(z_--z_+)BAe^{-x} \Big( \frac{1}{6x^2}-\frac{1}{12x}+\frac{1}{12}+\frac{xe^x}{12}E(x)\Big).
 \end{align}
\end{subequations}
\\
\\
\textbf{D.} \textit{First approximation of $\Sigma_+$ and $\Sigma_-$.}\\
With $S$ according to (63), $I$ according to (65) and $y$ according to (54) from (59a and b) solutions for $\Sigma_+$ and $\Sigma_-$ can be found. The differential equations for $\Sigma_+$ and $\Sigma_-$ become
\begin{subequations}
\begin{align}
\nonumber
 \frac{d^2\Sigma_\pm}{dx^2}+\frac{2}{x}\frac{d\Sigma_\pm}{dx}- \frac{2\Sigma_\pm}{x^2}\mp \frac{z_\pm}{z_++z_-} \Big\{ P \Big( \frac{1}{x} +\frac{1}{x^2}\Big) e^{-x}\\
 \nonumber
 +(z_--z_+)Ae^{-x}\Big(\frac{1}{2}+\frac{3}{4x}+\frac{3}{4x^2}\Big) + (z_--z_+)BA e^{-x}\frac{1}{2x^3}\Big\}
 \\
 \nonumber
-\frac{z_\pm^2}{z_++z_-}\Big(1-\frac{2B}{x^3}\Big) Ae^{-x}\Big(\frac{1}{x}+\frac{1}{x^2}\Big) =0,\end{align}
\end{subequations}
with as solution
\begin{subequations}
\begin{align}
\nonumber
 \tag{66} \Sigma_\pm=C_1x+\frac{C}{x^2} \; \pm \; \frac{z_\pm}{z_++z_-}  P \Big( \frac{1}{x} +\frac{1}{x^2}\Big) e^{-x}
 \; \pm \; \frac{z_\pm(z_--z_+)}{z_++z_-}Ae^{-x} \Big(\frac{7}{4x^2}+\frac{7}{4x}+\frac{1}{2}\Big)\\
 \nonumber
 \pm \frac{z_\pm(z_--z_+)}{z_++z_-}BAe^{-x} \Big(\frac{1}{6x^2}-\frac{1}{12x}+\frac{1}{12}+\frac{xe^x}{12}E(x)\Big)\\
 \nonumber
 + \frac{z_\pm^2}{z_++z_-}Ae^{-x} \Big\{\frac{1}{x^2}+\frac{1}{x}-B\Big(\frac{1}{2x^3}-\frac{1}{6x^2}+\frac{1}{12x}-\frac{1}{12}-\frac{xe^x}{12}E(x)\Big)\Big\}.
 \end{align}
\end{subequations}
$C_1$ must be equal to zero, since $\Sigma_+$ and $\Sigma_-$ approach zero for $x\rightarrow \infty$. $C$ has the same value for both equations since $\Sigma_+-\Sigma_-= S$ and $S$ does not contain a term proportional to $1/x^2$.
\\
\\
\textbf{E.} \textit{First approximation of} $\frac{u_r}{E \cos \theta}$ \\
The here found solutions in the first order approach for $S$, $\Sigma_+$, $\Sigma_-$ and $I$ are substituted in (60), to determine the 2nd order approximation of $S$. In addition a solution to the 1st order approximation of $\frac{u_r}{E \cos \theta}$ must be provided. It can be determined from (59e) with
\begin{equation}
\nonumber
 \tag{61'} I=x+\frac{B}{x^2} \text{    and}
\end{equation}
\begin{equation}
\nonumber
 \tag{54} y=\frac{Ae^{-x}}{x}.
\end{equation}
The execution of the integrations does not exhibit any particular difficulties and gives as result
\begin{subequations}
\begin{align}
\tag{67} \nonumber \frac{u_r}{E\cos\theta} = \ifRED\else \frac{2kT\varepsilon_0 \varepsilon_r}{3\eta e}\fi\Big[\Big\{-\frac{Ae^{-\kappa a}}{\kappa a}-\frac{BAe^{-\kappa a}}{\kappa a}\Big(\frac{1}{8 \kappa a}-\frac{5}{24}-\frac{\kappa a}{48}+\frac{\kappa^2 a^2}{48}\Big) \\
\nonumber
+\frac{BA}{4}\Big(1-\frac{\kappa^2 a^2}{12}
\big) E(\kappa a)\Big\}\\
\nonumber
+\frac{1}{x^3}\Big\{Ae^{-\kappa a}(3+3\kappa a+ \kappa^2 a^2)\\
\nonumber
+BAe^{-\kappa a}\Big(\frac{1}{5}-\frac{\kappa a}{20}+\frac{\kappa^2 a^2}{60}-\frac{\kappa^3 a^3}{120}+\frac{\kappa^4 a^4}{120}+\frac{\kappa^5 a^5 e^{\kappa a}}{120}E(\kappa a)\Big) \Big\}\\
\nonumber
+\Big\{-3Ae^{-x}\Big(\frac{1}{x^3}+\frac{1}{x^2}\Big)-BAe^{-x}\Big(\frac{1}{5x^3}-\frac{7}{40x^2}+\frac{9}{40x}+\frac{1}{80}-\frac{x}{80}\Big)
-\frac{BA}{4}\Big(1-\frac{x^2}{20}\Big) E(x)\Big\} \Big].
\end{align}
\end{subequations}
For $x\rightarrow \infty$, $\frac{u_r}{E\cos \theta}=-U$. If we call the coefficient of $1/x^3$ in (67) $F$, then (67) can be written in the form
\begin{subequations}
\begin{align}
\tag{67'} \nonumber \frac{u_r}{E \cos\theta} = -U +\frac{F}{x^3}-\ifRED\else \frac{2kT\varepsilon_0 \varepsilon_r}{\eta e}\fi\frac{Ae^{-x}}{x}\Big\{\Big(\frac{1}{x^2}+\frac{1}{x}\Big)\\
\nonumber
+B\Big(\frac{1}{15x^2}-\frac{7}{120 x}+\frac{3}{40}+\frac{x}{240}-\frac{x^2}{240}+\frac{1}{12}\Big(x-\frac{x^3}{20}\Big) e^x E(x)\Big) \Big\}.
\end{align}
\end{subequations}
\\
\textbf{F.} \textit{Second approximation of $S$.}\\
By substituting the values of $S$ from (63), $\Sigma_\pm$ from (66), $I$ from (65) and $\frac{u_r}{E \cos \theta}$ from (67') in (60), the final differential equation for $S$ will be found
\begin{subequations}
\begin{align}
\nonumber \tag{68}\\
\nonumber
\frac{d^2S}{dx^2} + \frac{2}{x}\frac{dS}{dx}-\frac{2S}{x^2}-S - (z_--z_+)PAe^{-2x}\Big(\frac{1}{x^3}+\frac{1}{x^2}\Big)\\
\nonumber
-(z_--z_+)^2A^2 e^{-2x} \Big(\frac{3}{4x^3}+\frac{3}{4x^2}+\frac{1}{2x}\Big)-(z_--z_+)^2BA^2e^{-2x}\frac{1}{2x^4}\\
\nonumber
+(z_++z_-)CAe^{-x}\Big(\frac{2}{x^5}+\frac{2}{x^4}+\frac{1}{x^3}\Big)-(z_--z_+)PAe^{-2x}\Big(\frac{2}{x^5}+\frac{4}{x^4}+\frac{4}{x^3}+\frac{2}{x^2}\Big)\\
\nonumber
+(z_+^2-z_+z_-+z_-^2)A^2 e^{-2x}\Big(\frac{2}{x^5}+\frac{4}{x^4}+\frac{4}{x^3}+\frac{2}{x^2}\Big)\\
\nonumber
-(z_+^2-z_+z_-+z_-^2)BA^2e^{-2x}\Big\{\frac{3}{2x^6}+\frac{5}{3x^5}+\frac{7}{12x^4}-\frac{1}{6x^3}+\frac{1}{6x^2}-\frac{1}{12x}\\
\nonumber
+\big(\frac{1}{12x^2}+
\frac{1}{12x}-\frac{1}{12}\Big) e^x E(x)\Big\}\\
\nonumber
-(z_--z_+)^2 A^2 e^{-2x}\Big(\frac{7}{2x^5}+\frac{7}{x^4}+\frac{7}{x^3}+\frac{4}{x^2}+\frac{1}{x}\Big)
\end{align}
\end{subequations}
\begin{subequations}
\begin{align}
\nonumber
-(z_--z_+)^2BA^2e^{-2x}\Big\{\frac{1}{3x^5}+\frac{5}{12x^4}+\frac{1}{6x^3}-\frac{1}{6x^2}+\frac{1}{12x}\\
\nonumber
-\Big(\frac{1}{12x^2}+\frac{1}{12x}-\frac{1}{12}\Big) e^x E(x)\Big\}+(z_--z_+)Ae^{-x}\Big(\frac{1}{x^2}+\frac{1}{x}\Big)\\
\nonumber
-(z_--z_+)BAe^{-x}\Big(\frac{2}{x^5}+\frac{2}{x^4}\Big) - (z_--z_+)PAe^{-2x}\Big(\frac{2}{x^5}+\frac{4}{x^4}+\frac{3}{x^3}+\frac{1}{x^2}\Big)\\
\nonumber
-(z_--z_+)^2A^2 e^{-2x}\Big(\frac{7}{2x^5}+\frac{7}{x^4}+\frac{21}{4x^3}+\frac{9}{4x^2}+\frac{1}{2x}\Big)\\
\nonumber
-(z_--z_+)^2BA^2e^{-2x}\Big\{\frac{1}{3x^5}+\frac{5}{12x^4}-... -\frac{1}{12x^2}-\frac{1}{12}\Big(\frac{1}{x^2}+\frac{1}{x}\Big)e^x E(x)\Big\}\\
\nonumber
+(z_+^2-z_+z_-+z_-^2)A^2 e^{-2x}\Big(\frac{1}{x^3}+\frac{1}{x^2}\Big)-(z_+^2 - z_+ z_- +z_-^2)BA^2e^{-2x}\Big(\frac{2}{x^6}+\frac{2}{x^5}\Big)\\
\nonumber
-(z_--z_+)^2A^2 e^{-2x}\Big(1-\frac{2B}{x^3}\Big)\Big\{\frac{1}{4}\Big(\frac{1}{x^2}+\frac{1}{x}\Big)e^x E(x) +
\frac{1}{4}\Big(\frac{1}{x^2}+\frac{1}{x}\Big)e^{3x}E(3x)\Big\}
\end{align}
\end{subequations}
\begin{subequations}
\begin{align}
\nonumber
 \qquad \qquad + \ifRED\else \frac{z_+/\lambda_+ + z_-/\lambda_-}{(z_++z_-) e}\fi UAe^{-x}\Big(\frac{1}{x^2}+\frac{1}{x}\Big)- \ifRED\else \frac{z_+/\lambda_+ + z_-/\lambda_-}{(z_++z_-) e}\fi FAe^{-x}\Big(\frac{1}{x^5}+\frac{1}{x^4}\Big)\\
\nonumber
+\ifRED\else \frac{z_+/\lambda_+ + z_-/\lambda_-}{(z_++z_-) e} \cdot \frac{2 \varepsilon_0 \varepsilon_r kT}{\eta e}\fi
A^2 e^{-2x}\Big\{\frac{1}{x^5}+\frac{2}{x^4}+\frac{1}{x^3}\\
\nonumber
+B\Big(\frac{1}{15x^5}+\frac{1}{120x^4}+\frac{1}{60x^3}+\frac{19}{240x^2}.... - \frac{1}{240}\\
\nonumber
+\frac{1}{12}\Big(\frac{1}{x^2}+\frac{1}{x}-\frac{1}{20}-\frac{x}{20}\Big) e^x E(x)\Big) \Big\}=0. 
\end{align}
\end{subequations}
From equation (63) it appears\footnote{After all $S$ is proportional to the additional charge density in the double layer.}, that in a solution of a non symmetric electrolyte ($z_-\neq z_+$) already in the first approximation relaxation effects appear, while for a symmetric electrolyte these effects only appear in the second order approximation. We shall for this reason from this moment onward only perform the calculation of the second order approximation for symmetric electrolytes and thus remove the terms proportional to $(z_--z_+)^2A^2e^{-2x}$. The terms proportional to $(z_--z_+)PAe^{-2x}$ are for the time being kept, since we do not know anything about the magnitude of $P$ (see \BLUE{last pages before Chapter IV-C} for more info). By furthermore combining all terms proportional to $(z_--z_+)PAe^{-2x}$, (68) can be written in a slightly more compact manner.
\begin{subequations}
\begin{align}
\nonumber \tag{68'}
\frac{d^2S}{dx^2} + \frac{2}{x}\frac{dS}{dx}-\frac{2S}{x^2}-S =V(x)=- (z_--z_+)Ae^{-x}\Big(\frac{1}{x^2}+\frac{1}{x}\Big)\\
\nonumber
+(z_--z_+)BA e^{-x} \Big(\frac{2}{x^5}+\frac{2}{x^4}\Big)+(z_--z_+)PAe^{-2x}\Big(\frac{4}{x^5}+\frac{8}{x^4}+\frac{8}{x^3}+\frac{4}{x^2}\Big)\\
\nonumber
-(z_++z_-)CAe^{-x}\Big(\frac{2}{x^5}+\frac{2}{x^4}+\frac{1}{x^3}\Big)\\
\nonumber
-(z_+^2-z_+z_-+z_-^2)A^2 e^{-2x}\Big(\frac{2}{x^5}+\frac{4}{x^4}+\frac{4}{x^3}+\frac{2}{x^2}\Big)\\
\nonumber
+(z_+^2-z_+z_-+z_-^2)BA^2e^{-2x}\Big\{\frac{3}{2x^6}+\frac{5}{3x^5}+\frac{7}{12x^4}\\
\nonumber
-\frac{1}{6x^3} +\frac{1}{6x^2}-\frac{1}{12x}+\Big(\frac{1}{12x^2}+
\frac{1}{12x}-\frac{1}{12}\Big) e^x E(x)\Big\}\\
\nonumber
-(z_+^2-z_- z_++z_+^2) A^2 e^{-2x}\Big(\frac{1}{x^3}+\frac{1}{x^2}\Big)
\end{align}
\end{subequations}
\begin{subequations}
\begin{align}
\nonumber
+(z_+^2-z_-z_++z_-^2)BAe^{-2x}\Big(\frac{2}{x^6}+\frac{2}{x^5}\Big)\\
\nonumber
-Ge^{-x}\Big(\frac{2}{x^2}+\frac{2}{x}\Big) +He^{-x}
\Big( \frac{4}{x^5}+\frac{4}{x^4}\Big)\\
\nonumber
\qquad \quad +Je^{-2x}\Big\{\frac{4}{x^5}+\frac{8}{x^4}+\frac{4}{x^3}+B\Big(\frac{4}{15x^5}+\frac{1}{30x^4}+\frac{1}{15x^3}+\frac{19}{60x^2}...-\frac{1}{60}\\
\nonumber
+\frac{1}{3}\Big(\frac{1}{x^2}+\frac{1}{x}-\frac{1}{20}-\frac{x}{20}\Big) e^x E(x)\Big) \Big\},
\end{align}
\end{subequations}
in which
\begin{subequations}
\begin{align}
\nonumber
\tag{69} &G= \frac{z_+/\lambda_++z_-/\lambda_-}{(z_++z_-)e}
\frac{UA}{2} \qquad \qquad \qquad
\\
\nonumber
&H=\frac{z_+/\lambda_++z_-/\lambda_-}{(z_++z_-)e}
\frac{FA}{4} \qquad \qquad \qquad
\\
\nonumber
&J=-\frac{z_+/\lambda_++z_-/\lambda_-}{(z_++z_-)e}
\frac{2 \varepsilon_0 \varepsilon_r kT}{\eta e}\frac{A^2}{4} \qquad
.
\end{align}
\end{subequations}
The solution for $S$ is then given by
\begin{subequations}
\begin{align}
\nonumber
S&=P\Big(\frac{1}{x^2}+\frac{1}{x}\Big) e^{-x}-Q\Big(\frac{1}{x^2}-\frac{1}{x}\Big) e^x\\
\nonumber
&+\frac{1}{2} \int^x V(\xi) \cdot \xi^2
\Big\{\Big(\frac{1}{\xi^2}-\frac{1}{\xi}\Big) e^\xi\Big(\frac{1}{x^2}+\frac{1}{x}\Big)e^{-x} -\Big(\frac{1}{\xi^2}+\frac{1}{\xi}\Big)e^{-\xi}\Big(\frac{1}{x^2}-\frac{1}{x}\Big) e^x \Big\} d\xi,
\end{align}
\end{subequations}
which after doing the integration and noting that $Q=0$ (see also the remark after equation (63)), gives
\begin{subequations}
\begin{align}
\nonumber
\tag{70}
\qquad \quad S=P\Big( \frac{1}{x^2} +\frac{1}{x}\Big) e^{-x} + (z_- - z_+)Ae^{-x} \Big( \frac{3}{4x^2} +\frac{3}{4x} +\frac{1}{2}\Big) \\
\nonumber
+(z_- - z_+) BAe^{-x} \frac{1}{2x^3} + (z_- - z_+) PA e^{-2x} \Big\{ \frac{1}{x^3} + \frac{1}{x^2}\\
\nonumber
-\frac{1}{4} \Big( \frac{1}{x^2} + \frac{1}{x}\Big) e^x E(x) + \frac{1}{4} \Big( \frac{1}{x^2} - \frac{1}{x}\Big) e^{3x} E(3x) \Big\}\\
\nonumber
-(z_+ + z_-)CA e^{-x} \Big(\frac{1}{2x^3} + \frac{1}{2x^2}\Big)
\end{align}
\end{subequations}
\begin{subequations}
\begin{align}
\nonumber
-(z_+^2 -z_+z_- +z_-^2)A^2e^{-2x} \Big\{ \frac{1}{2x^3} +\frac{1}{2x^2}\\
\nonumber
-\frac{1}{8}\Big( \frac{1}{x^2} +
\frac{1}{x}\Big) e^x E(x) +\frac{1}{8}\Big( \frac{1}{x^2} -\frac{1}{x}\Big) e^{3x} E(3x) \Big\} \\
\nonumber
+(z_+^2 - z_+ z_- +z_-^2)BA^2 e^{-2x} \Big\{ \frac{3}{20x^4}  -\frac{1}{30x^3} - \frac{23}{40x^2}\\
\nonumber
-\frac{1}{16x} +\frac{1}{48} -\Big( \frac{3}{32x^2} +\frac{3}{32x} +\frac{1}{24} -\frac{x}{48} \Big) e^x E(x) \\
\nonumber 
\qquad \qquad \quad -\frac{13}{30} \Big( \frac{1}{x^2} +\frac{1}{x}\Big) e^x E(x) + \frac{63}{100}\Big( \frac{1}{x^2} - \frac{1}{x}\Big) e^{3x} E(3x) \Big\}
\end{align}
\end{subequations}
\begin{subequations}
\begin{align}
\nonumber
+(z_+^2 - z_+z_- +z_-^2) A^2e^{-2x} \Big\{ \frac{1}{4}\Big( \frac{1}{x^2} +\frac{1}{x}\Big) e^x E(x)\\
\nonumber
-\frac{1}{4} \Big( \frac{1}{x^2} -\frac{1}{x}\Big) e^{3x} E(3x) \Big\} \\
\nonumber
+(z_+^2 - z_+ z_- +z_-^2)BA^2 e^{-2x} \Big\{ \frac{1}{5x^4} - \frac{1}{10x^3} +\frac{7}{30x^2} \\
\nonumber
\qquad \qquad \quad +\frac{19}{120} \Big( \frac{1}{x^2} +\frac{1}{x}\Big) e^x E(x) - \frac{9}{40}\Big( \frac{1}{x^2} - \frac{1}{x}\Big) e^{3x} E(3x) \Big\}
\end{align}
\end{subequations}
\begin{subequations}
\begin{align}
\nonumber
+Ge^{-x} + He^{-x} \cdot \frac{1}{x^3} +J e^{-2x}\Big\{ \frac{1}{x^3} +\frac{1}{x^2} +\frac{3}{4} \Big( \frac{1}{x^2} +\frac{1}{x}\Big) e^x E(x) \\
\nonumber
-\frac{3}{4} \Big( \frac{1}{x^2} -\frac{1}{x}\Big) e^{3x}E(3x) \Big\}\\
\nonumber
+BJe^{-2x} \Big\{ \frac{1}{15x^3} -\frac{143}{180x^2} - \frac{121}{720x} + \frac{1}{240} + \frac{x}{360}\\
\nonumber
-\Big( \frac{25}{96x^2} + \frac{25}{96x} +\frac{1}{6} -\frac{x}{144} - \frac{x^2}{360} \Big) e^x E(x)\\
\nonumber
-\frac{5}{12} \Big( \frac{1}{x^2} +\frac{1}{x}\Big) e^x E(x) + \frac{27}{32} \Big( \frac{1}{x^2} - \frac{1}{x}\Big) e^{3x} E(3x) \Big\}.
\end{align}
\end{subequations}
\textbf{G.} \textit{Second approximation of $I$.}\\
This value of $S$ is now substituted in (59d) and gives a differential equation for $I$, from which the solution can be found once more according to (61)
\begin{subequations}
\begin{align}
\nonumber
\tag{71} I= x +\frac{B}{x^2} +P\Big( \frac{1}{x^2} +\frac{1}{x}\Big) e^{-x} +(z_--z_+) A e^{-x}\Big( \frac{7}{4x^2} + \frac{7}{4x} +\frac{1}{2}\Big) \\
\nonumber
+(z_- - z_+) BA e^{-x} \Big( \frac{1}{6x^2} -\frac{1}{12x} +\frac{1}{12} +\frac{xe^x}{12}E(x) \Big) \\
\nonumber
-(z_- - z_+) PA e^{-2x} \Big\{ \frac{1}{4} \Big( \frac{1}{x^2} + \frac{1}{x}\Big) e^x E(x) - \frac{1}{4}\Big( \frac{1}{x^2} -\frac{1}{x}\Big) e^{3x} E(3x) \Big\}\\
\nonumber
- (z_+ + z_-) CA e^{-x} \Big( \frac{1}{3x^2} +\frac{1}{12x} - \frac{1}{12} - \frac{xe^x}{12} E(x) \Big) 
\end{align}
\end{subequations}
\begin{subequations}
\begin{align}
\nonumber
 +(z_+^2 - z_+ z_- +z_-^2) A^2e^{-2x} \Big\{ \frac{1}{8} \Big( \frac{1}{x^2} +\frac{1}{x} \Big) e^x E(x) \\
\nonumber
-\frac{1}{8} \Big( \frac{1}{x^2} -\frac{1}{x}\Big) e^{3x}E(3x) \Big\}\\
\nonumber
\qquad \quad \qquad - (z_+^2 - z_+ z_- +z_-^2) BA^2 e^{-2x} \Big\{ + \frac{35}{72x^2} + \frac{49}{720x} - \frac{143}{720}\\
\nonumber
+\Big( \frac{1}{96x^2} +\frac{1}{96x} - \frac{1}{24} - \frac{x}{48}\Big) e^x E(x) - \frac{16}{45} xe^{2x} E(2x) \\
\nonumber
+\frac{13}{30}\Big( \frac{1}{x^2} + \frac{1}{x}\Big) e^x E(x) - \frac{63}{160}\Big( \frac{1}{x^2}- \frac{1}{x}\Big) e^{3x} E(3x) \Big\}
\end{align}
\end{subequations}
\begin{subequations}
\begin{align}
\nonumber
+ (z_+^2 - z_+ z_- + z_-^2) A^2 e^{-2x} \Big\{ \frac{1}{4x^2} +\frac{1}{4}\Big( \frac{1}{x^2} +\frac{1}{x}\Big) e^x E(x) \\
\nonumber
-\frac{1}{4}\Big( \frac{1}{x^2}- \frac{1}{x}\Big) e^{3x} E(3x) \Big\} \\
\nonumber
\quad \qquad +(z_+^2 - z_+ z_- + z_-^2) BA^2 e^{-2x}\Big\{ \frac{1}{6x^2} +\frac{1}{15x} -\frac{2}{15} - \frac{4xe^{2x}}{15} E(2x) \\
\nonumber
+\frac{19}{120}\Big( \frac{1}{x^2} +\frac{1}{x}\Big) e^x E(x) - \frac{9}{40} \Big( \frac{1}{x^2} - \frac{1}{x}\Big) e^{3x}E(3x) \Big\}
\end{align}
\end{subequations}
\begin{subequations}
\begin{align}
\nonumber +Ge^{-x} \Big( \frac{2}{x^2} +\frac{2}{x} + 1\Big) \\
\nonumber +H e^{-x} \Big( \frac{1}{3x^2} - \frac{1}{6x} +\frac{1}{6} +\frac{xe^x}{6} E(x) \Big) \\
\nonumber
\qquad +Je^{-2x} \Big\{\frac{1}{x^2} +\frac{3}{4} \Big( \frac{1}{x^2} +\frac{1}{x}\Big) e^x E(x) -\frac{3}{4} \Big( \frac{1}{x^2} -\frac{1}{x}\Big) e^{3x} E(3x) \Big\}\\
\nonumber
-BJe^{-2x} \Big\{ \frac{83}{90x^2} +\frac{137}{720x} -\frac{143}{720} -\frac{x}{360} +\Big( \frac{41}{96x^2} + \frac{41}{96x}\\
\nonumber
+ \frac{1}{12} -\frac{17x}{720} -\frac{x^2}{360}\Big) e^x E(x) - \frac{16xe^{2x}}{45} E(2x) \\
\nonumber
+\frac{5}{12} \Big( \frac{1}{x^2} + \frac{1}{x}\Big) e^x E(x) - \frac{27}{32}\Big( \frac{1}{x^2} - \frac{1}{x}\Big) e^{3x} E(3x) \Big\}.
\end{align}
\end{subequations}
\textbf{H.} \textit{Second approximation of $\Sigma_+$ and $\Sigma_-$.}\\
The now obtained relationships for $\Sigma_\pm$ (66), $S$ (70), $I$ (71) and $\frac{u_r}{E \cos \theta}$ (67) are used in (59a and b), which gives the following differential equations for $\Sigma_+$ and $\Sigma_-$. (Terms proportional to $(z_--z_+)A^2e^{-2x}$ are being neglected).
\marginnote{$kT \nabla^2 \delta n $ \\
$\delta n \psi^0$ }[1cm]
\begin{subequations}
\begin{align}
\nonumber
\tag{72} \frac{d^2 \Sigma_\pm}{dx^2} +\frac{2}{x} \frac{d \Sigma_\pm}{dx} - \frac{2 \Sigma_\pm}{x^2} \\
\nonumber
\pm z_\pm CAe^{-x}\Big( \frac{2}{x^5} +\frac{2}{x^4} +\frac{1}{x^3}\Big) + \frac{z_\pm^2}{z_++z_-}PAe^{-2x} \Big(\frac{2}{x^5} +\frac{4}{x^4} +\frac{4}{x^3} +\frac{2}{x^2}\Big) \\
\nonumber
\pm \frac{z_\pm^3}{z_++z_-}A^2 e^{-2x}\Big( \frac{2}{x^5} +\frac{4}{x^4} +\frac{4}{x^3} +\frac{2}{x^2}\Big) \\
\nonumber
\mp \frac{z_\pm^3}{z_++z_-}BA^2e^{-2x} \Big( \frac{3}{2x^6} + \frac{5}{3x^5} +\frac{7}{12x^4} -\frac{1}{6x^3} +\frac{1}{6x^2} -\frac{1}{12x} \\
\nonumber
+\frac{1}{12} \Big( \frac{1}{x^2} +\frac{1}{x} -1 \Big) e^x E(x) \Big)
\end{align}
\end{subequations}
\marginnote{$n^0 \nabla^2 \delta \psi $ }[1cm]
\begin{subequations}
\begin{align}
\nonumber
\mp \frac{z_\pm}{z_++z_-} Pe^{-x} \Big( \frac{1}{x^2} + \frac{1}{x} \Big) \\
\nonumber
\mp \frac{z_\pm (z_- - z_+)}{z_+ + z_-} A e^{-x} \Big( \frac{3}{4x^2} +\frac{3}{4x} +\frac{1}{2}\Big) \mp \frac{z_\pm (z_- -z_+)}{z_++z_-} BAe^{-x}\frac{1}{2x^3}\\
\nonumber
\mp \frac{z_\pm (z_- - z_+)}{z_+ + z_-} PAe^{-2x} \Big\{ \frac{1}{x^3} +\frac{1}{x^2} -\frac{1}{4} \Big( \frac{1}{x^2} +\frac{1}{x}\Big) e^x E(x)\\
\nonumber
+\frac{1}{4} \Big( \frac{1}{x^2}- \frac{1}{x}\Big) e^{3x} E(3x) \Big\} \pm z_\pm CAe^{-x} \Big( \frac{1}{2x^3}+ \frac{1}{2x^2} \Big) \\
\nonumber
\qquad \qquad \pm \frac{z_\pm (z_+^2 - z_+ z_- + z_-^2)}{z_+ +z_-} A^2 e^{-2x} \Big\{ \frac{1}{2x^3} +\frac{1}{2x^2}-\frac{1}{8} \Big( \frac{1}{x^2} +\frac{1}{x}\Big) e^x E(x) \\
\nonumber
+\frac{1}{8} \Big( \frac{1}{x^2}- \frac{1}{x} \Big) e^{3x} E(3x) \Big\}
\end{align}
\end{subequations}
\begin{subequations}
\begin{align}
\nonumber
\qquad \qquad \mp \frac{z_\pm (z_+^2 - z_+ z_- + z_-^2)}{z_+ + z_-} BA^2 e^{-2x} \Big\{ \frac{3}{20x^4} -\frac{1}{30x^3} -\frac{23}{40x^2} -\frac{1}{16x} +\frac{1}{48}\\
\nonumber
-\Big( \frac{3}{32x^2} +\frac{3}{32x} +\frac{1}{24} -\frac{x}{48}\Big) e^x E(x)\\
\nonumber
-\frac{13}{30} \Big( \frac{1}{x^2} + \frac{1}{x}\Big) e^x E(x) + \frac{63}{160} \Big( \frac{1}{x^2} -\frac{1}{x} \Big) e^{3x} E(3x) \Big\} \\
\nonumber
\mp \frac{z_\pm(z_+^2 - z_+ z_- +z_-^2)}{z_++ z_-} A^2 e^{-2x} \Big\{ \frac{1}{4}\Big( \frac{1}{x^2} + \frac{1}{x} \Big) e^x E(x) \\
\nonumber
- \frac{1}{4} \Big( \frac{1}{x^2} - \frac{1}{x}\Big) e^{3x} E(3x) \Big\}
\end{align}
\end{subequations}
\begin{subequations}
\begin{align}
\nonumber
 \mp \frac{z_\pm(z_+^2 - z_+ z_- +z_-^2)}{z_+ + z_-} BA^2e^{-2x} \Big\{ \frac{1}{5x^4} - \frac{1}{10x^3} + \frac{7}{30x^2}\\
\nonumber
+\frac{19}{120}\Big( \frac{1}{x^2} + \frac{1}{x}\Big) e^x E(x) -\frac{9}{40} \Big( \frac{1}{x^2} - \frac{1}{x}\Big) e^{3x} E(3x) \Big\} \\
\nonumber
\qquad \qquad \qquad \mp \frac{z_\pm}{z_+ + z_-} G e^{-x} \mp \frac{z_\pm}{z_++z_-} H e^{-x} \frac{1}{x^3} \mp \frac{z_\pm}{z_++z_-} J e^{-2x} \Big\{ \frac{1}{x^3} + \frac{1}{x^2}\\
\nonumber
+\frac{3}{4} \Big( \frac{1}{x^2} +\frac{1}{x} \Big) e^x E(x) - \frac{3}{4} \Big( \frac{1}{x^2} - \frac{1}{x}\Big) e^{3x} E(3x) \Big\}
\end{align}
\end{subequations}
\begin{subequations}
\begin{align}
\nonumber
\qquad \qquad \qquad \qquad \quad \mp \frac{z_\pm}{z_+ + z_-} BJ e^{-2x} \Big\{ \frac{1}{15x^3} -\frac{143}{180x^2} -\frac{121}{720x} +\frac{1}{240} +\frac{x}{360}\\
\nonumber
-\Big( \frac{25}{96x^2} +\frac{25}{96x} +\frac{1}{6} -\frac{x}{144} -\frac{x^2}{360} \Big) e^x E(x)\\
\nonumber
-\frac{5}{12} \Big( \frac{1}{x^2} +\frac{1}{x} \Big) e^x E(x) + \frac{27}{32}\Big( \frac{1}{x^2}- \frac{1}{x}\Big) e^{3x} E(3x) \Big\}
\end{align}
\end{subequations}
\marginnote{$\frac{\partial n^0}{\partial r} \frac{\partial \delta \psi}{\partial r}$  }[1cm]
\begin{subequations}
\begin{align}
\nonumber
+\frac{z_\pm^2}{z_++z_-} PAe^{-2x}\Big( \frac{1}{x^3} + \frac{1}{x^2} \Big) \\
\nonumber
-\frac{z_\pm^2}{z_++z_-} Ae^{-x} \Big( \frac{1}{x^2} + \frac{1}{x} \Big) + \frac{z_\pm^2}{z_++z_-} BA e^{-x} \Big( \frac{2}{x^5} + \frac{2}{x^4}\Big) \\
\nonumber
+\frac{z_\pm^2}{z_++z_-} PA e^{-2x} \Big( \frac{2}{x^5} + \frac{4}{x^4} + \frac{3}{x^3} + \frac{1}{x^2} \Big) \\
\nonumber
\qquad \qquad \qquad \quad \pm \frac{z_\pm^3}{z_++z_-} A^2 e^{-2x} \Big( \frac{1}{x^3} + \frac{1}{x^2} \Big) \mp \frac{z_\pm^3}{z_++z_-} BA^2 e^{-2x} \Big( \frac{2}{x^6} + \frac{2}{x^5}\Big)  
\end{align}
\end{subequations}
\marginnote{$u n^0$  }[1cm]
\begin{subequations}
\begin{align}
\nonumber
\qquad \qquad \qquad \quad \pm \ifRED\else \frac{z_\pm /\lambda_\pm}{(z_++z_-)e}\fi UAe^{-x} \Big( \frac{1}{x^2} + \frac{1}{x} \Big) \pm \ifRED\else \frac{z_\pm /\lambda_\pm}{(z_++z_-)e}\fi FAe^{-x} \Big( \frac{1}{x^5} + \frac{1}{x^4} \Big) \\
\nonumber
\pm \ifRED\else \frac{z_\pm /\lambda_\pm}{(z_++z_-)e} \frac{2\varepsilon_0 \varepsilon_r kT}{\eta e}\fi A^2 e^{-2x} \Big\{ \frac{1}{x^5} +\frac{2}{x^4} + \frac{1}{x^3}\\
\nonumber
+B \Big( \frac{1}{15x^5} +\frac{1}{120x^4} +\frac{1}{60x^3} + \frac{19}{240x^2} .... -\frac{1}{240}\\
\nonumber
+\frac{1}{12}\Big( \frac{1}{x^2} +\frac{1}{x} -\frac{1}{20}-\frac{x}{20} \Big) e^x E(x) \Big) \Big\} = 0.
\end{align}
\end{subequations}
In order not to loose the connection with reality in these very long and obscure equations, in the margins is indicated with which terms of equation (10') \BLUE{just before (57c)} certain terms in (72) and the following equations correspond.

The solutions of equations (72) are (here we have already taken into account the remark after (66) that $C_1=0$)
\marginnote{$kT \nabla^2 \delta n $ \\
$\delta n \psi^0$ }[1cm]
\begin{subequations}
\begin{align}
\nonumber
\tag{73} \Sigma_\pm = \frac{C}{x^2}\\
\nonumber
\qquad \qquad \qquad \qquad \qquad \mp z_\pm CA e^{-x} \Big( \frac{1}{2x^3} + \frac{1}{6x^2} - \frac{1}{12x} +\frac{1}{12} +\frac{xe^x}{12} E(x) \Big)
\end{align}
\end{subequations}
\marginnote{$n^0 \nabla^2 \delta \psi $ }[3.5cm]
\begin{subequations}
\begin{align}
\nonumber
-\frac{z_\pm^2}{z_++z_-} PAe^{-2x} \Big( \frac{1}{2x^3} +\frac{1}{2x^2}\Big) \\
\nonumber
\mp \frac{z_\pm^3}{z_++z_-} A^2 e^{-2x} \Big( \frac{1}{2x^3}+\frac{1}{2x^2}\Big) \pm \frac{z_\pm^3}{z_++z_-} BA^2 e^{-2x} \Big\{ \frac{3}{20x^4}\\
\nonumber
-\frac{1}{30x^3} -\frac{4}{45x^2} +\frac{1}{180x} -\frac{8}{45} -\frac{16xe^{2x}}{45} E(2x) - \frac{1}{12} \Big( \frac{1}{x^2} +\frac{1}{x} +1\Big) e^x E(x) \Big\}\\
\nonumber
\pm \frac{z_\pm}{z_++ z_-}Pe^{-x} \Big( \frac{1}{x^2} +\frac{1}{x}\Big) \\
\nonumber
\pm \frac{z_\pm(z_- - z_+)}{z_++z_-} Ae^{-x} \Big( \frac{7}{4x^2} + \frac{7}{4x} +\frac{1}{2} \Big)
\end{align}
\end{subequations}
\begin{subequations}
\begin{align}
\nonumber
\pm \frac{z_\pm(z_--z_+)}{z_++z_-} BA e^{-x} \Big(\frac{1}{6x^2} -\frac{1}{12x} +\frac{1}{12} +\frac{xe^x}{12}E(x) \Big)\\
\nonumber
\qquad \mp \frac{z_\pm(z_--z_+)}{z_++z_-} PAe^{-2x} \Big\{ \frac{1}{4} \Big( \frac{1}{x^2} + \frac{1}{x}\Big) e^x E(x) - \frac{1}{4} \Big( \frac{1}{x^2} -\frac{1}{x} \Big) e^{3x} E(3x) \Big\}\\
\nonumber
\mp CA e^{-x} \Big( \frac{1}{3x^2} +\frac{1}{12x} - \frac{1}{12} -\frac{xe^x}{12} E(x) \Big)\\
\nonumber
\pm \frac{z_\pm(z_+^2 -z_+z_- +z_-^2)}{z_+ + z_-} A^2 e^{-2x} \Big\{ \frac{1}{8} \Big( \frac{1}{x^2}+ \frac{1}{x}\Big) e^x E(x) \\
\nonumber
-\frac{1}{8} \Big( \frac{1}{x^2}- \frac{1}{x}\Big) e^{3x} E(3x) \Big\}
\end{align}
\end{subequations}
\begin{subequations}
\begin{align}
\nonumber
\qquad \qquad \qquad \qquad \mp \frac{z_\pm(z_+^2 -z_+z_- +z_-^2)}{z_++z_-} BA^2 e^{-2x} \Big\{ \frac{35}{72x^2} +\frac{49}{720x} - \frac{143}{720}\\
\nonumber
-\frac{16xe^{2x}}{45} E(2x) + \Big( \frac{1}{96x^2} +\frac{1}{96x} -\frac{1}{24} -\frac{x}{48} \Big) e^x E(x)\\
\nonumber
+\frac{13}{30} \Big( \frac{1}{x^2} + \frac{1}{x}\Big) e^x E(x) - \frac{63}{100}\Big( \frac{1}{x^2}- \frac{1}{x}\Big) e^{3x} E(3x) \Big\}\\
\nonumber
\pm \frac{z_\pm(z_+^2 -z_+ z_- +z_-^2)}{z_+ + z_-} A^2 e^{-2x}\Big\{ \frac{1}{4x^2}\\
\nonumber
+\frac{1}{4} \Big( \frac{1}{x^2}+\frac{1}{x} \Big) e^x E(x) - \frac{1}{4}\Big( \frac{1}{x^2}-\frac{1}{x} \Big)e^{3x} E(3x) \Big\}
\end{align}
\end{subequations}
\begin{subequations}
\begin{align}
\nonumber
\pm \frac{z_\pm (z_+^2 - z_+ z_- + z_-^2)}{z_+ + z_-} BA^2 e^{-2x} \Big\{ \frac{1}{6x^2} +\frac{1}{15x} -\frac{2}{15}\\
\nonumber
\qquad \quad -\frac{4xe^{2x}}{15} E(2x) + \frac{19}{120} \Big(\frac{1}{x^2} + \frac{1}{x}\Big) e^x E(x) - \frac{9}{40} \Big( \frac{1}{x^2}-\frac{1}{x}\Big) e^{3x} E(3x) \Big\} \\
\nonumber
\pm \frac{z_\pm}{z_++z_-} G e^{-x} \Big( \frac{2}{x^2} + \frac{2}{x} +1 \Big)  \pm \frac{z_\pm}{z_+ + z_-} He^{-x} \Big( \frac{1}{3x^2}\\
\nonumber
- \frac{1}{6x} +\frac{1}{6} +\frac{xe^x}{6} E(x) \Big) \pm \frac{z_\pm}{z_+ +z_-} Je^{-2x} \Big\{ \frac{1}{x^2}\\
\nonumber
+\frac{3}{4}\Big( \frac{1}{x^2} +\frac{1}{x} \Big) e^x E(x) - \frac{3}{4} \Big( \frac{1}{x^2} - \frac{1}{x} \Big) e^{3x} E(3x) \Big\}
\end{align}
\end{subequations}
\begin{subequations}
\begin{align}
\nonumber
\mp -\frac{z_\pm}{z_++z_-} BJ e^{-2x} \Big\{ \frac{83}{90x^2} +\frac{137}{720x} -\frac{143}{720} -\frac{x}{360}\\
\nonumber
\qquad \qquad \quad +\Big( \frac{41}{96x^2} + \frac{41}{96x} +\frac{1}{12} -\frac{17x}{720} -\frac{x^2}{360} \Big) e^x E(x) - \frac{16 x e^{2x}}{45} E(2x) \\
\nonumber
+\frac{5}{12}\Big( \frac{1}{x^2} + \frac{1}{x} \Big) e^x E(x) - \frac{27}{32} \Big( \frac{1}{x^2} - \frac{1}{x}\Big) e^{3x} E(3x) \Big\}
\end{align}
\end{subequations}
\marginnote{$\frac{\partial n^0}{\partial r} \frac{\partial \delta \psi}{\partial r}$  }[1cm]
\begin{subequations}
\begin{align}
\nonumber
-\frac{z_\pm^2}{z_++z_-} PA e^{-2x} \frac{1}{4x^2}\\
\nonumber
+\frac{z_\pm^2}{z_++z_-}Ae^{-x} \Big( \frac{1}{x^2} +\frac{1}{x}\Big) - \frac{z_\pm^2}{z_++z_-}BAe^{-x} \Big( \frac{1}{2x^3}-\frac{1}{6x^2}\\
\nonumber
\qquad \qquad \qquad +\frac{1}{12x} -\frac{1}{12}-\frac{xe^x}{12}E(x) \Big) -\frac{z_\pm^2}{z_++z_-}PAe^{-2x} \Big( \frac{1}{2x^3}+\frac{1}{4x^2}\Big) \\
\nonumber
\mp \frac{z_\pm^3}{z_++z_-} A^2e^{-2x} \frac{1}{4x^2} \pm \frac{z_\pm^3}{z_++z_-} BA^2 e^{-2x}
\Big( \frac{1}{5x^4} - \frac{1}{10x^3}\\
\nonumber
+\frac{1}{15x^2} -\frac{1}{15x} + \frac{2}{15} +\frac{4xe^{2x}}{15} E(2x) \Big) 
\end{align}
\end{subequations}
\marginnote{$u n^0$  }[1cm]
\begin{subequations}
\begin{align}
\nonumber
\mp \ifRED\else \frac{z_\pm /\lambda_\pm}{(z_++z_-)e}\fi UA e^{-x} \Big( \frac{1}{x^2} + \frac{1}{x} \Big)\\
\nonumber
\pm \ifRED\else \frac{z_\pm /\lambda_\pm}{(z_++z_-)e}\fi FA e^{-x}\Big( \frac{1}{4x^3} -\frac{1}{12x^2} +\frac{1}{24x} -\frac{1}{24} - \frac{xe^x}{24} E(x) \Big)\\
\nonumber
\mp \ifRED\else \frac{z_\pm /\lambda_\pm}{(z_++z_-)e} \frac{2\varepsilon_0 \varepsilon_r kT}{\eta e}\fi A^2 e^{-2x} \Big\{ \frac{1}{4x^3} + B\Big( \frac{1}{60x^3}+ \frac{23}{720 x^2} + \frac{1}{180x}\\
\nonumber
\qquad \qquad \qquad -\frac{7}{144} -\frac{4x e^{2x}}{45}E(2x) + \Big( \frac{1}{24x^2} + \frac{1}{24x}- \frac{1}{48} - \frac{x}{240} \Big) e^x E(x) \Big) \Big\}. 
\end{align}
\end{subequations}
Now the ion distributions ($\Sigma_+$ and $\Sigma_-$), the charge density $S$ and the field $I$ around the moving particle are known, except for the still to be determined integration constants $P$ from (70) and (63, $B$ from (71) and (61') and $C$ from (73) and (66). These must be determined from the boundary conditions on the boundary of the sphere and fluid (the boundary conditions at infinity are already used to determine the constants $Q$ from (63), $B_1$ from (61) and $C_1$ from (66)).
\\
\\
\textbf{I.} \textit{The taking into account of the boundary conditions.}\\
From the boundary conditions 4, 5 and 6 \BLUE{(just before Chapter II-B)}, the field inside the sphere can be determined in addition to a relationship between the constants $P$, $B$ and $C$. 

If the potential inside the sphere is called $(\delta \psi_i + \psi^0_i)$ and the charge density on the sphere surface is $\alpha=\frac{n\varepsilon_i}{4\pi a^2}$, then these boundary conditions can be easily expressed as\\
condition 4. $(\delta \psi + \psi^0)_{r=a} = (\delta \psi_i + \psi^0_i)_{r=a}$ \\
condition 5. $\epsilon_0 \epsilon_r \Big(\frac{\partial}{\partial r}(\delta \psi +\psi^0)\Big)_{r=a} = \epsilon_0\epsilon_{ri} \Big(\frac{\partial}{\partial r}(\delta \psi_i +\psi^0_i)\Big)_{r=a} = \alpha$\\ .

Since at rest $\epsilon_0 \epsilon_r \Big(\frac{\partial \psi^0}{\partial r} \Big)_{r=a}=\varepsilon_0 \varepsilon_{ri}\Big(\frac{\partial \psi_i^0}{\partial r} \Big)_{r=a} + \alpha$ and $\psi_i^0= \psi_a^0=\zeta$ (see also (47)), conditions 4 and 5 can be reduced to 
\begin{align}
\nonumber \tag{74} 
\delta \psi_{r=a}=\big(\delta \psi_i\big)_{r=a} \hspace{1cm} \text{and} \hspace{1cm} \epsilon_r\Big(\frac{\partial \delta \psi}{dr}\Big)_{r=a} = \epsilon_{ri}\Big(\frac{\partial \delta \psi_i}{dr}\Big)_{r=a}
\end{align}
Inside the sphere, the charge density is zero everywhere, thus
\begin{equation}
\begin{aligned}
\nonumber
\nabla^2 \delta \psi_i = 0.
\hspace{2cm} 
\end{aligned}
\end{equation}
The potential inside the sphere can thus be given by
\begin{align}
\nonumber \tag{75}
\delta \psi_i= -E \Big( Lr + \frac{M}{r^2}\Big) \cos \theta. \hspace{2cm}
\end{align}
The coefficient $M$ must be zero, since otherwise in the center of the sphere the potential would become infinite (boundary condition 6 \BLUE{just before Chapter II-B}), thus
\begin{align}
\nonumber \tag{76}
\delta \psi_i = -ELr \cos\theta. \hspace{2cm}
\end{align}
From (76), (74) and (17) we then immediately get
\begin{equation}
\begin{aligned}
\nonumber
La=R_a \hspace{5 mm} \text{and} \hspace{5 mm} \epsilon_r\Big( \frac{dR}{dr}\Big)_{r=a} = \epsilon_{ri} L \hspace{1cm} \hspace{5 mm} \text{or}
\end{aligned}
\end{equation}
\begin{align}
\nonumber \tag{77}
\epsilon_r\Big(\frac{dR}{dr}\Big)_{r=a}=\epsilon_{ri} \frac{R_a}{a}. \hspace{2cm}
\end{align}
After introducing new variables $x=\kappa r$ and $I=\kappa R$ we get from (77)
\begin{align} \tag{77'}
\nonumber
\epsilon_r\Big(\frac{dI}{dx}\Big)_{x=\kappa a}=\epsilon_{ri}\Big(\frac{I}{x}\Big)_{x=\kappa a}. \hspace{2cm}
\end{align}
By substituting the value of $I$ from (71) into (77), we find one expression, to which the three constants $P$, $B$ and $C$ must obey. 

The other boundary conditions (7 and 8 \BLUE{just before Chapter II-B}) indicate something concerning the transport of charge carriers (ions) towards the boundary. We will now for the time being only look at an \textit{insulating} sphere and in chapter VI revisit the calculation of the electrophoretic velocity for a conducting sphere. For an insulating sphere, in the stationary state, the transport of anions and cations through the boundary sphere-fluid must be zero. According to equation (5), the ion transport is determined by the vectors $t_+$ and $t_-$. On the boundary of the sphere and the liquid the $r$-components of these vectors must disappear.\\
$r$-component of\\ $\Big(-n_+z_+e\lambda_+\nabla(\delta \psi + \psi^0)-kT\lambda_+\nabla n_+ + n_+u \Big)=0 $ and\\
$r$-component of\\ $\Big(+n_-z_-e\lambda_-\nabla(\delta \psi + \psi^0)-kT\lambda_-\nabla n_- + n_-u \Big)=0 $.\\
Using the expressions
\begin{equation} \nonumber \tag{7}
    n_\pm = n_\pm^0 + \delta n_\pm \qquad \qquad 
\end{equation}

\begin{equation} \nonumber \tag{9}
    n_{\pm}^0 = n_{\pm}^\infty \exp(\mp z_{\pm} e \psi^0/kT) 
\end{equation}
and taking into account that $u$ becomes zero at the boundary of the sphere, and that terms proportional to $\delta n \nabla \delta \psi$ can be neglected, the boundary conditions can be written as
\begin{align}
\nonumber \tag{78}
\lambda_\pm\Big\{\mp n_\pm^0 z_\pm e\frac{\partial \delta \psi}{\partial r} \mp \delta n_\pm z_\pm e\frac{d \psi^0}{d r}-kT\frac{\partial \delta n_\pm}{\partial r}\Big\}_{r=a}=0 
.
\end{align}
By introducing new variables $x$ (51), $y$ (52), $I$ (58) and $\Sigma_\pm$ (58) these relationships can be transformed into
\begin{align}
\tag{79}
\nonumber
\Big\{e^{\mp z \pm y} \frac{dI}{dx} -(z_+ + z_-) \Sigma_\pm \frac{dy}{dx} \mp \frac{z_+ + z_-}{z_\pm}\frac{d\Sigma_\pm}{dx}\Big\}_{x=\kappa a}=0,
\end{align}
which after developing the $e$-powers becomes
\begin{subequations}
\begin{align}
\tag{79'}
\nonumber
\Big\{ \Big(1 \mp z_\pm \frac{Ae^{-x}}{x} + \frac{z_\pm^2}{2} \frac{A^2 e^{-2x}}{x^2}\Big) \frac{dI}{dx}&\\
\nonumber 
+(z_+ + z_-) \Sigma_\pm A e^{-x}\Big(\frac{1}{x^2}+\frac{1}{x}&\big) \mp \frac{z_++ z_-}{z_\pm} \frac{\Sigma_\pm}{dx}\Big\}_{x=\kappa a}=0.
\end{align}
\end{subequations}
In these relationships the values for $I$ from (71) and $\Sigma_+$ and $\Sigma_-$ from (73) are used, in which all terms, which are proportional to the third or higher power of $\frac{Ae^{-x}}{x}$ are again neglected\footnote{In this we have taken into account that the constants $C$, $U$, $G$, $F$ and $H$ are also of the order $\frac{Ae^{-x}}{x}$.}. 

In the workings, many terms originating from the term $dI/dx$, are cancelling against those from $\mp \frac{z_++z_-}{z_\pm} \frac{d\Sigma_\pm}{dx}$\footnote{After all from (59) $I$ contains a term $x\int \frac{dx}{x^4}\int x^3 S dx$, and $\Sigma_\pm$ contains a term $\pm \frac{z_\pm}{z_+ + z_-} x \int \frac{dx}{x^4} \int x^3 S dx$, which exactly compensate each other in (79').}, which makes the formulation of the boundary conditions remarkably shorter. As in the assumption introduced \BLUE{just before (68')}, terms proportional to $(z_- - z_+) A^2 e^{-2x}$ are again neglected. From (79') we derive in this manner, in which for $x$ the value $\kappa a$ must be substituted:
\marginnote{$kT \nabla^2 \delta n $ \\
$n^0 \nabla^2 \delta \psi $ \\
$\frac{\partial n^0}{\partial r} \frac{\partial \delta \psi}{\partial r}$  }[2cm]
\begin{subequations}
\begin{align}
\nonumber
\tag{80} 0 = 1 - \frac{2B}{x^3} \mp z_\pm A e^{-x} \Big( \frac{1}{x} - \frac{2B}{x^4}\Big) + \frac{z_\pm^2}{2} A^2 e^{-2x}
\Big( \frac{1}{x^2}-\frac{2B}{x^5}\Big)\\
\nonumber
\pm z_\pm P A e^{-2x} \Big( \frac{2}{x^4} +\frac{2}{x^3} +\frac{1}{x^2}\Big)\\
\nonumber
\qquad \qquad \qquad +(z_+ + z_-)CA e^{-x} \Big( \frac{1}{x^4} + \frac{1}{x^3}\Big) \pm z_\pm PA e^{-2x} \Big( \frac{1}{x^4} +\frac{2}{x^3}+ \frac{1}{x^2}\Big) \\
\nonumber
+ z_\pm A^2 e^{-2x} \Big( \frac{1}{x^4} +\frac{2}{x^3} +\frac{1}{x^2}\Big) \\
\nonumber
-z_\pm^2 B A^2 e^{-2x} \Big\{ \frac{1}{2x^5} + \frac{1}{3x^4} - \frac{1}{12 x^3}....\\
\nonumber
-\frac{1}{12x} - \Big( \frac{1}{12x} + \frac{1}{12}\Big) e^x E(x) \Big\} \pm \frac{z_+ + z_-}{z_\pm} \frac{2C}{x^3}
\end{align}
\end{subequations}
\marginnote{$kT \nabla^2 \delta n $ }[-1.2cm]
\marginnote{
$\delta n \psi^0$ }[3cm]
\begin{subequations}
\begin{align}
\nonumber
-(z_+ + z_-) CA e^{-x}\Big( \frac{3}{2x^4} +\frac{5}{6x^3} +\frac{1}{12x^2} -\frac{1}{12x} - \frac{e^x}{12} E(x) \Big)\\
\nonumber
\mp z_\pm P A e^{-2x} \Big( \frac{3}{2x^4} +\frac{2}{x^3} + \frac{1}{x^2} \Big)\\
\nonumber
-z_\pm^2 A^2 e^{-2x} \Big( \frac{3}{2x^4} + \frac{3}{x^3} + \frac{1}{x^2} \Big)\\
\nonumber
\qquad \qquad  + z_\pm^2 B A^2 e^{-2x} \Big\{ \frac{3}{5x^5} + \frac{1}{5x^4} - \frac{29}{180 x^3} -\frac{4}{45x^2} + \frac{17}{180x} +\frac{16 e^{2x}}{45}E(2x) \\
\nonumber
-\Big( \frac{1}{6x^3} +\frac{1}{6x^2} +\frac{1}{12x} +\frac{1}{12} \Big) e^x E(x) \Big\}
\end{align}
\end{subequations}
\marginnote{
$\frac{\partial n^0}{\partial r} \frac{\partial \delta \psi}{\partial r}$  }[1cm]
\begin{subequations}
\begin{align}
\nonumber
\mp z_\pm P A e^{-2x} \Big( \frac{1}{2x^3} + \frac{1}{2x^2}\Big) \pm z_\pm A e^{-x} \Big( \frac{2}{x^3} + \frac{2}{x^2} + \frac{1}{x}\Big) \\
\nonumber
\mp z_\pm BA e^{-x} \Big( \frac{3}{2x^4} + \frac{1}{6x^3} - \frac{1}{12x^2} + \frac{1}{12x} +\frac{e^x}{12}E(x) \Big)\\
\nonumber
\qquad \qquad \qquad \quad \mp z_\pm PA e^{-2x} \Big( \frac{3}{2x^4} +\frac{3}{2x^3} + \frac{1}{2x^2}\Big) - z_\pm^2 A^2 e^{-2x}
\Big(\frac{1}{2x^3} + \frac{1}{2x^2} \Big)
\end{align}
\end{subequations}
\begin{subequations}
\begin{align}
\nonumber
+z_\pm^2 B A^2 e^{-2x} \Big( \frac{4}{5x^5} + \frac{1}{10 x^4}-\frac{1}{15 x^3} + \frac{1}{15 x^2} -\frac{2}{15 x} - \frac{4 e^{2x}}{15} E(2x) \Big)\\
\nonumber
- \ifRED\else \frac{1}{\lambda_\pm e}\fi U A e^{-x} \Big( \frac{2}{x^3} + \frac{2}{x^2} + \frac{1}{x}\Big)\\
\nonumber 
+ \ifRED\else \frac{1}{\lambda_\pm e}\fi F A e^{-x} \Big( \frac{3}{4x^4} + \frac{1}{12x^3} -\frac{1}{24x^2} + \frac{1}{24x} +\frac{e^x}{24} E(x) \Big) \\
\nonumber
-\ifRED\else \frac{1}{\lambda_\pm e} \frac{2 \varepsilon_0 \varepsilon_r kT}{\eta e}\fi A^2 e^{-2x} \Big\{ \frac{3}{4x^4} + \frac{1}{2x^3} +B\Big(\frac{1}{20x^4} + \frac{1}{18 x^3} + \frac{1}{36x^2}\\
\nonumber
+\frac{23}{720x} - \frac{1}{240} + \frac{4 e^{2x}}{45} E(2x) + \Big( \frac{1}{12x^3} + \frac{1}{12 x^2} + \frac{1}{24x} - \frac{1}{60} - \frac{x}{240}\Big) e^x E(x) \Big) \Big\}.
\end{align}
\end{subequations}
Itturns out that terms with $PA^{-2x}$ also cancel out against each other, such that the two constants $B$ and $C$ from (80) can be determined. With the third boundary condition (77) $P$ can be calculated as well. 

In order to provide a more orderly equation for $B$ and $C$, we subtract the two equations (80) from each other and find then,
\begin{subequations}
\begin{align}
\nonumber
\tag{81} 0=(z_+ + z_-) \frac{Ae^{-x}}{x}
+(z_+ + z_-) BA e^{-x} \frac{2}{x^4} +\frac{(z_+ + z_-)^2}{z_+ z_-} \frac{2C}{x^3}\\
\nonumber
+(z_+ + z_-) \frac{Ae^{-x}}{x} \Big(\frac{2}{x^3}+\frac{2}{x^2} +\frac{1}{x}\Big)\\
\nonumber
-(z_+ + z_-) BAe^{-x} \Big(\frac{3}{2x^4}+\frac{1}{6x^3}-\frac{1}{12x^2}+\frac{1}{12x}+\frac{e^x}{12}E(x) \Big)\\
\nonumber 
+ \Big( \text{terms of order } \frac{e^{-2x}}{x^2}\Big)
\end{align}
\end{subequations}
It is not necessary to elaborate further on terms with $\frac{e^{-2x}}{x^2}$, since in the final electrophoresis formula $C$ only needs to be known to first order. From (81) follows, after substituting $x=\kappa a$
\begin{subequations}
\begin{align}
\nonumber
\tag{82} C=-\frac{z_+ z_-}{z_+ + z_-} \frac{Ae^{-\kappa a}}{\kappa a} \Big\{ \kappa a + \kappa^2 a^2\\
\nonumber
+\frac{2B}{\kappa^3 a^3}\Big(\frac{\kappa^3 a^3}{8} - \frac{\kappa^4 a^4}{24}+\frac{\kappa^5 a^5}{48} -\frac{\kappa^6 a^6}{48} - \frac{\kappa^7 a^7 e^{\kappa a}}{48}E(\kappa a)\Big) \Big\}.
\end{align}
\end{subequations}
Although this equation is valid for any value of $\kappa a$, it is not suitable for calculations, when $\kappa a$ is much larger than 5. In order to calculate for larger values of $\kappa a$ for
\begin{equation}
\nonumber
\frac{\kappa^3 a^3}{8} - \frac{\kappa^4 a^4}{24}+\frac{\kappa^5 a^5}{48} -\frac{\kappa^6 a^6}{48} - \frac{\kappa^7 a^7 e^{\kappa a}}{48}E(\kappa a)
\end{equation}
as well with enough accuracy, we will use the \BLUE{after (55a)} mentioned series for $E(x)$. With this series we then find (82') for $C$, which is valid for large values of $\kappa a$.
\begin{equation}
\nonumber
\tag{82'} C= - \frac{z_+ z_-} {z_+ + z_-} \frac{Ae^{-\kappa a}}{\kappa a}\Big\{ \kappa^2 a^2 + \kappa a +\frac{2B}{\kappa^3 a^3}\Big(\frac{\kappa^2 a^2}{2}-\frac{5 \kappa a}{2}+15 -\frac{105}{\kappa a}+ ...\Big) \Big\}
\end{equation}
Since concerning $B$ the contributions of order $\frac{e^{-2x}}{x^2}$ must be known, the equations (80) are multiplied with $\frac{z_+}{z_++z_-}$ (cations) and $\frac{z_-}{z_++z_-}$ (anions) respectively and added together, such that the term $\pm \frac{z_++z_-}{z_\pm} \frac{2C}{x^3}$, which has an uncertainty of order $\frac{e^{-2x}}{x^2}$ cancels. Then the following equation appears
\begin{subequations}
\begin{align}
\nonumber 
\tag{83} \qquad \qquad \qquad \qquad \quad 0 = 1 - \frac{2B}{x^3} + (z_- - z_+) A e^{-x} \Big( \frac{1}{x} - \frac{2B}{x^4}\Big)\\
\nonumber
+ \frac{z_+^2 - z_+ z_- + z_-^2}{2} A^2 e^{-2x} \Big( \frac{1}{x^2} - \frac{2B}{x^5}\Big) \\
\nonumber
+(z_+ + z_-) CA e^{-x} \Big( \frac{1}{x^4} + \frac{1}{x^3}\Big)
\end{align}
\end{subequations}
\begin{subequations}
\begin{align}
\nonumber
+(z_+^2 - z_+ z_- + z_-^2) A^2 e^{-2x} \Big( \frac{1}{x^4} + \frac{2}{x^3} +\frac{1}{x^2}\Big)\\
\nonumber
+(z_+^2 - z_+ z_- + z_-^2)BA^2 e^{-2x}
\Big\{ \frac{1}{2x^5} + \frac{1}{3x^4} - \frac{1}{12 x^3}.....\\
\nonumber
-\frac{1}{12 x} -\Big(\frac{1}{12x} + \frac{1}{12}\Big) e^x E(x) \Big\} - (z_+ + z_-) CA e^{-x} \Big( \frac{3}{2x^4} + \frac{5}{6x^3}
\end{align}
\end{subequations}
\begin{subequations}
\begin{align}
\nonumber
+\frac{1}{12 x^2} - \frac{1}{12 x} - \frac{e^x}{12} E(x)\Big)\\
\nonumber
-(z_+^2 - z_+ z_- +z_-^2) A^2 e^{-2x} \Big( \frac{3}{2x^4} + \frac{2}{x^3} + \frac{1}{x^2}\Big)\\
\nonumber
+(z_+^2 - z_+ z_- +z_-^2) BA^2 e^{-2x} \Big\{ \frac{3}{5x^5} + \frac{1}{5x^4} - 
\frac{29}{180x^3} - \frac{4}{45x^2}\\
\nonumber
\quad +\frac{17}{180 x} +\frac{16 e^{2x}}{45} E(2x) - \Big( \frac{1}{6x^3} + \frac{1}{6x^2} +\frac{1}{12x} +\frac{1}{12}\Big) e^x E(x) \Big\}
\end{align}
\end{subequations}
\begin{subequations}
\begin{align}
\nonumber
-(z_- - z_+)Ae^{-x}\Big( \frac{2}{x^3} + \frac{2}{x^2} +\frac{1}{x}\Big)\\
\nonumber
\qquad +(z_- - z_+) BAe^{-x} \Big( \frac{3}{2x^4} + \frac{1}{6x^3} - \frac{1}{12 x^2} +\frac{1}{12 x} +\frac{e^x}{12}E(x) \Big)\\
\nonumber
-(z_+^2 - z_+ z_- +z_-^2)A^2 e^{-2x} \Big( \frac{1}{2x^3} + \frac{1}{2x^2}\Big) \\
\nonumber
+(z_+^2 - z_+ z_- +z_-^2)BA^2 e^{-2x} \Big(\frac{4}{5x^5} + \frac{1}{10 x^4} - \frac{1}{15 x^3}\\
\nonumber
+ \frac{1}{15 x^2} +\frac{2}{15 x} - \frac{4 e^{2x}}{15} E(2x) \Big) 
\end{align}
\end{subequations}
\begin{subequations}
\begin{align}
\nonumber 
-\ifRED\else \frac{z_+/\lambda_+ + z_-\lambda_-}{(z_+ + z_-) e}\fi UAe^{-x} \Big(\frac{2}{x^3} + \frac{2}{x^2} + \frac{1}{x} \Big)\\
\nonumber 
+\ifRED\else \frac{z_+/\lambda_+ + z_-\lambda_-}{(z_+ + z_-) e}\fi FA e^{-x} \Big( \frac{3}{4x^4} + \frac{1}{12x^3} - \frac{1}{24x^2} +\frac{1}{24x} +\frac{e^x}{24} E(x) \Big)\\
\nonumber 
-\ifRED\else \frac{z_+/\lambda_+ + z_-\lambda_-}{(z_+ + z_-) e} \frac{2 \varepsilon_0 \varepsilon_r kT}{\eta e}\fi A^2 e^{-2x} \Big\{ \frac{3}{4x^4} + \frac{1}{2x^3} +B \Big( \frac{1}{20 x^4} + \frac{1}{18 x^3}\\
\nonumber
\frac{1}{36x^2} + \frac{23}{720 x^2} - \frac{1}{240} + \frac{4e^{2x}}{45} E(2x) \\
\nonumber
+\Big( \frac{1}{12x^3} + \frac{1}{12x^2} + \frac{1}{24 x} - \frac{1}{60} - \frac{x}{240} \Big) e^x E(x) \Big) \Big\}.
\end{align}
\end{subequations}

Now by substituting the values of $\kappa$, $\kappa a$, for $C$ the value from (82'), and for $U$ and $F$ the values from (67') and collect similar terms, we can write down the next expression, from which $B$ can be calculated\footnote{$y_a = \frac{Ae^{-\kappa a}}{\kappa a}$. The higher order terms from (55b) can here be neglected. }: 
\begin{subequations}
\begin{align}
\nonumber
\tag{84} \qquad \qquad \frac{2B}{\kappa^3 a^3} \Big\{ 1 + (z_-- z_+) y_a \Big( \frac{1}{4} - \frac{\kappa a}{12} + \frac{\kappa^2 a^2}{24} - \frac{\kappa^3 a^3}{24} - \frac{\kappa^4 a^4 e^{\kappa a}}{24} E(\kappa a) \Big)\\
\nonumber
+(z_+^2 - z_+ z_- + z_-^2) y_a^2 \Big( \frac{1}{20} + \frac{\kappa a}{60} + \frac{13 \kappa^2 a^2}{180} + \frac{\kappa^3 a^3}{90} - \frac{\kappa^4 a^4}{45}\\
\nonumber
- \frac{2 \kappa^5 a^5 e^{2 \kappa a}}{45} E(2\kappa a)  + \Big( \frac{\kappa^2 a^2}{12} + \frac{\kappa^3 a^3}{12} \Big)  e^{\kappa a} E(\kappa a) \Big) 
\end{align}
\end{subequations}
\begin{subequations}
\begin{align}
\nonumber
-z_+ z_- y_a^2 \Big( \frac{1}{2 \kappa^3 a^3} - \frac{1}{6 \kappa^2 a^2} + \frac{1}{12 \kappa a} - \frac{1}{12} - \frac{\kappa a e^{\kappa a}}{12} E(\kappa a) \Big)\\
\nonumber
\times \Big( \frac{\kappa^3 a^3}{8} - \frac{\kappa^4 a^4}{24} + \frac{\kappa^5 a^5}{48} - \frac{\kappa^6 a^6}{48} - \frac{\kappa^7 a^7 e^{\kappa a}}{48}E(\kappa a) \Big)\\
\nonumber
\qquad \qquad \qquad \qquad +\ifRED\else \frac{z_+/\lambda_+ + z_-/\lambda_-}{(z_+ + z_-)e} \frac{2\varepsilon_0 \varepsilon_r kT}{3\eta e}\fi y_a^2 \Big( \frac{2}{\kappa^2 a^2} + \frac{2}{\kappa a} +1 \Big) \Big( \frac{\kappa^2 a^2}{16} - \frac{5 \kappa^3 a^3}{48}\\
\nonumber
- \frac{\kappa^4 a^4}{96} + \frac{\kappa^5 a^5}{96} - \Big( \frac{\kappa^4 a^4}{8} - \frac{\kappa^6 a^6}{96} \Big) e^{\kappa a} E(\kappa a) \Big) 
\end{align}
\end{subequations}
\begin{subequations}
\begin{align}
\nonumber
-\ifRED
\else \frac{z_+/\lambda_+ + z_-/\lambda_-}{(z_+ + z_-)e} \frac{2\varepsilon_0 \varepsilon_r kT}{3\eta e}\fi y_a^2
\Big( \frac{3}{4 \kappa^3 a^3} + \frac{1}{12 \kappa^2 a^2} - \frac{1}{24 \kappa a} + \frac{1}{24}\\
\nonumber
+ \frac{\kappa a e^{\kappa a}}{24} E(\kappa a) \Big) \Big( \frac{\kappa^4 a^4}{10} - \frac{\kappa^5 a^5}{40}+ \frac{\kappa^6 a^6}{120} -\frac{\kappa^7 a^7}{240} +\frac{\kappa^8 a^8}{240} + \frac{\kappa^9 a^9 e^{\kappa a}}{240}E(\kappa a) \Big)\\
\nonumber
+\ifRED\else \frac{z_+/\lambda_+ + z_-/\lambda_-}{(z_+ + z_-)e} \frac{2\varepsilon_0 \varepsilon_r kT}{3\eta e}\fi y_a^2 \Big( \frac{3 \kappa a}{40} + \frac{\kappa^2 a^2}{12}+ \frac{\kappa^3 a^3}{24} + \frac{23 \kappa^4 a^4}{480} - \frac{\kappa^5 a^5}{160}\\
\nonumber
+ \frac{2 \kappa^5 a^5 e^{2\kappa a}}{15}E(2\kappa a) + \Big( \frac{\kappa^2 a^2}{8} + \frac{\kappa^3 a^3}{8} + \frac{\kappa^4 a^4}{16} - \frac{\kappa^5 a^5}{40} - \frac{\kappa^6 a^6}{160} \Big) e^{\kappa a} E(\kappa a) \Big) \Big\}
\end{align}
\end{subequations}
\begin{subequations}
\begin{align}
\nonumber
\qquad =1 
-(z_- - z_+) y \Big(\frac{2}{\kappa^2 a^2} + \frac{2}{\kappa a} \Big) - (z_+^2 - z_+ z_- + z_-^2) y_a^2 \Big(\frac{1}{2 \kappa^2 a^2} + \frac{1}{2\kappa a}\Big)\\
\nonumber
+ z_+ z_- y_a^2 \Big(\frac{1}{2 \kappa^3 a^3} - \frac{1}{6 \kappa^2 a^2} + \frac{1}{12 \kappa a} - \frac{1}{12} - \frac{\kappa a e^{\kappa a}}{12} E(\kappa a)\Big) (\kappa a + \kappa^2 a^2)
\end{align}
\end{subequations}
\begin{subequations}
\begin{align}
\nonumber
-\ifRED\else \frac{z_+/\lambda_+ + z_-/\lambda_-}{(z_+ + z_-)e} \frac{2\varepsilon_0 \varepsilon_r kT}{3\eta e}\fi y_a^2 \Big(\frac{2}{\kappa^2 a^2} + \frac{2}{\kappa a} + 1\Big) \\
\nonumber
\qquad \qquad \qquad \qquad +\ifRED\else \frac{z_+/\lambda_+ + z_-/\lambda_-}{(z_+ + z_-)e} \frac{2\varepsilon_0 \varepsilon_r kT}{3\eta e}\fi y_a^2 \Big(\frac{3}{4\kappa^3 a^3} +\frac{1}{12\kappa^2 a^2}- \frac{1}{24\kappa a} + \frac{1}{24} \\
\nonumber
+ \frac{\kappa a e^{\kappa a}}{24} E(\kappa a)\Big) ( 3 \kappa a + 3 \kappa^2 a^2 + \kappa^3 a^3)\\
\nonumber
-\ifRED\else \frac{z_+/\lambda_+ + z_-/\lambda_-}{(z_+ + z_-)e} \frac{2\varepsilon_0 \varepsilon_r kT}{3\eta e}\fi y_a^2 \Big(\frac{9}{4\kappa^2 a^2} +\frac{3}{2\kappa a} \Big).
\end{align}
\end{subequations}
It is again desired, to make use of the series $E(x)$ in order to calculate $B$ for large values of $\kappa a$:
\begin{subequations}
\begin{align}
\nonumber
\tag{84'} \frac{2B}{\kappa^3 a^3} \Big\{ 1 + (z_-- z_+) y_a \Big( \frac{1}{\kappa a} - \frac{5}{\kappa^2 a^2} + \frac{30}{\kappa^3 a^3} - \frac{210}{\kappa^4 a^4}....\Big)\\
\nonumber
 + (z_+^2 - z_+ z_- + z_-^2) y_a^2
\Big(\frac{1}{4 \kappa a} -\frac{5}{4 \kappa^2 a^2} + \frac{57}{8 \kappa^3 a^3} - \frac{93}{2 \kappa^4 a^4}.... \Big)\\
\nonumber
\qquad \qquad - z_+ z_- y_a^2 \Big(\frac{2}{\kappa^4 a^4} - \frac{10}{\kappa^5 a^5} + \frac{60}{\kappa^6 a^6} - \frac{420}{\kappa^7 a^7} .... \Big)
\times \Big( \frac{\kappa^2 a^2}{2}- \frac{5 \kappa a}{2} +15 - \frac{105}{\kappa a}....\Big) 
\end{align}
\end{subequations}
\begin{subequations}
\begin{align}
\nonumber
\qquad \qquad +\ifRED\else \frac{z_+/\lambda_+ + z_-/\lambda_-}{(z_+ + z_-)e} \frac{2\varepsilon_0 \varepsilon_r kT}{3\eta e}\fi y_a^2 \Big(1 + \frac{2}{\kappa a} + \frac{2}{\kappa^2 a^2}\Big) 
\times \Big( \frac{1}{2} - \frac{9}{2 \kappa a} + \frac{75}{2 \kappa^2 a^2} - \frac{330}{\kappa^3 a^3}\Big) \\
\nonumber
-\ifRED\else \frac{z_+/\lambda_+ + z_-/\lambda_-}{(z_+ + z_-)e} \frac{2\varepsilon_0 \varepsilon_r kT}{3\eta e}\fi y_a^2 \Big(\frac{1}{\kappa^3 a^3} -\frac{1}{\kappa^4 a^4}+ \frac{5}{\kappa^5 a^5} - \frac{30}{\kappa^6 a^6}....\Big) \\
\nonumber
\times \Big( \frac{\kappa^3 a^3}{2} - 3 \kappa^2 a^2 + 21 \kappa a - 168 \Big)\\
\nonumber
 +\ifRED\else \frac{z_+/\lambda_+ + z_-/\lambda_-}{(z_+ + z_-)e} \frac{2\varepsilon_0 \varepsilon_r kT}{3\eta e}\fi y_a^2 \Big(\frac{3}{4\kappa a} -\frac{9}{\kappa^2 a^2} + \frac{765}{8\kappa^3 a^3} - \frac{2109}{2\kappa^4 a^4}....\Big)\Big\} =
\end{align}
\end{subequations}
\begin{subequations}
\begin{align}
\nonumber
= 1 
- (z_- - z_+) y_a \Big( \frac{2}{\kappa a} + \frac{2}{\kappa^2 a^2}\Big) - (z_+^2 - z_+ z_- + z_-^2) y_a^2 \Big( \frac{1}{2 \kappa a} + \frac{1}{2\kappa^2 a^2}\Big)\\
\nonumber
+z_+ z_- y_a^2 \Big( \frac{2}{\kappa^4 a^4} - \frac{10}{\kappa^5 a^5} + \frac{60}{\kappa^6 a^6} - \frac{420}{\kappa^7 a^7}.... \Big) (\kappa^2 a^2 + \kappa a)\\
\nonumber
-\ifRED\else \frac{z_+/\lambda_+ + z_-/\lambda_-}{(z_+ + z_-)e} \frac{2\varepsilon_0 \varepsilon_r kT}{3\eta e}\fi y_a^2 \Big( 1 + \frac{2}{\kappa a} + \frac{2}{\kappa^2 a^2} \Big)\\
\nonumber
+\ifRED\else \frac{z_+/\lambda_+ + z_-/\lambda_-}{(z_+ + z_-)e} \frac{2\varepsilon_0 \varepsilon_r kT}{3\eta e}\fi y_a^2 \Big( \frac{1}{\kappa^3 a^3} - \frac{1}{\kappa^4 a^4} + \frac{5}{\kappa^5 a^5} - \frac{30}{\kappa^6 a^6}.... \Big)
\times (\kappa^3 a^3 + 3 \kappa^2 a^2 + 3 \kappa a)\\
\nonumber
-\ifRED\else \frac{z_+/\lambda_+ + z_-/\lambda_-}{(z_+ + z_-)e} \frac{2\varepsilon_0 \varepsilon_r kT}{3\eta e}\fi y_a^2 \Big( \frac{3}{2\kappa a} + \frac{9}{4\kappa^2 a^2} \Big).
\end{align}
\end{subequations}

From (84) and (84') we see that for very small $y_a$ (thus for very small $\zeta$), $\frac{2B}{\kappa^3 a^3}$ is about 1 (similar to the value of Henry) and that for larger $y_a$, terms proportional to $y_a$ and $y_a^2$ appear.

It it still worth and can be illuminating, from the equation for $B$, which as we will see is directly related to the moment in the double layer and the dielectric constant of the sol, to highlight the terms that give the greatest contributions for $\kappa a  \rightarrow 0$ and $\kappa \rightarrow \infty$. We then find:  

For $\kappa a \rightarrow 0$ 
\begin{subequations}
\begin{align}
\nonumber \tag{85a}
\frac{2B}{\kappa^3 a^3}= 1-(z_--z_+) y_a \frac{2}{\kappa^2 a^2}- (z_+^2 - z_+ z_- + z_-^2) y_a^2 \Big(\frac{1}{2 \kappa^2 a^2} + \frac{1}{2\kappa a} \Big)\\
\nonumber
+ z_+ z_- y_a^2 \Big(\frac{1}{2\kappa^2 a^2}+ \frac{1}{3 \kappa a}\Big) - \ifRED\else \frac{z_+/\lambda_+ + z_-/\lambda_-}{(z_+ + z_-)e} \frac{2\varepsilon_0 \varepsilon_r kT}{3\eta e}\fi y_a^2 \frac{2}{ \kappa^2 a^2}.
\end{align}
\end{subequations}

For $\kappa a \rightarrow \infty$
\begin{subequations}
\begin{align}
\nonumber \tag{85b}
\frac{2B}{\kappa^3 a^3}= 1-(z_--z_+) y_a \frac{3}{\kappa a}- (z_+^2 - z_+ z_- + z_-^2) y_a^2 \frac{3}{4 \kappa a}\\
\nonumber
+ z_+ z_- y_a^2 \frac{3}{\kappa^2 a^2} - \ifRED\else \frac{z_+/\lambda_+ + z_-/\lambda_-}{(z_+ + z_-)e} \frac{2\varepsilon_0 \varepsilon_r kT}{3\eta e}\fi y_a^2 \frac{9}{4 \kappa a}.
\end{align}
\end{subequations}

The term $\frac{z_+/ \lambda_+ + z_- /\lambda_-}{(z_+ + z_-) e} \frac{2\varepsilon_0 \varepsilon_r kT}{3 \eta e} $ has a value of about 0.1 - 0.5. It turns out that the correction terms on $B$ in $\kappa a$ are all of the same order and it was thus indeed necessary to take into account all these terms.  

Now that the values of the constants $B$ and $C$ are known, with the help of (77), the value of $P$ can be calculated as well. For completeness sake, we shall perform this calculation, although to calculate the electrophoretic velocity we don't need to know $P$ (see also \BLUE{the discussion after (87)}).

In order to do so, $I$ (from (71)) is used in (77). It turns out that it is sufficient to take only the terms of zeroth and first order and thus neglect all contributions with $y_a^2$. Thus we find:
\begin{subequations}
\begin{align}
\nonumber
D\Big\{1 -\frac{2B}{x^3} -Pe^{-x}\Big(\frac{2}{x^3}+ \frac{2}{x^2} + \frac{1}{x}\Big) - (z_- - z_+)A e^{-x} \Big(\frac{7}{2x^3} +\frac{7}{2x^2} +\frac{7}{4x}+\frac{1}{2} \Big )\\
\nonumber
-(z_- - z_+)BAe^{-x} \Big(\frac{1}{3x^3} + \frac{1}{12x^2} -\frac{1}{12x} -\frac{e^x}{12} E(x) \Big) \Big\}_{x=\kappa a}\\
\nonumber
=D_i\Big\{1 + \frac{B}{x^3} +Pe^{-x}\Big(\frac{1}{x^3} +\frac{1}{x^2} \Big) + (z_- - z_+) Ae^{-x}\Big( \frac{7}{4x^3} + \frac{7}{4x^2} + \frac{1}{2x} \Big)\\
\nonumber
+(z_- - z_+)BAe^{-x} \Big(\frac{1}{6x^3} - \frac{1}{12 x^2} + \frac{1}{12 x} + \frac{e^x}{12}E(x) \Big) \Big\}_{x=\kappa a}
\end{align}
\end{subequations}
For $\frac{2B}{\kappa^3 a^3}$ according to (84) in the first approximation: 
\begin{equation}
\begin{aligned}
\nonumber
\frac{2B}{\kappa^3 a^3}= 1-(z_- - z_+) y_a \Big( \frac{2}{\kappa^2 a^2} + \frac{2}{\kappa a} + \frac{1}{4} - \frac{\kappa a}{12}+ \frac{\kappa^2 a^2}{24} - \frac{\kappa^3 a^3}{24} - \frac{\kappa^4 a^4 e^{\kappa a}}{24}E(\kappa a) \Big).
\end{aligned}
\end{equation}
Then we find for P:
\begin{subequations}
\begin{align}
\nonumber
\tag{86} \frac{Pe^{-\kappa a}}{\kappa a}=- \frac{D(z_--z_+) y_a \Big(\frac{3}{2 \kappa^2 a^2} +\frac{3}{2 \kappa a} +\frac{3}{2} + \frac{3 \kappa a}{4}\Big)}{D\Big(\frac{2}{\kappa^2 a^2} + \frac{2}{\kappa a} +1 \Big)+ D_i \Big(\frac{1}{\kappa^2 a^2} + \frac{1}{\kappa a} \Big)}\\
\nonumber 
-\frac{D_i \Big\{\frac{3}{2} + (z_- - z_+) y_a \Big( \frac{3}{4\kappa^2 a^2} + \frac{3}{4\kappa a} + \frac{3}{8} + \frac{\kappa a}{8} -\frac{\kappa^2 a^2}{16} +\frac{\kappa^3 a^3}{16} +\frac{\kappa^4 a^4 e^{\kappa a}}{16}E(\kappa a)\Big) \Big\}}{D\Big(\frac{2}{\kappa^2 a^2} + \frac{2}{\kappa a} + 1\Big) + D_i \Big(\frac{1}{\kappa^2 a^2}+\frac{1}{\kappa a} \Big)}.
\end{align}
\end{subequations}
Approximation for large $\kappa a$
\begin{subequations}
\begin{align}
\nonumber
\tag{86'} \frac{Pe^{-\kappa a}}{\kappa^2 a^2}=- \frac{D(z_--z_+) y_a \Big(\frac{3\kappa a}{4} +\frac{3}{2} +\frac{3}{2\kappa a} + \frac{3}{2 \kappa^2 a^2}\Big)}{D\Big(\kappa a + 2 +\frac{2}{\kappa a} \Big)+ D_i \Big(1 + \frac{1}{\kappa a} \Big)}\\
\nonumber 
-\frac{D_i \Big\{\frac{3}{2} + (z_- - z_+) y_a \Big( \frac{3}{4} - \frac{3}{4\kappa a} + \frac{33}{4 \kappa^2 a^2} - \frac{45}{\kappa^3 a^3} ....\Big) \Big\}}{D\Big(\kappa a + 2 + \frac{2}{\kappa a} \Big) + D_i \Big(1+\frac{1}{\kappa a} \Big)}.
\end{align}
\end{subequations}
Since always $D\gg D_i$, for small or large $\kappa a$, $\frac{Pe^{-\kappa a}}{\kappa a}$ is of the order $y_a$ and $y_a \kappa a$ respectively.

\subsection*{C - THE ELECTROPHORESIS FORMULA}

Finally from the equation
\begin{equation}
\nonumber
 \tag{87} \frac{U}{E} = \frac{2 \varepsilon_0 \varepsilon_r kT}{3 \eta e} \int_\infty^{\kappa a} \Big\{\frac{I}{x} \frac{dy}{dx} - 2x \int_\infty^x\Big( \frac{1}{x'^2} \frac{dI}{dx'} - \frac{I}{x'^3} \Big)\frac{dy}{dx'} dx' \Big\} dx
\end{equation}
after supplying $I$ from (71) and $y$ from (55b) and taking into account the in (82), (84) and (86) found integration constants $C$, $B$ and $P$ the final electrophoresis formula can be calculated. All terms of the first and second order are taken into account (proportional to $y_a$ and $y_a^2$), while terms of third order (proportional to $y_a^3$) are partly taken into account, i.e. as far as they are non zero for a symmetric electrolyte. Thus terms proportional to $(z_- - z_+) y_a^3$ and  $(z_- - z_+)^2 y_a^3$ will be deleted. 

In the determination of the integrals 2 terms appear proportional to $(z_- - z_+)P e^{-\kappa a} y_a^2$, which cancel each other out. Moreover these terms themselves are already proportional to $(z_- - z_+) y_a^3$ (Eq.(86)), such that they can be neglected. It thus turns out that, at least in the by us used approximation, the value of $P$ does not have any influence on the electrophoretic velocity.

(88) gives the result of the calculation of $U$. In order to show the connection with the electrophesis formula of Henry clearly, the term $\frac{kT y_a}{e}$ is replaced by $\zeta$, while the relationship between $y_a$ and $A$ from (55b) has been deduced. 

In the margins it is indicated from which terms of $I$ the various terms of (88) originate from and also, if that makes sense, which parts of (10') \BLUE{(which can be found after (55b))} correspond to the terms of (88)\footnote{The value of the integrals $\int \frac{e^{-x}E(x)}{x} dx$, $\int \frac{e^x E(3x)}{x} dx$, $\int \frac{e^x E(4x)}{x} dx$,  $\int \frac{e^{-x} E(2x)}{x} dx$ and $\int \frac{e^{-2x}E(x)}{x} dx$ can be determined with enough accuracy through graphical integration.\\
}.
\marginnote{$I=x$ entirely \\
$I=\frac{B}{x^2}$ first approx. y (55)\\
$I=\frac{B}{x^2}$ second approx. y (55)}[1cm]
\begin{subequations}
\begin{align}
\nonumber
\tag{88} \frac{U}{E}= \ifRED\else \frac{2\varepsilon_0 \varepsilon_r \zeta}{3\eta}\fi \Big[1 + \frac{2B}{\kappa^3 a^3}\Big\{ \frac{\kappa^2 a^2}{16} - \frac{5 \kappa^3 a^3}{48} - \frac{\kappa^4 a^4}{96} + \frac{\kappa^5 a^5}{96}\\
\nonumber
- \Big( \frac{\kappa^4 a^4}{8} - \frac{\kappa^6 a^6}{96}\Big) e^{\kappa a}E(\kappa a) \Big\}\\
\nonumber
+ (z_+ - z_+) y_a \frac{2B}{\kappa^3 a^3} \Big\{ \frac{\kappa a e^{\kappa a}}{4} E(\kappa a) - \frac{\kappa a e^{3 \kappa a}}{4}E(3 \kappa a) \Big\}\\
\nonumber
\qquad \qquad \qquad \times \Big\{ \frac{\kappa^2 a^2}{16} - \frac{5 \kappa^3 a^3}{48} - \frac{\kappa^4 a^4}{96} +\frac{\kappa^5 a^5}{96} - \Big(\frac{\kappa^4 a^4}{8}- \frac{\kappa^6 a^6}{96}\Big) e^{\kappa a} E(\kappa a)\Big\}
\end{align}
\end{subequations}
\begin{subequations}
\begin{align}
\nonumber
+(z_--z_+) y_a \frac{2B}{\kappa^3 a^3}\Big\{  \frac{\kappa^2 a^2}{60} - \frac{7 \kappa^3 a^3}{240} + \frac{193 \kappa^4 a^4}{1440} + \frac{17 \kappa^5 a^5}{1440} - \frac{83 \kappa^6 a^6}{2880}\\
\nonumber
+ \Big( \frac{7 \kappa^5 a^5}{24} - \frac{83 \kappa^7 a^7}{1440} \Big) e^{2\kappa a} E(2 \kappa a) \\
\nonumber
-\Big( \frac{\kappa^3 a^3}{64} - \frac{5 \kappa^5 a^5}{192} - \frac{\kappa^5 a^5}{384} + \frac{\kappa^6 a^6}{384}\Big) e^{\kappa a} E(\kappa a)\\
\nonumber 
+\Big(\frac{\kappa^3 a^3}{64} - \frac{5 \kappa^4 a^4}{192} - \frac{\kappa^5 a^5}{384} + \frac{\kappa^6 a^6}{384} \Big) e^{3\kappa a} E(3 \kappa a)\\
\nonumber
+\Big(\frac{\kappa^5 a^5}{32} - \frac{\kappa^7 a^7}{384}\Big) e^{2\kappa a}\int_\infty^{\kappa a} \frac{e^{-x}E(x)}{x} dx - \Big( \frac{\kappa^5 a^5}{32} - \frac{\kappa^7 a^7}{384}\Big) e^{2\kappa a} \int_\infty^{\kappa a} \frac{e^x E(3x)}{x}dx \Big\}
\end{align}
\end{subequations}
\marginnote{
$I=\frac{B}{x^2}$ third approx. y (55)}[1cm]
\begin{subequations}
\begin{align}
\nonumber
\qquad \qquad \qquad + (z_+^2 - z_+ z_- + z_-^2) y_a^2 \frac{2B}{\kappa^3 a^3} \Big\{ \frac{\kappa^2 a^2 e^{4 \kappa a}}{3} E(4\kappa a) - \frac{\kappa^2 a^2 e^{2\kappa a}}{6} E(2\kappa a) \Big\}\\
\nonumber
\times \Big\{\frac{\kappa^2 a^2}{16} - \frac{5 \kappa^3 a^3}{48} -\frac{\kappa^4 a^4}{96} + \frac{\kappa^5 a^5}{96}- \Big( \frac{\kappa^4 a^4}{8} - \frac{\kappa^6 a^6}{96}\Big) e^{\kappa a} E(\kappa a) \Big\} \\
\nonumber
+(z_+^2 - z_+ z_- + z_-^2) y_a^2 \frac{2B}{\kappa^3 a^3} \Big\{ \frac{\kappa^2 a^2}{288} - \frac{\kappa^3 a^3}{160} + \frac{3 \kappa^4 a^4}{640}\\
\nonumber
- \frac{427 \kappa^5 a^5}{5760} - \frac{59 \kappa^6 a^6}{11520} + \frac{79 \kappa^7 a^7}{3840} - \Big( \frac{15 \kappa^6 a^6}{64} - \frac{79 \kappa^8 a^8}{1280} \Big) e^{3 \kappa a} E(3 \kappa a) \\
\nonumber
+ \Big(\frac{\kappa^4 a^4}{96} - \frac{5 \kappa^5 a^5}{288} - \frac{\kappa^6 a^6}{576} + \frac{\kappa^7 a^7}{576}\Big) e^{2 \kappa a} E(2 \kappa a) \\
\nonumber
- \Big( \frac{\kappa^4 a^4}{48} + \frac{5 \kappa^5 a^5}{144} - \frac{\kappa^6 a^6}{288} -  \frac{\kappa^7 a^7}{288} \Big) e^{4 \kappa a} E(4 \kappa a)
\end{align}
\end{subequations}
\marginnote{
$I$ higher terms}[2cm]
\begin{subequations}
\begin{align}
\nonumber
+ \Big(\frac{\kappa^6 a^6}{24} -\frac{\kappa^8 a^8}{288} \Big) e^{3 \kappa a} \int_\infty^{\kappa a} \frac{e^x E(4x)}{x}dx - \Big(\frac{\kappa^6 a^6}{48}-\frac{\kappa^8 a^8}{576} \Big) e^{3 \kappa a} \int_\infty^{\kappa a} \frac{e^{-x} E(2x)}{x}dx \Big\}\\
\nonumber
+ (z_- - z_+) y_a \Big( \frac{1}{8} - \frac{\kappa a}{4} - \frac{\kappa^2 a^2 e^{2\kappa a}}{2} E(2 \kappa a) \Big)
\end{align}
\end{subequations}
\marginnote{
First appr. y \\
$\frac{\partial n}{\partial r} \frac{\partial \delta \psi}{\partial r}$
\\ \; 
$\delta n \; \psi^0$ }[1cm]
\begin{subequations}
\begin{align}
\nonumber
-(z_-- z_+) y_a \frac{2B}{\kappa^3 a^3} \Big( \frac{\kappa^2 a^2}{120} - \frac{\kappa^3 a^3}{40} - \frac{\kappa^4 a^4}{90} + \frac{\kappa^5 a^5}{90} - \frac{\kappa^6 a^6}{45}\\
\nonumber 
-\frac{\kappa^4 a^4 e^{\kappa a}}{24} E(\kappa a) - \frac{2 \kappa^7 a^7 e^{2\kappa a}}{45} E(2 \kappa a) \Big)
\\
\nonumber
- z_+ z_- y_a^2 \Big\{ \kappa a + \kappa^2 a^2 + \frac{2B}{\kappa^3 a^3}\Big(\frac{\kappa^3 a^3}{8} - \frac{\kappa^4 a^4}{24} + \frac{\kappa^5 a^5}{48} - \frac{\kappa^6 a^6}{48}\\
\nonumber
\qquad \qquad -\frac{\kappa^7 a^7 e^{\kappa a}}{48}E(\kappa a) \Big\} \Big\{ \frac{1}{60 \kappa a} - \frac{1}{120} + \frac{31 \kappa a}{180} - \frac{\kappa^2 a^2}{180} + \frac{\kappa^3 a^3}{90} + \frac{\kappa a e^{\kappa a}}{12} E(\kappa a) \\
\nonumber
+ \Big( \frac{\kappa^2 a^2}{3} + \frac{\kappa^4 a^4}{45}\Big) e^{2 \kappa a} E(2\kappa a)   \Big\} + 
\end{align}
\end{subequations}
\begin{subequations}
\begin{align}
\nonumber
+(z_+^2 - z_+ z_- + z_-^2) y_a^2 \Big\{ \frac{1}{60} - \frac{\kappa a}{80} + \frac{89 \kappa^2 a^2}{720} + \frac{\kappa^3 a^3}{480} - \frac{\kappa^4 a^4}{160}\\
\nonumber
+\Big( \frac{\kappa^2 a^2}{12} + \frac{3 \kappa^3 a^3}{8} - \frac{3 \kappa^5 a^5}{160}\Big) e^{3 \kappa a} E(3\kappa a) \Big\} \\
\nonumber
\qquad \qquad \qquad -(z_+^2 -z_+ z_- + z_-^2) y_a^2 \frac{2B}{\kappa^3 a^3}\Big\{\frac{\kappa^2 a^2}{960} - \frac{7 \kappa^3 a^3}{2880} - \frac{97 \kappa^4 a^4}{11520}- \frac{131 \kappa^5 a^5}{1280}
\end{align}
\end{subequations}
\marginnote{
$\frac{\partial n}{\partial r} \frac{\partial \delta \psi}{\partial r}$\\
}[-1cm]
\begin{subequations}
\begin{align}
\nonumber 
+ \frac{3 \kappa^6 a^6}{512} - \frac{9 \kappa^7 a^7}{512} - \frac{\kappa^4 a^4 e^{\kappa a}}{192} E(\kappa a) - \frac{8 \kappa^5 a^5 e^{2 \kappa a}}{45} E(2\kappa a)\\
\nonumber
+ \Big(\frac{21 \kappa^5 a^5}{160} - \frac{189 \kappa^6 a^6}{640} - \frac{27 \kappa^8 a^8}{512} \Big) e^{3 \kappa a} E(3 \kappa a) \Big\}\\
\nonumber
+(z_+^2 - z_+z_- +z_-^2) y_a^2\\
\nonumber
\qquad \times \Big\{\frac{\kappa a}{32} + \frac{19 \kappa^2 a^2}{288} - \frac{\kappa^3 a^3}{192} +\frac{\kappa^4 a^4}{64} + \Big(\frac{\kappa^2 a^2}{6} + \frac{3 \kappa^3 a^3}{16} + \frac{3 \kappa^5 a^5}{64}\Big) e^{3\kappa a} E(3 \kappa a)
\end{align}
\end{subequations}
\marginnote{$u n$  
}[2.5cm]
\begin{subequations}
\begin{align}
\nonumber
-(z_+^2 - z_+ +z_-^2) y_a^2 \frac{2B}{\kappa^3 a^3}\Big\{ \frac{\kappa^2 a^2}{720} - \frac{\kappa^3 a^3}{240} + \frac{\kappa^4 a^4}{60} + \frac{\kappa^5 a^5}{16}- \frac{3 \kappa^6 a^6}{320}\\
\nonumber
\qquad +\frac{9 \kappa^7 a^7}{320} + \frac{2 \kappa^5 a^5 e^{2\kappa a}}{15} E(2 \kappa a) - \Big( \frac{3 \kappa^5 a^5}{40} - \frac{27 \kappa^6 a^6}{160} - \frac{27 \kappa^8 a^8}{320}\Big) e^{3\kappa a} E(3\kappa a)\Big\} \\
\nonumber
+\ifRED\else \frac{z_+/\lambda_+ + z_-/\lambda_-}{(z_+ + z_-)e} \frac{2\varepsilon_0 \varepsilon_r kT}{3\eta e}\fi y_a^2 \Big\{1 + \frac{2B}{\kappa^3 a^3}\Big(\frac{\kappa^2 a^2}{16} - \frac{5 \kappa^3 a^3}{48} - \frac{\kappa^4 a^4}{96} + \frac{\kappa^5 a^5}{96}\\
\nonumber
-\Big(\frac{\kappa^4 a^4}{8} - \frac{\kappa^6 a^6}{96}\Big) e^{\kappa a} E(\kappa a)\Big\} \Big\{ \frac{1}{8} - \frac{\kappa a}{4} - \frac{\kappa^2 a^2 e^{2 \kappa a}}{2} E(2\kappa a) \Big\}
\end{align}
\end{subequations}
\begin{subequations}
\begin{align}
\nonumber
-\ifRED\else \frac{z_+/\lambda_+ + z_-/\lambda_-}{(z_+ + z_-)e} \frac{2\varepsilon_0 \varepsilon_r kT}{3\eta e}\fi y_a^2 \Big\{ 3 + 3\kappa a + \kappa^2 a^2\\
\nonumber
+\frac{2B}{\kappa^3 a^3}\Big( \frac{\kappa^3 a^3}{10} - \frac{\kappa^4 a^4}{40} + \frac{\kappa^5 a^5}{120} -\frac{\kappa^6 a^6}{240} + \frac{\kappa^7 a^7}{240} + \frac{\kappa^8 a^8 e^{\kappa a}}{240} E(\kappa a) \Big) \Big\}\\
\nonumber
\qquad \times \Big\{\frac{1}{120} - \frac{\kappa a}{40} - \frac{\kappa^2 a^2}{90} + \frac{\kappa^3 a^3}{90} - \frac{\kappa^4 a^4}{45} - \frac{\kappa^2 a^2 e^{\kappa a}}{24} E(\kappa a) - \frac{2 \kappa^5 a^5 e^{2\kappa a}}{45} E(2 \kappa a) \Big\}\\
\nonumber
+ \ifRED\else \frac{z_+/\lambda_+ + z_-/\lambda_-}{(z_+ + z_-)e} \frac{2\varepsilon_0 \varepsilon_r kT}{3\eta e}\fi y_a^2 \Big\{ \frac{1}{40} - \frac{9 \kappa a}{80} - \frac{\kappa^2 a^2}{80}\\
\nonumber
+\frac{3 \kappa^3 a^3}{160} - \frac{9\kappa^4 a^4}{160} - \Big( \frac{3 \kappa^2 a^2}{8} + \frac{27 \kappa^5 a^5}{160}\Big) e^{3\kappa a} E(3 \kappa a) \Big\}+
\end{align}
\end{subequations}
\begin{subequations}
\begin{align}
\nonumber
+\ifRED\else \frac{z_+/\lambda_+ + z_-/\lambda_-}{(z_+ + z_-)e} \frac{2\varepsilon_0 \varepsilon_r kT}{3\eta e}\fi y_a^2 \frac{2B}{\kappa^3 a^3}\Big\{ \frac{\kappa^3 a^3}{1200} + \frac{611 \kappa^4 a^4}{38400} - \frac{6913 \kappa^5 a^5}{115200}\\
\nonumber
\qquad \qquad \qquad +\frac{173 \kappa^6 a^6}{76800} - \frac{559 \kappa^7 a^7}{76800}+ \Big( \frac{\kappa^4 a^4}{128} - \frac{47 \kappa^5 a^5}{1920} + \frac{\kappa^6 a^6}{3840} - \frac{\kappa^7 a^7}{1920}\Big) e^{\kappa a} E(\kappa a)\\
\nonumber
- \frac{2 \kappa^5 a^5 e^{2\kappa a}}{15} E(2 \kappa a) + \Big(\frac{27 \kappa^5 a^5}{128} - \frac{45 \kappa^6 a^6}{256} - \frac{559 \kappa^8 a^8}{25600}\Big) e^{3\kappa a} E(3\kappa a)\\
\nonumber
-\Big( \frac{3 \kappa^6 a^6}{64} + \frac{\kappa^8 a^8}{960}\Big) e^{3\kappa a} \int_\infty^{\kappa a} \frac{e^{-2 x} E(x)}{x}dx \Big\} \Big]. 
\end{align}
\end{subequations}
If now with the help of the series for $E(x)$, see \BLUE{after (55a)}, the various approximated terms are calculated for large $\kappa a$, then equation (88)\footnote{\BLUE{Note of the translator, a missing bracket "$)$" has been added in the 8th line from the end of (88) in the term $+\frac{2B}{\kappa^3 a^3}\Big( \frac{\kappa^3 a^3}{10} - \frac{\kappa^4 a^4}{40} + \frac{\kappa^5 a^5}{120} -\frac{\kappa^6 a^6}{240} + \frac{\kappa^7 a^7}{240} + \frac{\kappa^8 a^8 e^{\kappa a}}{240} E(\kappa a) \Big)$ in order to be consistent with (83): 10th line from the end.}} takes on the following shape:
\begin{subequations}
\begin{align}
\nonumber
\tag{88'} \qquad \qquad \qquad  \frac{U}{E}=\ifRED\else \frac{2\varepsilon_0 \varepsilon_r \zeta}{3 \eta}\fi\Big[1 +\frac{2B}{\kappa^3 a^3}\Big(\frac{1}{2}-\frac{9}{2\kappa a}+ \frac{75}{2 \kappa^2 a^2}- \frac{330}{\kappa^3 a^3}....\Big)\\
\nonumber
+(z_- - z_+) y_a \frac{2B}{\kappa^3 a^3}\Big( -\frac{1}{6}+\frac{2}{9\kappa a}-\frac{13}{27 \kappa^2 a^2} +\frac{40}{27 \kappa^3 a^3}.....\Big)\\
\nonumber
\times \Big( \frac{1}{2}-\frac{9}{2\kappa a} +\frac{75}{2 \kappa^2 a^2}-\frac{330}{\kappa^3 a^3}..... \Big)\\
\nonumber
+(z_- - z_+)y_a \frac{2B}{\kappa^3 a^3}\Big( \frac{1}{12}-\frac{35}{72 \kappa a}+\frac{269}{108 \kappa^2 a^2} - \frac{2815}{216 \kappa^3 a^3}....\Big)\\
\nonumber
+(z_+^2 - z_+ z_- + z_-^2) y_a^2 \frac{2B}{\kappa^3 a^3}\Big(-\frac{1}{48}+\frac{1}{32 \kappa a}- \frac{7}{128 \kappa^2 a^2}+
\end{align}
\end{subequations}
\begin{subequations}
\begin{align}
\nonumber
+\frac{15}{128 \kappa^3 a^3}.....\Big) \Big(\frac{1}{2}-\frac{9}{2\kappa a}+ \frac{75}{2 \kappa^2 a^2}- \frac{330}{\kappa^3 a^3}.....\Big)\\
\nonumber
\qquad +(z_+^2 - z_+ z_- + z_-^2) y_a^2 \frac{2B}{\kappa^3 a^3}\Big( \frac{1}{96} -\frac{3}{64\kappa a} +\frac{491}{2304 \kappa^2 a^2}-\frac{1371}{1728 \kappa^3 a^3}....\Big)\\
\nonumber
+(z_- - z_+)y_a \Big( \frac{1}{8 \kappa a} - \frac{3}{16 \kappa^2 a^2}+ \frac{3}{8 \kappa^3 a^3} - \frac{15}{16 \kappa^4 a^4} .... \Big)\\
\nonumber
-(z_- - z_+) y_a \frac{2B}{\kappa^3 a^3} \Big( \frac{1}{8 \kappa a} - \frac{3}{2 \kappa^2 a^2} + \frac{57}{4 \kappa^3 a^3} -\frac{525}{4 \kappa^4 a^4}....\Big)\\
\nonumber
-z_+ z_- y_a^2 \Big\{ \kappa^2 a^2 + \kappa a + \frac{2B}{\kappa^3 a^3}\Big( \frac{\kappa^2 a^2}{2} - \frac{5 \kappa a}{2} + 1 
\end{align}
\end{subequations}
\begin{subequations}
\begin{align}
\nonumber
-\frac{105}{\kappa a}....\Big) \Big\} \Big\{ \frac{1}{8 \kappa^3 a^3}- \frac{15}{16 \kappa^4 a^4} +\frac{51}{8 \kappa^5 a^5} -\frac{729}{16\kappa^6 a^6}....\Big\}\\
\nonumber
+(z_+^2 - z_+ z_- +z_-^2) y_a^2 \Big(\frac{1}{54 \kappa a} - \frac{2}{81 \kappa^2 a^2} +\frac{19}{486 \kappa^3 a^3}- \frac{1}{54 \kappa^4 a^4}.....\Big)\\
\nonumber
-(z_+^2 - z_+ z_- +z_-^2) y_a^2 
\frac{2B}{\kappa^3 a^3}\Big( \frac{1}{54 \kappa a}-\frac{71}{324 \kappa^2 a^2} + \frac{1979}{972 \kappa^3 a^3} - \frac{53347}{2916 \kappa^4 a^4}....\Big)\\
\nonumber
+(z_+^2 - z_+ z_- + z_-^2) y_a^2 \Big(\frac{1}{108 \kappa a} -\frac{7}{324 \kappa^2 a^2}+ \frac{49}{972 \kappa^3 a^3} - \frac{95}{729 \kappa^4 a^4}....\Big)\\
\nonumber
-(z_+^2 - z_+ z_- + z_-^2) y_a^2 \frac{2B}{\kappa^3 a^3}\Big( \frac{1}{108 \kappa a} - \frac{25}{324 \kappa^2 a^2} +\frac{959}{1944 \kappa^3 a^3} - \frac{4333}{1458 \kappa^4 a^4}.....\Big)\\
\nonumber
+ \ifRED\else \frac{z_+/\lambda_+ + z_-/\lambda_-}{(z_+ + z_-)e} \frac{2\varepsilon_0 \varepsilon_r kT}{3\eta e}\fi y_a^2
\end{align}
\end{subequations}
\begin{subequations}
\begin{align}
\nonumber
\times \Big\{1 + \frac{2B}{\kappa^3 a^3}\Big( \frac{1}{2}- \frac{9}{2 \kappa a} + \frac{75}{2 \kappa^2 a^2}- \frac{330}{\kappa^3 a^3}.... \Big) \Big\} \Big\{ \frac{1}{8 \kappa a}\\
\nonumber
-\frac{3}{16 \kappa^2 a^2} + \frac{3}{8 \kappa^3 a^3}- \frac{15}{16 \kappa^4 a^4}.... \Big\}\\
\nonumber
- \ifRED\else \frac{z_+/\lambda_+ + z_-/\lambda_-}{(z_+ + z_-)e} \frac{2\varepsilon_0 \varepsilon_r kT}{3\eta e}\fi y_a^2 \Big\{ \kappa^2 a^2 + 3 \kappa a + 3 + \frac{2B}{\kappa^3 a^3}\Big( \frac{\kappa^2 a^2}{2}\\
\nonumber
-3 \kappa a +21 - \frac{168}{\kappa a}....\Big) \Big\} \Big\{ \frac{1}{8 \kappa^3 a^3} - \frac{3}{2 \kappa^4 a^4}+ \frac{57}{4 \kappa^5 a^5} -\frac{525}{4 \kappa^6 a^6}....\Big\}\\
\nonumber
+\ifRED\else \frac{z_+/\lambda_+ + z_-/\lambda_-}{(z_+ + z_-)e} \frac{2\varepsilon_0 \varepsilon_r kT}{3\eta e}\fi y_a^2 \Big( \frac{1}{36 \kappa^2 a^2} - \frac{5}{54 \kappa^3 a^3} + \frac{23}{81 \kappa^4 a^4} - \frac{74}{81 \kappa^5 a^5}....\Big) \\
\nonumber
+\ifRED\else \frac{z_+/\lambda_+ + z_-/\lambda_-}{(z_+ + z_-)e} \frac{2\varepsilon_0 \varepsilon_r kT}{3\eta e}\fi y_a^2 \frac{2B}{\kappa^3 a^3}\Big( \frac{1}{72 \kappa^2 a^2}- \frac{8}{27 \kappa^3 a^3} + \frac{2903}{648 \kappa^4 a^4} - \frac{16043}{105 \kappa^5 a^5}....\Big) \Big]. 
\end{align}
\end{subequations}

With this our goal has been reached and we indeed found a formula for the electrophoretic velocity, which takes into account the relaxation effect and can be applied for any value of $\kappa a$, i.e for any electrolyte concentration or particle radius.

Inspecting formulas (88) and (88') we note that so many of the correction terms have a positive sign, i.e they accelerate the electrophoresis. However, one must keep in mind, that in $\frac{2B}{\kappa^3 a^3}$ terms
with $y_a$ and $y_a^2$ also appear. If one combines the corresponding terms of $U$ (88) and $\frac{2B}{\kappa^3 a^3}$ (84), most of these terms appear to be decelerating. As an example we shall treat, in the case of large $\kappa a$, the correction term which is proportional to $(z_- - z_+) y_a$. According to (84')
\begin{subequations}
\begin{align}
\nonumber
\frac{2B}{\kappa^3 a^3}\Big\{1+(z_--z_+) y_a \Big(\frac{1}{\kappa a} -\frac{5}{\kappa^2 a^2}+ \frac{30}{\kappa^3 a^3} - \frac{210}{\kappa^4 a^4}....\Big) + \text{etc.}\Big\}\\
\nonumber
=1 -(z_--z_+) y_a \Big(\frac{2}{\kappa a} + \frac{2}{\kappa^2 a^2}\Big) +\text{etc.} 
\end{align}
\end{subequations}
with the neglect of quantities of the order $(z_- - z_+)^2 y_a^2$ (thus with the same accuracy, where the whole calculation is done for non symmetric electrolytes), we can write for this
\begin{equation}
\begin{aligned}
\nonumber
\frac{2B}{\kappa^3 a^3}=1-(z_--z_+) y_a \Big( \frac{3}{\kappa a}- \frac{3}{\kappa^2 a^2}+\frac{30}{\kappa^3 a^3}-\frac{210}{\kappa^4 a^4}..... \Big) + \text{etc.}
\end{aligned}
\end{equation}
Substituting in (88') gives:
\begin{subequations}
\begin{align}
\nonumber
\frac{U}{E}=\ifRED\else \frac{2 \varepsilon_0 \varepsilon_r \zeta}{3 \eta}\fi \Big[ 1 +\frac{1}{2}-\frac{9}{2\kappa a} + \frac{75}{2 \kappa^2 a^2} - \frac{330}{\kappa^3 a^3}.....\\
\nonumber
-(z_--z_+) y_a \Big(\frac{3}{\kappa a} -\frac{3}{\kappa^2 a^2}+\frac{30}{\kappa^3 a^3}-\frac{210}{\kappa^4 a^4}...\Big)\Big(\frac{1}{2}-\frac{9}{2\kappa a}+\frac{75}{2 \kappa^2 a^2}- \frac{330}{\kappa^3 a^3}... \Big)\\
\nonumber
+(z_- - z_+) y_a \Big(-\frac{1}{6}+\frac{2}{9\kappa a}-\frac{13}{27 \kappa^2 a^2}+\frac{40}{27 \kappa^3 a^3}....\Big)\Big(\frac{1}{2}
-\frac{9}{2\kappa a}+ \frac{75}{2 \kappa^2 a^2} -\frac{330}{\kappa^3 a^3} ..... \Big)\\
\nonumber
+(z_- - z_+) y_a \Big(\frac{1}{12}-\frac{35}{72 \kappa a}+\frac{269}{108 \kappa^2 a^2}-\frac{2815}{216 \kappa^3 a^3}....\Big)+
\end{align}
\end{subequations}
\begin{subequations}
\begin{align}
\nonumber
+(z_-- z_+) y_a \Big( \frac{1}{8 \kappa a} -\frac{3}{16 \kappa^2 a^2}+\frac{3}{8\kappa^3 a^3}-\frac{15}{16\kappa^4 a^4}...\Big)\\
\nonumber
\qquad \qquad \qquad \qquad -(z_--z_+)y_a\Big(\frac{1}{8 \kappa a}- \frac{3}{2 \kappa^2 a^2} + \frac{57}{4 \kappa^3 a^3}-\frac{525}{4 \kappa^4 a^4}....\Big) + \text{etc.}\Big] \\
\nonumber
=\ifRED\else \frac{2 \varepsilon_0 \varepsilon_r \zeta}{3 \eta}\fi \Big[\frac{3}{2}-\frac{9}{2\kappa a}+\frac{75}{2\kappa^2 a^2}-\frac{330}{\kappa^3 a^3}-(z_-- z_+)y_a \Big( \frac{9}{8\kappa a}
-\frac{181}{16 \kappa^2 a^2}+\frac{305}{3 \kappa^3 a^3}.....\Big) + \text{etc.}\Big].
\end{align}
\end{subequations}

This example shows very clearly that it is necessary to take into account the higher order terms in equation (55b) for $y(x)$. the second approximation of $y$ namely gives rise to a correction $+(z_- - z_+) y_a \Big( \frac{3}{8 \kappa a} - \frac{5}{\kappa^2a^2} ....\Big)$, which is of the same order of magnitude as the other correction terms. 

By combining all terms in (84) and (88) in this manner, we get the following electrophoresis velocity formula

\begin{subequations}
\begin{align}
\nonumber
\frac{U}{E}=\ifRED\else \frac{2 \varepsilon_0 \varepsilon_r \zeta}{3 \eta}\fi \Big[f_1(\kappa a) - (z_- - z_+)y_a f_2(\kappa a) + z_+ z_- y_a^2 g_3(\kappa a)\\
\nonumber
-(z_+^2 -z_+ z_- +z_-^2) y_a^2 g_4(\kappa a)- \ifRED\else \frac{z_+/\lambda_+ + z_-/\lambda_-}{(z_+ + z_-)e} \frac{2\varepsilon_0 \varepsilon_r kT}{3\eta e}\fi y_a^2 f_4(\kappa a) \Big].
\end{align}
\end{subequations}
In this relationship the term $(z_- - z_+)^2 y_a^2 g(\kappa a)$ and all contributions with higher power in $y_a$ are being neglected. The terms $g_3(\kappa a)$ and $g_4(\kappa a)$ can thus only be used in those cases, where $z_+ = z_-$ because if $z_+ \neq z_-$, they are of the same order of magnitude as the neglected term $(z_- - z_+)^2 y_a^2 g(\kappa a)$.

We therefore split the electrophoretic formula in a formula (89a), valid for symmetric electrolytes, and a formula (89b), valid for non symmetric elecrolytes.\\

\noindent
Symmetric electrolyte\footnote{\BLUE{This equation is identical to the one found by Ohshima et al. for a large $\kappa a$ approximation (H.Ohshima, T.W.Healy and L.R.White J.Chem. Soc., Faraday Trans. 2, 1983, 79, 1613-1628). In their notation (their Eq. 63): $E_m = \frac{3}{2}\overline{\zeta}-\frac{9\overline{\zeta}}{2\kappa a} -\frac{(\overline{\zeta})^3}{2 \kappa a} -\frac{9(m_+ + m_-)}{16 \kappa a}(\overline{\zeta})^3$, where $E_m$ is the dimensionless electrophoretic velocity and $\overline{\zeta}$ is the dimensionless zeta-potential.}}
\begin{equation}
\nonumber
\frac{U}{E} = \frac{2 \varepsilon_0 \varepsilon_r\zeta}{3 \eta } \Big[f_1(\kappa a) - z^2 y_a^2 f_3(\kappa a) - \frac{1/\lambda_+ + 1/\lambda_-}{2 e} \frac{2 \varepsilon_0 \varepsilon_r kT}{3 \eta e}y_a^2 f_4(\kappa a) \Big] \tag{89a}
\end{equation}

\noindent
Non-symmetric electrolyte
\begin{equation}
\nonumber
\frac{U}{E} = \frac{2 \varepsilon_0 \varepsilon_r\zeta}{3 \eta } \Big[f_1(\kappa a) - (z_- - z_+)y_a f_2(\kappa a) - \frac{z_+/\lambda_+ + z_-/\lambda_-}{(z_+ + z_-) e} \frac{2 \varepsilon_0 \varepsilon_r kT}{3 \eta e}y_a^2 f_4(\kappa a) \Big]. \tag{89b}
\end{equation}
In the formula for the non symmetric electrolytes is, despite the fact that contributions proportional to $z_+ z_- y_a^2$, $(z_+^2 - z_+ z_- + z_-^2) y_a^2$ and $(z_- - z_+)^2 y_a^2$ were neglected, the term $\frac{z_+/\lambda_+ + z_-/\lambda_-}{(z_+ + z_-) e} \frac{2 \varepsilon_0 \varepsilon_r kT}{3 \eta e}y_a^2 f_4(\kappa a)$ 
 is \textit{not} deleted, since this term, which describes the influence of the ionic flow on the ion distribution, for small $\kappa a$ dominates over the others (connection to the theory of Onsager) and because at least it is clear that with non symmetric electrolytes except for the term, proportional to $(z_- - z_+) y_a$ a negative correction term, proportional to $y_a^2$ exists.  

\newpage
\section{THE ELECTROPHORETIC VELOCITY OF NON-CONDUCTING PARTICLES}

In the previous chapter it was deduced that the electrophoresis formula for a non-conducting sphere in a solution of a symmetric or non-symmetric electrolyte could be respectively summarized with the equations:\footnote{\BLUE{In this chapter, we have written all equations, except (89a) in modern notation. In order to get back Overbeek's original equations, it suffices to change back $E$ into $X$, $\epsilon_0 \epsilon_r$ into $D/(4\pi)$, $\lambda_\pm$ into $1/\varrho_\pm$ and $e$ into $\varepsilon$ in all $U/E$ formulas (see as an example 89a)).}}\\
\\
symmetric electrolyte
\begin{subequations}
\begin{align}
\nonumber \tag{89a}
\frac{U}{E} = \frac{2 \varepsilon_0 \varepsilon_r\zeta}{3 \eta } \Big[f_1(\kappa a) - z^2 \Big(\frac{e \zeta}{kT}\Big)^2 f_3(\kappa a) - \frac{1/\lambda_+ + 1/\lambda_-}{2 e} \frac{2 \varepsilon_0 \varepsilon_rkT}{3 \eta e}\Big(\frac{e \zeta}{kT}\Big)^2 f_4(\kappa a) \Big]
\end{align}
\end{subequations}
non-symmetric electrolyte
\begin{subequations}
\begin{align}
\nonumber
\frac{U}{E} = \frac{2 \epsilon_0 \epsilon_r\zeta}{3  \eta } \Big[f_1(\kappa a) - (z_- - z_+)\Big(\frac{e \zeta}{kT}\Big) f_2(\kappa a)   \nonumber - \frac{z_+/\lambda_+ + z_-/\lambda_-}{(z_+ + z_-) e} \frac{2 \epsilon_0 \epsilon_r kT}{3  \eta e}\Big(\frac{e \zeta}{kT}\Big)^2 f_4(\kappa a) \Big]. \tag{89b}
\end{align}
\end{subequations}
Here $f_1(\kappa a)$ is identical with $f(\kappa a)$ from Henry\footnote{D.C. HENRY, Proc. Roy. Soc. London. \textbf{A 133}, 106 (1931)} and describes, together with $\zeta$ from the term $\frac{D\zeta}{6\pi \eta}$, the influence of the electrophoretic drag. 

The relaxation effect is expressed through the correction terms $f_2(\kappa a)$ to $f_4(\kappa a)$, which can be considered to be the first terms of a series of increasing powers of $\Big( \frac{e \zeta}{kT}\Big)$ . In the equation for symmetric electrolytes (89a), which does not contain a contribution proportional to $\Big( \frac{e \zeta}{kT}\Big)^1$, the term proportional to $\Big( \frac{e \zeta}{kT}\Big)^2$ is entirely taken into account. In (89b) some of the terms proportional to $\Big( \frac{e \zeta}{kT}\Big)^2$ have been neglected, i.e. those who represent the influence of the electric field on the ion distribution (see also chapter IV). This influence is already expressed in the term $(z_- - z_+)\frac{e \zeta}{kT} f_2(\kappa a)$ and it would require a very considerable amount of calculation to get the second approximation entirely. The last term of (89b) describes the influence of the fluid flow on the ion distribution, which only expresses itself in the second order approximation. We should however be mindful, that in (89b) a contribution proportional to $\Big( \frac{e \zeta}{kT}\Big)^2$ has been neglected and thus (89b) is less good than (89a). 

The physical meaning of the terms can be described as follows. $f_2$ and $f_3$ reflect the influence of the electric field on the ion distribution and originate from the terms 
$n^0_\pm z_\pm e\lambda_\pm \nabla \delta \psi$ and
$\delta n_\pm z_\pm e\lambda_\pm \nabla \psi^0$ of (10), while $f_4$ describes the influence of the fluid flow ($u n_\pm^0$ from (10)).
The influence of the diffusion $\big(kT\lambda_\pm \nabla  \delta n_\pm \big)$ is in each of the terms $f_2$ to $f_4$ partly incorporated. 

$f_2$, $f_3$ and $f_4$ are positive for an $\kappa a$, the correction terms $f_3$ and $f_4$ will thus always lead to a reduction of the electrophoretic velocity; the sign of the correction term $(z_- - z_+)\frac{e \zeta}{kT} f_2(\kappa a)$ depends on the sign of $(z_- - z_+)$. If the valence of the counter ion is larger than that of the ion which has the same sign as the particle, then this term will give a reduction, in the opposite case it will lead to an increase in electrophoretic velocity.

In the first instance we will now rewrite equation (89) for very small and very large $\kappa a$. For very \textit{small} $\kappa a$, $\{f_1(\kappa a)-1\}$ and $f_3(\kappa a)$ approach faster to zero than all other terms (as $\kappa^2 a^2$ vs. $\kappa a$), such that remains
\begin{equation}
\nonumber
\tag{89I} \frac{U}{E} = \frac{2 \epsilon_0 \epsilon_r\zeta}{3 \eta } \Big\{1 - (z_- - z_+)\frac{e \zeta}{kT} \frac{\kappa a}{6}  - \frac{z_+/\lambda_+ + z_-/\lambda_-}{(z_+ + z_-) e} \frac{2 \epsilon_0 \epsilon_r kT}{3 \eta e}\Big(\frac{e \zeta}{kT}\Big)^2 \frac{\kappa a}{6} \Big\}.
\end{equation}
where both formulas (89a and b) are combined. If we substitute $6\pi \eta a = 1/\lambda_+ $ and $\zeta = \frac{z_+ e}{4\pi \epsilon_0 \epsilon_r a} $ in this equation then it transforms into
\begin{equation}
\nonumber
 \frac{U}{E} = \frac{2 \epsilon_0 \epsilon_r\zeta}{3 \eta } \Big\{1 -  \frac{z_- /\lambda_+}{z_+ e} \frac{z_-/\lambda_+ + z_+/\lambda_-}{(z_+ + z_-) /\lambda_+} \frac{2 \epsilon_0 \epsilon_r kT}{3 \eta e}\Big(\frac{e \zeta}{kT}\Big)^2 \frac{\kappa a}{6} \Big\}.
\end{equation}
which is entirely comparable to the formulas that Debye and H\"uckel and Onsager deduced for the velocity of ions. See also chapter III .

In the other extreme case of \textit{large} $\kappa a$, (89a and b) transform into\\
\\
(89aII) symmetric electrolyte
\begin{equation}
\nonumber
\frac{U}{E} = \frac{2 \epsilon_0 \epsilon_r \zeta}{3 \eta } \Big\{1 + \frac{1}{2} -\frac{9}{2 \kappa a}- z^2 \Big(\frac{e \zeta}{kT}\Big)^2 \frac{1}{2 \kappa a} - \frac{1/\lambda_+ + 1/\lambda_-}{2 e} \frac{2 \epsilon_0 \epsilon_r kT}{3 \eta e}\Big(\frac{e \zeta}{kT}\Big)^2 \frac{9}{8 \kappa a} \Big\}
\end{equation}
(89bII) non-symmetric electrolyte
\begin{subequations}
\begin{align}
\nonumber
\frac{U}{E} = \frac{2 \epsilon_0 \epsilon_r \zeta}{3 \eta } \Big\{1 + \frac{1}{2} -\frac{9}{2 \kappa a} - (z_- - z_+)\frac{e \zeta}{kT} \frac{9}{8 \kappa a}  r - \frac{z_+/\lambda_+ + z_-/\lambda_-}{(z_+ + z_-) e} \frac{2 \epsilon_0 \epsilon_r kT}{3 \eta \varepsilon}\Big(\frac{e \zeta}{kT}\Big)^2 \frac{9}{8 \kappa a} \Big\}.
\end{align}
\end{subequations}
The correction terms are proportional to $1/(\kappa a)$, and thus approach zero for very large values of $\kappa a$. We thus come to the curious conclusion that the Smoluchowski equation \BLUE{(1)}
\begin{equation}
\nonumber
\frac{U}{E}=\frac{\varepsilon_0 \varepsilon_r\zeta}{\eta}\qquad 
\end{equation}
is indeed valid, even if one takes into account the distortion of the double layer, provided the double layer is thin! 
In chapter III we have pointed out that Hermans\footnote{J.J. HERMANS, Phil. Mag. VII, \textbf{26}, 650 (1938).} reaches a different conclusion, and that he finds correction terms for the relaxation effect, which are about $\kappa a$ times as big as ours. We can now prove that this difference is mainly due to the fact that Hermans did not take into account the interaction of the electrophoretic drag and the relaxation effect and thus $U$ was not calcalated according to our (87) and (48), but according to the electrophresisformula given before\footnote{\BLUE{Translator's note: This is now the footnote that appears just before equation (48a).}}.

We have, with the help of (70) calculated the value of $f(r)$ with the above mentioned footnote, according to $I=\kappa R = \kappa r + \frac{\kappa \lambda a^3}{2 r^3} + \kappa f(r)$, and substituted this value in the electrophoresis equation of this footnote\footnote{It is necessary to include the term $P e^{-x} \Big( \frac{1}{x}+\frac{1}{x^2}\Big)$, since this will now give a contribution to the electrophoretic velocity, which is not the case if one uses the electrophoresis formula (87) or (48).}. Then we find
\begin{equation}
\nonumber
\frac{U}{E}=\frac{2 \epsilon_0 \epsilon_r \zeta}{3\eta} \Big\{1 + \frac{1}{2}-(z_--z_+) \frac{e \zeta}{kT}(1 - ....) + \text{higher order terms of} \Big( \frac{e \zeta}{kT}\Big) \Big\}.
\end{equation}
So the relaxation effect is found to be a factor $\kappa a$ too large and it is certainly not allowed to consider the influence of the electrophoretic drag and the relaxation drag as two separate corrections for large $\kappa a$. The electrophoretic drag is so large\footnote{After all, $k_1= Q E \; $, the electrophoretic drag $k_3=(4\pi \epsilon_0 \epsilon_r \zeta a - Q) E \; $ and $Q=4\pi \epsilon_0 \epsilon_r \zeta a(1 + \kappa a)$  so $k_3=- \frac{\kappa a}{1+\kappa a} Q E $. This means that for large $\kappa a$ the electrophoretic drag is almost as big as the driving force $k_1$, such that the electrophoretic drag alone will already reduce the electrophoretic velocity with a factor $1/(\kappa a)$.} there, that it changes noticeably with a relative minor change of the relaxation in the field.

At small $\kappa a$, where the electrophoretic drag is small, this interaction does not play any role of great significance, such that from equations (87), and (48) the same electrophoretic velocity is calculated, as the one from the electrophoresis equation from the previously mentioned footnote. 

We have thus shown, that in the cases of very dilute and very compact double layers the corrections regarding the relaxation effect are very small. 

In the intermediate region, from the viewpoint of colloid chemistry the most interesting, this is not the case however, as is most clearly illustrated from a graphical representation of the various corrections terms. 
\ifFIG\begin{figure}
    \centering
    \includegraphics[width=0.5\textwidth]{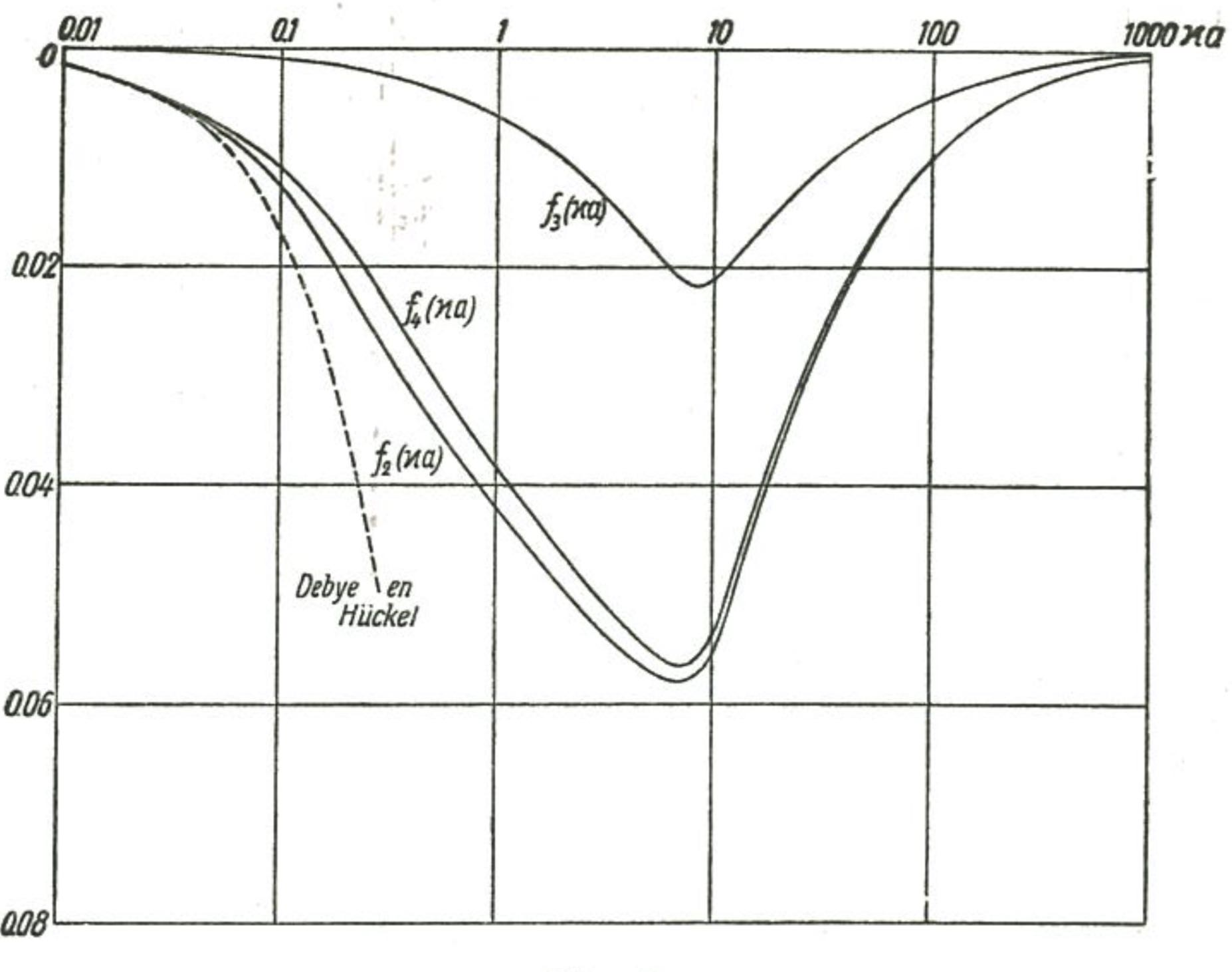}
    \caption{Values of $f_2(\kappa a)$, $f_3(\kappa a)$ and $f_4(\kappa a)$ from equation (89).}
    \label{fig:9}
\end{figure}\fi

\begin {table} \label{Table3}
\caption*{TABLE III Values of the various correction terms from the electrophoresis formula (89) 
 } \label{tab:title} 
\begin{tabular}{| l | l | l | l | l | l | l | l | l | l | }
  \hline
  $\kappa a$ & 0.1 & 0.3 & 1 & 3 & 5 & 10 & 20 & 50 & 100\\
  \hline
  $f_1(\kappa a)$ & 1.000545  & 1.00398 & 1.0267 & 1.1005 & 1.163 & 1.25 & 1.34 & 1.424 & 1.458\\
  $f_2(\kappa a)$ & 0.0125 & 0.0279 & 0.0411 & 0.053 & 0.057 & 0.056 & 0.04 & 0.0188 & 0.0102\\ 
  $f_3(\kappa a)$ & 0.00090  & 0.0044 & 0.0116 & 0.020 & 0.022 & 0.021 & 0.0145 & 0.00796 & 0.00444\\ 
  $f_4(\kappa a)$ & 0.0107  & 0.0218 & 0.0387 & 0.00515 & 0.0546 & 0.055 & 0.04 & 0.0177 & 0.00992\\ 
  \hline  
\end{tabular}
\end{table}

In Fig.\ref{fig:9}  and Table III\footnote{The values of $f(\kappa a)$ for $\kappa a=10$ and $\kappa a=20$ are less accurate than the others, since the series for such large values of $\kappa a$ converge badly and the application of the series for $\kappa a$ does not give accurate results, since $E(x)$ (see chapter IV) is not known with enough accuracy.} \footnote{After checking it appeared that some errors had crept in for the values of $f_3(\kappa a)$ between $\kappa a=0.3$ and $\kappa a=5$. This (by the way not very important) error was corrected in Tables III and V. It was however not possible to change the corresponding figures 9 and 10, such that they do not correspond at certain points with the values given. The shape and relative position of the curves does not change however. } the values of $f_1$ to $f_4$ are given as a function of $\kappa a$. In the area between $\kappa a=0.1$ and $\kappa a=100$ the value of the $f(\kappa a)'s$ reach values varying from 0.005 to 0.05. They will be multiplied by a factor  $\frac{e \zeta}{kT}$  and $\Big(\frac{e \zeta}{kT}\Big)^2$ respectively, which for $\zeta$- potential of 0, 25 and 100 millivolts is 0.1, 4 and 16 respectively, and with a valence factor $\frac{z_+ /\lambda_+ + z_- /\lambda_-}{(z_+ + z_-) e}\frac{2 \epsilon_0 \epsilon_r kT}{3 \eta e}$ , which as seen from Table IV\footnote{The factor  $\frac{z_+ /\lambda_+ + z_-/ \lambda_-}{(z_+ + z_-) e}\frac{2 \varepsilon_0 \varepsilon_r kT}{3 \eta e}$ is given for two special cases: 1. both ion species have "normal" mobility. 2. the cation has the mobility of the H-ion, the anion has normal mobility.} range from 0-4 to 0.1-0.6 respectively.

\begin {table} \label{Table4}
\caption*{TABLE IV Valence and conductivity from formulas (89a and b). 
 } \label{tab:title} 
\begin{tabular}{| c | l | l | l | l |}
  \hline
  electrolytetype& & &  \multicolumn{2}{|c|}{ $\frac{z_+ /\lambda_+ + z_- /\lambda_-}{(z_++z_-)e}\frac{2 \epsilon_0 \epsilon_r kT}{3 \eta e}$ }\\
  - + & $(z_- + z_+) \hspace{1cm}$ & $z^2$ \hspace{1cm} & $\Lambda_+=\Lambda_-=70$ & $\Lambda_-=70$, $\Lambda_+=350$ \\
  \hline
  1-1 & 0  & 1 & 0.184 & 0.110 \\
  1-2 & -1 & - & 0.307 & 0.110 \\ 
  1-3 & -2 & - & 0.460 & 0.129 \\ 
  1-4 & -3 & - & 0.625 & 0.154 \\ 
  2-1 &  1 & - & 0.307 & 0.258 \\ 
  2-2 &  0 & 4 & 0.368 & 0.221 \\ 
  3-1 &  2 & - & 0.460 & 0.422 \\ 
  4-1 &  3 & - & 0.625 & 0.595 \\ 
  \hline  
\end{tabular}
\end{table}

It also turns out that the term $z^2\Big(\frac{e \zeta}{kT} \Big)^2 f_3(\kappa a)$, which for small $\kappa a$ can be neglected against the other correction terms, is of the same order of magnitude for large $\kappa a$ as
\begin{equation}
\nonumber
\frac{z_+ /\lambda_+ + z_-/ \lambda_-}{(z_+ + z_-) e}\frac{2 \varepsilon_0 \varepsilon_r kT}{3 \eta e}\Big(\frac{e \zeta}{kT} \Big)^2 f_4(\kappa a) \qquad 
\end{equation}
and in general it is thus not allowed to neglect this term. 

The dotted line in Fig.\ref{fig:9} indicates the effect of the relaxation effect according to Debye and H\"uckel. This shows that our calculation deviates considerable from that of Debye and H\"uckel if $\kappa a > 1/10$. This justifies our statement in chapter I, concerning the validity range of the method of Paine. 
\ifFIG\begin{figure}
    \centering
    \includegraphics[width=0.6\textwidth]{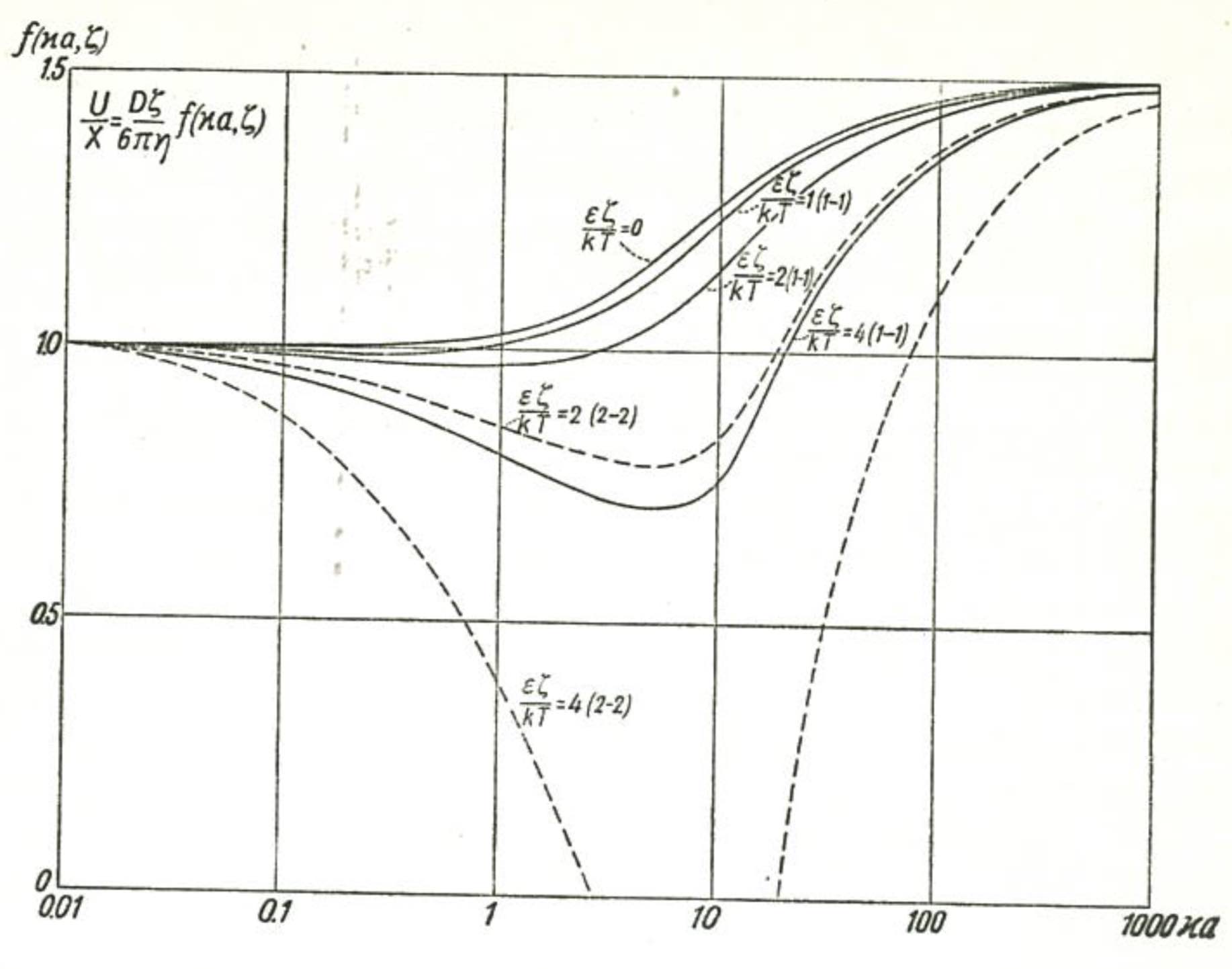}
    \caption{$f(\kappa a,\zeta)$ for symmetrical electrolytes.}
    \label{fig:10}
\end{figure}\fi

\ifFIG\begin{figure}
    \centering
    \includegraphics[width=0.6\textwidth]{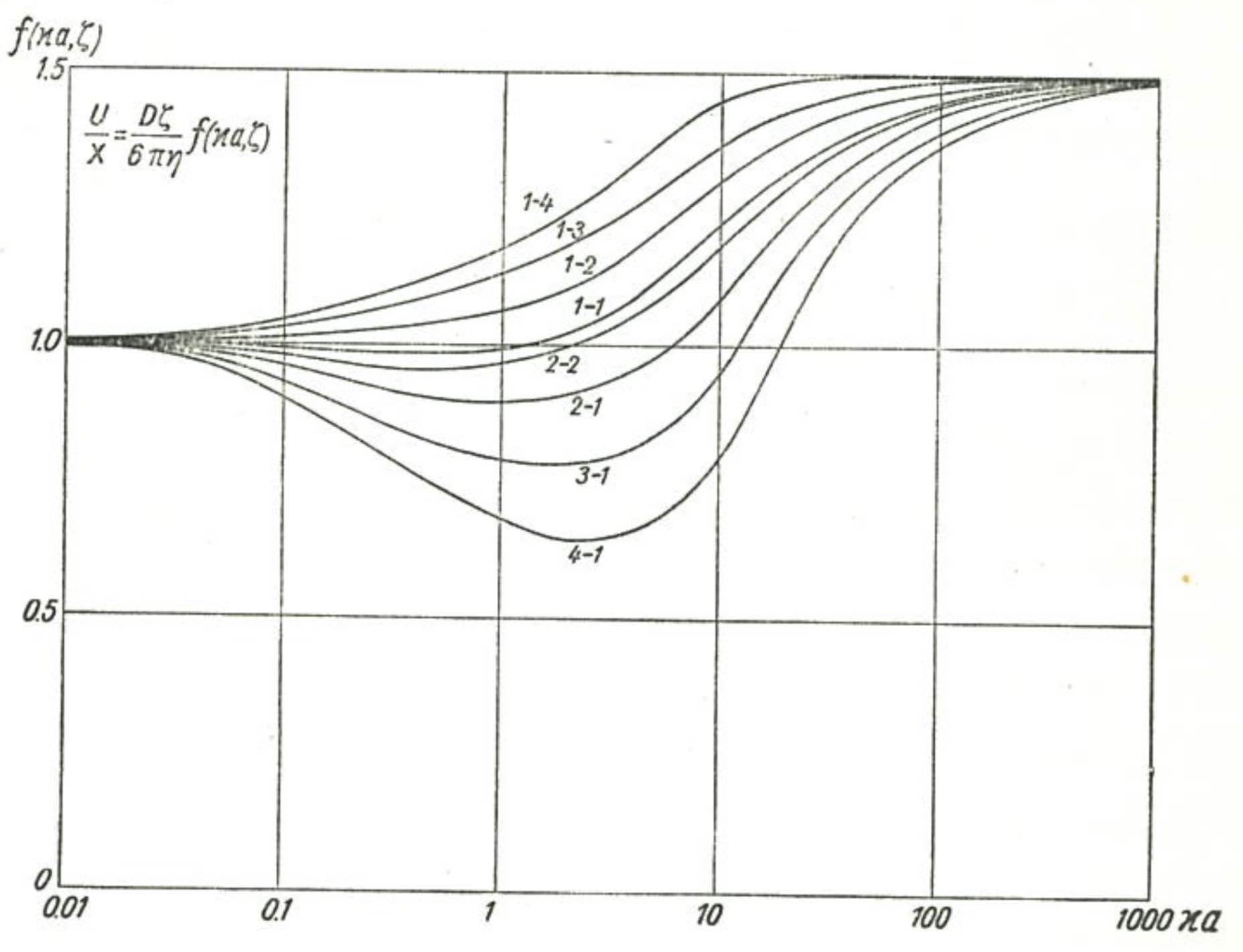}
    \caption{$f(\kappa a,\zeta)$ for non symmetrical electrolytes at $\frac{e \zeta}{kT}\ifRED \RED{\big[=\frac{\varepsilon \zeta}{kT}\big]}\fi =2$ ($\zeta$ = 50 mV).}
    \label{fig:11}
\end{figure}\fi
In Figs. \ref{fig:10} and \ref{fig:11} and Table V, the amount is indicated, for several values of $\frac{e \zeta}{kT}$  and $\kappa a$, with which the quantity $\frac{2 \varepsilon_0 \varepsilon_r \zeta}{3\eta}$ must be multiplied in order to give the electrophoretic velocity. We thereby have applied our formula to larger values of $\zeta$ as well, despite the fact, that our calculations will certainly no longer be correct. Since the equations (89a and b) can be considered as the beginning of a series of increasing powers of $\frac{e \zeta}{kT}$  , we can only be satisfied with the first calculated term in this series if $\frac{e \zeta}{kT}\ll 1$, i.e. if $\zeta \ll$25 mV.

\begin {table} \label{Table5}
\caption*{TABLE V $f(\kappa a, \zeta)$ for symmetrical electrolytes
 } \label{tab:title} 
\begin{tabular}{| l | l | l | l | l | l | l | l | l |}
  \hline
  electrolyte& $\frac{e \zeta}{kT}$ & & & & $\kappa a$ & &  &\\
  type &  & 0.01 & 0.1 & 1 & 5 & 10 & 100 & 1000 \\
  \hline
      &  0  & 1.000 & 1.000 & 1.027 & 1.163& 1.25& 1.458& 1.495 \\
  1-1 &  1 & 1.000 & 1.00 & 1.01& 1.13 & 1.22& 1.45& 1.495 \\ 
      &  2 & 0.999 & 0.99 & 0.95 & 1.03 & 1.12 & 1.43 & 1.493 \\
      &  4 & 0.995 & 0.95 & 0.73& 0.65 & 0.75& 1.38 & 1.484\\
      &  & & & & & & &       \\
      &  0 & 1.000 & 1.000 & 1.027 & 1.163 & 1.25 & 1.458 & 1.495 \\ 
  2-2 &  1 & 0.999 & 0.99 & 0.97 & 1.05 & 1.15 & 1.44 & 1.493 \\ 
      &  2 & 0.998 & 0.97 & 0.77 & 0.73 & 0.83 & 1.37 & 1.486 \\ 
      &  4 & 0.990 & 0.88 & 0.06 & -0.57 & -0.41 & 1.11 & 1.457 \\ 
  \hline  
\end{tabular}
\end{table}

In order to be able to apply the equations on cases where $\zeta >$25 mV, we will have to know, when the coefficients $(f(\kappa a))$ of the next terms are much smaller than that of the first term. But we don't know that. Only for small $\kappa a$ we know this, otherwise the theory of Onsager for strong electrolytes, where the first approximation on hit $\zeta$-potentials is applied, couldn't have turned out so precise.

Since for the colloid chemist it is of utmost importance to know which conclusion he can draw from electrophoresis measurements, we still tried to apply (89a and b) on high $\zeta$-potentials. The thus obtained results are in agreement to such an extent that we can apparently indeed calculate from (89a and b) the correct order of magnitude of the relaxation effect, even for high $\zeta$-potentials. Yet one should consider the here given numeral values with the necessary caution, for example it cannot be concluded from the fact that $f(\kappa a, \zeta)$ for 2-2 electrolytes can become negative at $\frac{e \zeta}{kt}=4$, that due to the relaxation effect alone the electrophoresis velocity can switch sign!

The electrophoretic velocity\footnote{\BLUE {It appears that the words "electrophoretic velocity" or rather its abbreviation are missing in the original Dutch version, they have been added in the translation.}} in Fig. \ref{fig:11} for non symmetric electrolytes is calculated using $f_1(\kappa a)$, $f_2(\kappa a)$ and $f_4(\kappa a)$, while in view of the earlier mentioned considerations, the term corresponding to $f_3(\kappa a)$ has been neglected.  These values will however approach the truth less well, than those from Table V and Fig.\ref{fig:10}.  

In Fig.\ref{fig:10} it can first be noticed that the theory of Henry is entirely confirmed for very small $\zeta$-potentials (curve with $\frac{e \zeta}{kT}=0$) and that the deviations for $\zeta$-potentials lower than 25 mV are minor. This conclusion is of theoretical interest, since we learn that for low $\zeta$-potentials the relaxation effect has very little influence, while the electrophoretic drag can be quite large there (since at least for large $\kappa a$, $\frac{2 \epsilon_0 \epsilon_r \zeta}{3 \eta} f_1(\kappa a)$ is much smaller than $\frac{Q}{6 \pi \eta a}$ ). From a practical point of view this is of importance, especially for the interpretation of the electrophoretic velocity of bio colloids, since there many cases of low electrophoretic velocity occur and from this we can calculate with certainty the $\zeta$-potential (and the charge). 

If however the $\zeta$-potential is larger than 25mV ($\frac{e \zeta}{kT}>1$), the relaxation effect has a noticeable influence on the electrophoretic velocity. This influence is largest, when the dimensions of the particle and the double layer are of the same order. Because the relaxation effect increases so rapidly with increasing $\zeta$-potential, an unlimited increase in $\zeta$ does not necessarily cause an unlimited increase of the electrophoretic velociy. This is very clear from Fig.\ref{fig:12}, where the electrophoretic velocity is illustrated as a function of $\zeta$ and $\kappa a$ for 1-1 electrolytes. While $U$ increases linearly with $\zeta$ for very large and very small $\kappa a$, the $U$-$\zeta$ curve for intermediate values of $\kappa a$ exhibits a maximum.
\ifFIG\begin{figure}
    \centering
    \includegraphics[width=0.45\textwidth]{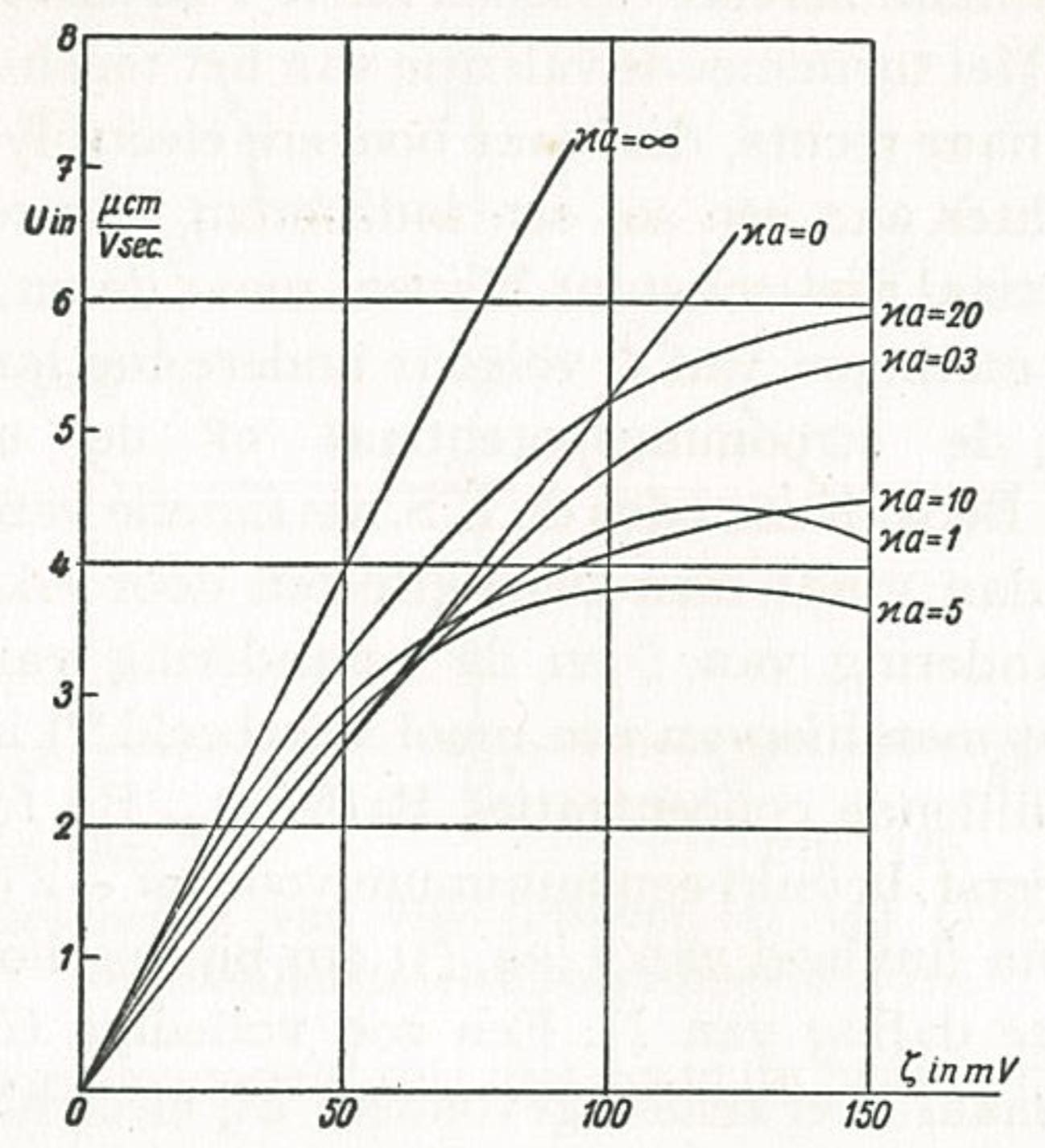}
    \caption{The electrophoretic velocity as a function of $\zeta$ at various values of $\kappa a$ (1-1 electrolyte).}
    \label{fig:12}
\end{figure}\fi

For the usual values of $\kappa a$ these maxima lie at electrophoretic velocities of the order of $5\frac{ \mu \text{cm}}{V \text{sec}}$, in good agreement with the by Von Hevesy observed maximum velocity of electrophoresis (see also Chapter I). The by Troelstra\footnote{S.A.TROELSTRA, Thesis Utrecht, 1941, Figures 38, 34 and 37} observed phenomenon, that the electrophoresis velocity of AgJ, at diminishing $p_J$, thus at increasing particle potential, goes through a maximum, is entirely consistent with this view. The height of this maximum $\Big( \frac{6 \mu \text{cm}}{V \text{sec}}\Big)$ as well as the potential at which it is found (total double layer potential  250 mV, $\zeta$-potential thus $<$ 250 mV) corresponds well with our curve for $\kappa a=20$ from Fig.\ref{fig:12}, while it follows from the values given by Troelstra of the particle radius (32 mm)\footnote{\BLUE{A value of 32 mm appears too large for a colloidal particle, it might have been 32 microns?}} and the electrolyte concentration (12 mmol KNO$_3$) that $\kappa a=11$ in his experiment.

From Figs.\ref{fig:10} and \ref{fig:11} it appears furthermore, that at constant $\zeta$ and at increasing \textit{electrolyte concentration} (increasing $\kappa a$), the electrophoretric velocity first diminishes, reaches a minimum between $\kappa a=1$ and $\kappa a=10$ and then increases again. With increasing valence of the counter ion this minimum shifts to the right, thus towards higher electrolyte concentration. 

\ifFIG\begin{figure}
    \centering
    \includegraphics[width=0.7\textwidth]{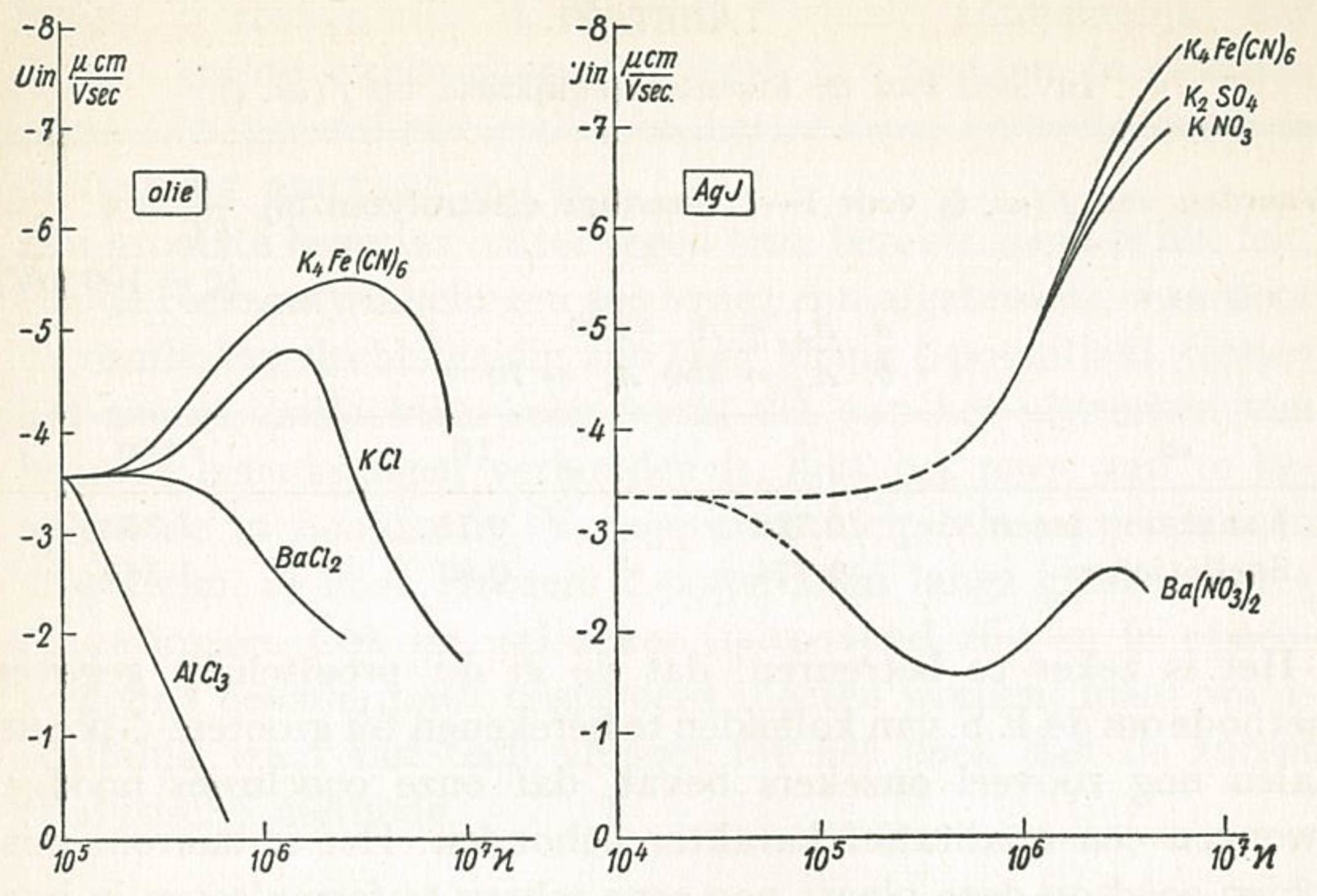}
    \caption{Electrophoresis velocities of oils (\BLUE{"olie"}, Powis) and AgJ (Troelstra) at various electrolyte concentrations.}
    \label{fig:13}
\end{figure}\fi
If one however adds to a sol an indifferent electrolyte, the $\zeta$-potential will not remain constant, but decreases, as is clear from measurements of $\zeta$ according to other methods, for example with the streaming potential or the method of De Bruyn\footnote{H.DE BRUYN, Thesis Utrecht 1938.}. If one thus determines the electrophoretic velocity as a function of the electrolyte concentration, one finds two effects which are intertwined, namely the real change of $\zeta$ and the change of the $f(\kappa a, \zeta)$. In Fig.\ref{fig:13}b a good example is shown concerning the electrophoresis velocity of AgJ at various concentrations of Ba(NO$_3$)$_2$. With increasing $\kappa a$, the electrophoretic velocity decreases first, reaches a minimum for $\kappa a\sim 2$ ($a=3 \times 10^{-6}$ and then increases (influence of $f(\kappa a, \zeta)$) only to decrease again at even larger values of $\kappa a$ (real decrease of $\kappa a$). Such a complete $U-c$ curve is however only seldomly found. For electrolytes with 1-valued counter ions, usually the first decrease is missing, such that we will have to assume that the electrophoretic velocity of the clean sol is roughly found in the minimum of the curves of Figs.\ref{fig:10} and \ref{fig:11}. We then observe curves with an increasing section (increasing $f_1(\kappa a)$ and decreasing of the relaxation effect) and a decreasing section (decreasing $\zeta$). See also Fig.13a obtained from the work of Powis\footnote{F.POWIS, Z. physik. Chem. \textbf{89}, 91 (1915).} for the electrophoretic velocity of oil-emulsions. If an electrophoresis curve is found to decrease continuously up to large concentrations, such as is often the case for multi valued counter ions, we can then indeed draw the conclusion that the decrease of $\zeta$ dominates over the other effects.

From Fig.\ref{fig:11} it is also clear, to which extend electrolytes of different \textit{valence type} influence the electrophoretic velocity. The predicted order corresponds well with experiments (Figs.\ref{fig:13}a and b). The other way around, such an ion spread only indicates something about the electrophoresis phenomenon, but not about the corresponding $\zeta$-potentials.

The \textit{radius of the particles} should also have an influence on the electrophoresis velocity. At equal $\zeta$-potentials, in general larger particles will travel faster than smaller ones, while perhaps in very dilute sols with small particles (small $\kappa a$) the opposite will happen. The measurements by Mooney\footnote{M.MOONEY, J. Phys. Chem. \textbf{35}, 331 (1931).} on relatively large oil droplets and by Kemp\footnote{J.KEMP, Trans. Far. Soc. \textbf{31}, 1347 (1935).} on quartz particles covered with gliadin are quantitatively in agreement with our theory. 

The \textit{mobility} of the ions has some influence on the relaxation correction. Very mobile ions such as $H^+$ and $OH^-$ (thus with small $1/\lambda_+$ and $1/\lambda_-$ respectively) cause a smaller relaxation effect than more immobile ions. See Table VI. 

This conclusion is not supported by facts, since in general acids decrease the electrophoretic velocity more than salts. Here a direct influence of the $H^+$ ions on the $\zeta$-potential must thus be assumed. 
\\
\begin {table} \label{Table6}
\caption*{TABLE VI: Influence of the ion mobility on $f(\kappa a, \zeta)$ } \label{tab:title}
\centering
\begin{tabular}{c}
\hline
Values of $f(\kappa,a$ for 1-1 valued electrolytes at $\frac{e\zeta}{kT}=4$, $(\zeta=100 mV)$.\\
\textit{a.} $\Lambda_+=\Lambda_-=70$\\
\textit{b.} $\Lambda_+=350$, $\Lambda_-=70$ \\
\begin{tabular}{ l | l | l | l}
  \hline
  $\kappa a$ & 1 & 10  & 100\\
  \hline
  \textit{a.} Slow ions & 0.728 & 0.75 & 1.358 \\
  \textit{b.} Fast ions & 0.774 & 0.83 & 1.369\\
-----------------------&
-----------------------&
-----------------------&
-----------------------
\end{tabular}

\end{tabular}

\end{table}

It is certainly regrettable, that the in this thesis given method to calculate the electrophoretic velocity for colloids at larger $\zeta$-potentials contains so many uncertainties, such that our conclusions necessarily can only be qualitative. It is therefore perhaps a good idea to formulate clearly once more to which extend our calculations are lacking.\\
1. The Brownian motion of the particle is not taken into account. This might lead to errors for sols with small particles.\\
2. The mutual interaction between sol particles is neglected, which is not allowed in highly concentrated sols. A sol can be considered concentrated, if the total charge of the sol particles is at least of the same order as the total charge of all other ions in the sol. \\
3. The ions in the double layer are considered as point charges, while only the Coulomb energy is accounted for. It should be necessary, in about the same manner as Stern\footnote{{O. STERN, Z. Elektrochem. \textbf{30}, 508 (1924).}} has indicated, to take into account the ion radius and adsorption forces.\\
4. Lately indications have shown that the structure of the double layer is more complex than we imagined up to now. Verlende\footnote{{Ed. VERLENDE, Proc. Kon. Ned. Akad. v. Westensch., Amsterdam \textbf{42 764} (1939. PhD Thesis Ghent 1940, page 50. }} finds that the conductivity of the double layer is 10-50 times larger than according to its charge. M. Klomp\'e\footnote{{M. KLOMP\'E, PhD. Thesis Utrecht (1941).} \BLUE{Marga Klomp\'e became the first female cabinet minister of the Netherlands in 1956 after being a member of parliament from 1948 onward.}} renders credible that a large electrophoretic velocity can exist, for cases where the potential difference in the diffuse part of the double layer is essentially zero.\\
5. The biggest objection against our calculations is the fact, that the series expansion was truncated early, such that the results are only valid for small $\zeta$-potentials. In view of the enormously tedious calculations that are associated with higher terms, it seems to be recommendable to solve the in chapter IV given principal equations through graphical means. This will certainly also be time consuming and will have to be done in successive approximations, but will probably still lead faster to the goal, than the pure analytic approach. 

In summarizing we have to conclude that this work does not enable us to indicate the relationship between the electrophoretic velocity and the $\zeta$-potential. We did however deepen our insight into the mechanism of electrophoresis and the phenomena related to the relaxation effect. \\
1. We have indicated clearly, which factors are important in the calculation of the electrophoretic velocity and how this calculation can be performed. \\
2. An electrolysis formula (48) has been deduced, which fully accounts for the electrophoretic drag and the relaxation effect.\\
3. From (48), after introducing several approximations, the formula (89) was deduced, which describes the relationship between the electrophoretic velocity and $\zeta$ for small $\zeta$-potentials accurately. This confirms the formula of Henry for $\zeta \rightarrow 0$ and also confirms that for $\zeta$-potentials below 25 mV only small corrections (a few percent) are needed. \\
4. If (89) is applied to larger $\zeta$-potentials, it turns out that the relaxation effect brings along important corrections on the electrophoretic velocity. The order of magnitude of these corrections, the dependence on $\kappa a$ and on the associated phenomena correspond to a large extend to experimental facts such as:\\
a.  the influence of the valence of co- and counter- ions, \\
b. the maximum value of the electrophoretic velocity (Von Hevesy),\\
c. maxima and minima in the $U-c$ curve, accompagnied by a monotoneous $\zeta - c$ curve,\\
d. an increase of the dielectric constant (see chapter VIII).

Since our results for higher $\zeta$-potentials keep bearing only a qualitative character, the colloid chemistry must for the time being draw not more than half quantitative conclusions from electrophoresis measurements, unless the electrophoretic velocity is small $< 1 \frac{\mu \text{cm}}{V \text{sec}}$. If one wants to obtain accurate values of the $\zeta$-potential, one is advised to get these according to other methods.

\newpage

\section{THE ELECTROPHORETIC VELOCITY OF CONDUCTING PARTICLES}

Before we apply our calculations from chapter IV on the electrophoretic velocity of conducting particles, we have to ask ourselves, if conducting particles behave as such during electrophoresis, Rightfully so, many authors (Henry, Verwey, Van Gils) have pointed out that this is doubtful, since during current transport through the boundary of the conducting particle - fluid, electrolysis will occur and this electrolysis will appear together with polarization phenomena. One can express this polarization as a surplus potential and one should now investigate, if this surplus potential is greater or smaller than the voltage difference between the front and the back of the particle that exists during electrophoris. Electrophoresis experiments are usually done at a potential difference of several Volts/cm and, since the dimensions of the colloidal particles are on the order of 10$^{-6}$ cm, the difference in potential between the front and the back will be several microvolts. The current density during electrophoresis experiments can vary between 10$^{-6}$ and 10$^{-1}$ Amp/cm$^2$ and the question is if the surplus potential at this current density at the material from which the conducting colloidal particle is made (Ag, Au, Pt, Hg, etc.) is larger or smaller than several \textit{micro}volts. During measurements on macro-electrodes at the aforementioned current densities, surplus potentials of a few to many \textit{milli}volts were found and from this one could immediately draw the conclusion that conducting particles still behave as insulators during electrophoresis. This surplus potential is however also dependent on the nature of the surface of the electrode and is strongly reduced if the number of active spots on the surface increases (raw, platinated electrode). The surface of a colloid particle has, in comparison to a macro-electrode, very many active spots and therefore the surplus potential can become considerably lower again. Thus with the above reasoning we cannot reach a definitive conclusion,

Experiments also do not guide us much further. In first approximation, the electrophoretic velocity of a conducing particle will be given by the curve ($\mu=\infty$) of Fig.3. This means at low and average $\kappa a$ there is an electrophoretic velocity of normal value, at very large $\kappa a$ it will be practically zero. The fact, that colloid metals exhibit a normal electrophoretic mobility, says, as long as $\kappa a$ is not too large, very little about the passing of current through the colloidal particles. Henry\footnote{{D.C. HENRY, Proc. Soc. London \textbf{A 133}, 122 (1931).}} cites one case, in which the polarization was deliberately kept very small and at which, even at a very large $\kappa a$, no electrophoresis was found (Ag-wire in a AgNO$_3$ solution). Opposed to these are the measurements of Bull and S\"ollner\footnote{{H.B. BULL and K. S\"OLLNER, Kolloid Z. \textbf{60}, 263 (1932).}}, who determined the electrophoretic velocity of rather rough Hg-emulsions in Hg$_2$(NO$_3)_2$-solution and found for $\kappa a=10-100$ a value of more than $3 \mu$ cm/Vsec. We would want to conclude that conducting colloids most likely behave as insulators during electrophoresis, but it cannot be excluded, that in certain cases they behave as conductors. This means that in general our boundary conditions for the current transport (7 and 8 \BLUE{just before Section II-B}) for a conducting particle remain valid unchanged, such that the constants $C$ and $B$ keep the values according to (82) and (84) and thus equation (89) can also be applied unchanged. In the boundary condition 5 \BLUE{just before Section II-B} one must take into account the infinitely large dielectric constant of a conductor, whereby the value of $P$ will be influenced. This however has no influence on the magnitude of the electrophoretic velocity. 

We also want to dedicate a few sentences to the electrophoretic velocity of particles, which do conduct the current well. In that case should not only the boundary condition 5, but also the boundary conditions 7 and 8 be changed, while the calculations of chapter IV can be used without any change. The taking into account of the boundary conditions results in unexpected difficulties, since two of the boundary conditions become identical and as a consequence one of the coefficients $B$, $C$ and $P$ will remain undetermined. After all, for a conductor the boundary conditions are:\\
1. The current is perpendicular to the sphere surface and is equal inside and outside the sphere. Since inside the sphere the conductivity is very high, the field is very low. Thus $\Phi_i=0$, thus $(\Phi)_{r=a}=0$, thus $(I)_{x=\kappa a}=0$. See eqs. (74), (17) and (58).\\
2. The current through the surface is transferred in a (given) ratio of cations and anions. It is for example imaginable, that only the $H^+$ ions (in the Pt-sol) or the $Ag^+$ ions (in a Ag-sol) are being discharged at the boundary of the sphere-fluid and that the anions are not participating at all. \\
3. The strengths of the field inside and outside the sphere are proportional to the dielectric constants.  Thus according to (77') $D\Big( \frac{dI}{dx}\Big)_{x=\kappa a}=D_i\Big( \frac{I}{x} \Big)_{x=\kappa a}$. Since for a good conductor the dielectric constant is very high, this condition demands that $(I)_{x=\kappa a}=0$, which is thus identical to the first condition. If one would like to be exact, then the charge density within the sphere should also be thought of as a volume one, and some kind of electron double layer would form. In doing this, the polarization of the metal roster and the finite value of the conductivity should also be taken into account and then conditions 1 and 3 would no longer coincide. This coinciding only happens in the limit of infinite large conductivity and dielectric constant. 

It does not seem very useful, to expand on this very complicated problem any further, as long as there is not more experimental proof that metal particles indeed act as conductors during electrophoresis. It thus appears wise, until further notice, to treat the electrophoretic velocity of conducting particles in the same manner as that of insulating particles, and thus apply our equation (89) unchanged.

\newpage
\section{CONDUCTIVITY AND TRANSPORT NUMBER}

If one totally ignores the interaction between sol particles and their counter ions, one can imagine the conductivity of a sol being composed out of three contributions. One contribution is given by the sol particles themselves, which with their electrophoresis velocity  $U/E$ transport their charge $Q$. The second contribution originates from the counter ions, which with their velocity $z_- e \lambda_-$ transport each a charge $z_- e$. The third contribution originates from the conductivity of the inter micellar fluid. We will only discuss from now on the first two contributions and assume that the third contribution is small, or that the measured conductivity has been corrected for this effect\footnote{The fact that this separation into three parts and the separate evaluation of the third term cannot be entirely consequent, is here ignored.}. 

The conductivity of a sol with $N$ particles per cm$^3$ (charge per particle $n e$, counter ion $z_-$ valence) is then given by
\begin{equation}
\nonumber
\tag{90} \lambda_{prim}=N\frac{U}{E} ne + N z_-e\lambda_- \frac{n}{z_-} z_- e=Nne  \Big( \frac{U}{E} + z_-e \lambda_-\Big) \hspace{1cm} 
\end{equation}
while the transport number of the counter ions, i.e. the fraction of the flow, which is being transported by the counter ions, is given by
\begin{equation}
\nonumber \frac{z_-e\lambda_-}{\frac{U}{E}+ z_-e\lambda_-}
\tag{91} \hspace{2cm} 
\end{equation}

If on the other hand the interaction between the sol particles and the counter ions is taken into account, then this means that, compared to (90), the contribution to the conductivity of the ions is becoming smaller, while those of the sol particles is unchanged. As a consequence of the relaxation the field in the double layer is lower than $E$, such that the ions which are close to the particle, contribute less to the conductivity than the ions that are at larger distances. Moreover the counterions are being hindered by the fluid flow around the particle with an amount, equal to the fluid flow at that position. 

Through these effects the transport number of the counter ions and the total conductivity are becoming smaller than according to the primitive calculation.

The magnitude of the relaxation drag of the counter ions can be taken into account easily on the grounds of the following reasoning: the double layer provides a force on the particle $k_4$ (see page 9), and thus inversely the same force acts on the counterions as well. 
From Figs. 10 and 11 on can see that the relaxation effect at $\kappa a=0.1$ is at most a few percent, while for $\kappa a =1$ at not to low $\zeta$- potentials this can be many tenths of percents. In the same way the contribution to the conductivity  of the counterions will be reduced and thus ends up at $f(\kappa a, \zeta)/f_1(\kappa a)$ of its original value (see equation (89))\footnote{Here we have ignored the interaction between relaxation and electrophoris, which is allowed here, since these considerations can only be applied anyway to cases, where the intermicellar fluid contains little electrolyte and thus $\kappa a$ is small.}.

The fluid flow imposes a velocity on the counter ions, which varies between $+U$ (near the particle) to zero (at large distances).

We can calculate the hereupon based decrease of the conductivity, if we remember that the fluid velocity in the direction of $\theta=0$ is $U-u_\theta \sin \theta + u_r \cos \theta$ with respect to a coordinate system at rest. Therefore the conductivity is reduced by
\begin{equation}
\nonumber
\lambda_{electroph.}=N \int_a^\infty \int_0^\pi \frac{U -u_\theta \sin \theta + u_r \cos \theta}{E} \rho(r) 2\pi r \sin \theta \text{ dr r} \; d\theta,
\end{equation}
which, with $U$ according to (48a) and $u_\theta$ and $u_r$ according to (49a and b) can be rewritten as
\begin{equation}
\nonumber
\lambda_{electroph.}=\frac{8\pi N\epsilon_0 \epsilon_r}{3 \eta} \int_a^\infty r^2 \rho(r) \int_\infty^r \xi dr dr. 
\end{equation}
To get an order of magnitude of this effect, we substitute
\begin{subequations}
\begin{align}
\nonumber
\rho(r)=-\epsilon_0 \epsilon_r\nabla^2 \psi^0 &\approx -\epsilon_0 \epsilon_r \kappa^2 \psi^0, \hspace{5 mm} \int_\infty^r\xi dr \approx \psi^0,
\hspace{5 mm}, \psi^0 = \zeta \frac{\kappa a e^{\kappa a- \kappa r}}{\kappa r} \hspace{5 mm} \text{and}\\
\nonumber
ne &= -4\pi \epsilon_0 \epsilon_r a^2 \Big( \frac{d\psi^0}{dr}\Big)_{r=a} = 4\pi \epsilon_0 \epsilon_r a \zeta (1+ \kappa a),
\end{align}
\end{subequations}
which is correct for small $\kappa a$ and $\psi^0$ , and  for not too large values of $\kappa a$ and $\psi^0$ will give a reasonable approximation. With this we find
\begin{equation}
\nonumber
\tag{92} \lambda_{electroph.}=\frac{2 N n e  \epsilon_0 \epsilon_r \zeta}{3 \eta} \frac{\kappa a}{2(1+\kappa a)} \approx Nn e \frac{U}{E} \frac{\kappa a}{2(1+\kappa a)}.
\end{equation}
In total the contribution to the conductivity of the free counterions will no longer be $Nn e^2 z_- \lambda_-$ , but $Nn e^2 z_- \lambda_- \frac{f(\kappa a,\zeta)}{f_1(\kappa a)}-Nn e \frac{U}{E} \frac{\kappa a}{(1+\kappa a)}$ .\\
Moreover, the possibility that a counterion is bound to the particle (non dissociated ions) must be taken into account and that consequently the "free" charge $Nn \varepsilon$ can be smaller than the total charge of the analytically determined counterions. 

May a few numerical examples illustrate the influence of the electrophoretic and relaxation drag of the counterions. 

we choose $\frac{f(\kappa a,\zeta)}{f_1(\kappa a)}=0.5$, $\frac{U}{E}=5\frac{\mu \text{cm}}{V \text{sec}}$ and $\kappa a=1$.

We consider two cases, namely $z_-e \lambda_-=5\frac{\mu \text{cm}}{V \text{sec}}$(slow counterions, corresponding to $\Lambda=52$) and $z_- e \lambda_-=34\frac{\mu \text{cm}}{V \text{sec}}$(fast counterions, corresponding to $\Lambda=350$).

In the first case, the average velocity of the ions is brought from 5 to $5\times 0.5 - 5 \times \frac{1}{4}=1.25$.

In the second case, the average velocity of the ions is brought from 34 to $34\times 0.5 - 5 \times \frac{1}{4}=15.75$.

We can thus observe that in each case the contribution to the conductivity can be remarkably reduced and that for fast counter ions the relaxation effect plays the most important role, while for slower counter ions the second term (electrophosis) can give an important contribution, while the velocity of the particle and the ion become of the same order of magnitude. 

It is thus not allowed to calculate the "free" charge of a sol according to the primitive equation (90). If one applies the above mentioned corrections to the contributions on the conductivity, one will find a free charge which is several times larger than according to the primitive calculation.

This also invokes that the total titratable charge surely must be considerably larger than the primitive free charge, which is also the case for various soles (\textit{AgJ, S, Fe$_2$O$_3$, Cr$_2$O$_3$})\footnote{\textit{S-sol} A.CHARIN, Acta Physicochimica U.R.S.S. \textbf{12}, 703 (1940).\\
\textit{Fe$_2$O$_3$-sol} and \textit{Cr$_2$O$_3$-sol} PAULI-VALK\'O, Elektromchemie der Kolloide, Springer 1929 page 275.\\ For a concentrated \textit{AgJ-sol} (10\%), the free charge according to (90) is about 1/5 of the total titrable charge. }. Where Pauli\footnote{W.PAULI and L.FUCHS, Kolloid Beih. \textbf{21}, 195, 412 (1925).\\ W.PAULI and E.FRIED, Kolloid Z. \textbf{36}, 138 (1925).} and coworkers concluded with acidoid silver and gold sols that the primitive free charge and the 
total charge are practically identical, the suspicion arises that the by them determined $H_+$ ions did not, or only partly, acted as counter ions for the sol particles, rather were freely present in the intermicellar fluid\footnote{To explain these large free charges A.J.Rutgers and J.Th.G.Overbeek (Z. physik. Chem. A 177, 29 (1936)) assumed that close to the particle (in the attached water layer) many conductometric  active counter ions would be present. When in this model the relaxation effect is included, it turns out that these ions are slowed down to such an extent that they virtually did not contribute to the conductivity, such that this explanation should be rejected.}. 

It is certainly of importance, to analyze the conduction of electric current in sols more closely and specially the drag of the counter ions with different speeds and compare to each other for example $H_+$ and $Na_+$, since there is a possibility, to obtain through experimental means further information concerning the influence of electrophoretic drag and the relaxation effect.

\newpage
\section{DIELECTRIC CONSTANT}

The dielectric constant of a colloid solution is in general not equal to that of the pure dispersion medium. The colloidal particles have a different (usually lower) dielectric constant and conductivity than the dispersion medium, such that the dielectric constant of the sol is being reduced. As a consequence of the deformation of the double layer a dipole is being formed at each particle, which is directed in the opposite direction to the imposed field. This will on the contrary increase the dielectric constant. Moreover the dipole moments of the sol particles themselves (protein!) and the hydration can have an influence on the dielectric constant. For hydrophobic sols we can ignore the two last mentioned effects and we can then ask ourselves how the dielectric constant will be affected by the first two effects. We limit ourselves to the dielectric constant measured at very low frequency, in which the non-symmetry of the double layer has had the time to fully establish itself. 

On a relatively large distance from the colloid particle the electric field is according to (17), (58) and (71) described by\footnote{\BLUE{There are two equations with number (92) in Overbeek's thesis.}}
\begin{equation}
\nonumber
\tag{92} \delta \psi = -E\Big(r + \frac{B}{\kappa^3 r^2} \Big)\cos\theta. \qquad  
\end{equation}
All other terms from $I$ (71) approach namely at large $r$ faster to zero than $B/r^2$. Each particle is thus behaving to the outside world as a dipole with moment
\begin{equation}
\nonumber
\tag{93} m=- 4\pi \varepsilon_0 \varepsilon_r \frac{B}{x^3} E. 
\end{equation}
We can then set the dielectric constant of the sol equal to the dielectric constant of the dispersion medium, in which $N$ dipoles each with moment $m$ are dispersed. The dielectric constant of the sol is then
\begin{equation}
\nonumber
\tag{94} \ifRED\else 4 \pi (\varepsilon_0 \varepsilon_r)_{sol}\fi=\ifRED\else 4\pi \varepsilon_0 \varepsilon_r\fi+ \frac{4\pi N m}{\ifRED\else E\fi} =\ifRED\else 4\pi \varepsilon_0 \varepsilon_r\fi- \frac{4 \pi N a^3}{3}\frac{3}{2}\frac{2B}{\kappa^3 a^3}\ifRED\else 4\pi \varepsilon_0 \varepsilon_r\fi = \ifRED\else 4\pi \varepsilon_0 \varepsilon_r\fi\Big(1 - \frac{3}{2}\delta \frac{2B}{\kappa^3 a^3}\Big),
\end{equation}
if $\delta$ represents the volume concentration of the dispersed phase. 

With the help of equation (84) for $\frac{2B}{\kappa^3 a^3}$ the dielectric constant of a sol can now directly be calculated as a function of concentration, $\zeta$-potential and $\kappa a$. In Table VII a number of values are shown of $\frac{D_{sol}-D)}{D}$ for a sol concentration of one volume percent in a 1-1 valence electrolyte with ion mobility $\Lambda_+=\Lambda_-=70$.
\begin {table} [!ht]\label{Table7}
\caption*{TABLE VII Increment of the dielectric constant at a sol concentration of 1 volume percent
 } \label{tab:title} 
\begin{tabular}{| l | l | l | l | l | l | l | }
  \hline
  $\zeta$ \hspace{3cm} &  & & & $\kappa a$ &  &\\
   & 0.1 & 0.3 & 1 & 5 & 20 & 100 \\
  \hline
   25 mV   &  +0.50  & +0.065 & -0.0028 & -0.0087 & -0.014 & -0.015  \\
  50 mV &  +2.32 & +0.29 & +0.031 & +0.0080 & -0.012 & -0.015 \\ 
  100 mV &  +8.40 & +0.99 & +0.128 & +0.054 & -0.005 & -0.015 \\
  \hline  
\end{tabular}
\end{table}

While for large $\kappa a$ the influence of the low dielectric constant (low conductivity) of the sol particles overshadows all double layer influences, at low $\kappa a$ and not to small $\zeta$-potentials considerable increase of the dielectric constant can be expected due to the distortion of the double layer. For very small $\kappa a$ equation (94) can no longer be applied, since the various double layers will penetrate each other and this was not taken into account in the derivation of (92), (93) and (94). In that case the dielectric constant increase will be smaller than according to our calculation and can be better approximated with the Debye and Falkenhagen\footnote{P.DEBYE and H.FALKENHAGEN, Physik. Z. \textbf{29}, 401 (1928).} theory. 

From equation (94) and Table VII it is clear that the increase of the dielectric constant is proportional to the concentration of the sol, increases strongly with increasing $\zeta$-potential and becomes smaller rapidly at increasing electrolyte concentration and even turns into a decrease. Form equation (84) for $\frac{2B}{\kappa^3 a^2}$ it follows that with increasing valence $(z)$ and decreasing ion mobility $(1/\varrho)$, the dielectric constant increase is becoming larger. 

These conclusions can be tested against the measurements of Kunst\footnote{H.KUNST, PhD. Thesis. Utrecht 1940.} He established at various frequencies the dielectric constant of a number of hydrophobic sols (AgBr, AgJ, AS$_2$S$_3$), and found that the dielectric constant is considerably above that of water. From his own numerical values it seems that even the lowest used frequency (1.5 $\times$ 10$^5$ sec$^{-1}$) still lies in the dispersion zone and that the static dielectric constant must thus be even greater than the measured value at the frequency 1.5 $\times$ 10$^5$.

From his measurements it follows that the increase in dielectric constant is proportional to the sol concentration and varies at 1 volume percent from 0.055-0.230. Also the order of magnitude corresponds thus well with the by us predicted value at $\kappa a\sim 1$. The predicted influence of the mobility of the counter ions is confirmed by the fact that the increase in dielectric constant of acid soles is about 10 percent smaller than those of neutral soles. 

The fact that all electrolytes increase the dielectric constant, is not in line with our expectations. For multi-valued electrolytes (increasing $z$) and potential determining electrolytes (increasing $\zeta$) this could still be understood, but not for single valued electrolytes such as KNO$_3$.

In general though we can say that our calculations are sufficiently supported by measurements, such that we can attribute these dielectric constant increases to the relaxation effect. The fact that the dispersion frequency agrees with this can be shown as follows. To build up the double layer time is needed, the relaxation time. If the duration of a period of the oscillating field, with which the dielectric constant is measured is getting smaller then the relaxation time, then the double layer will not have enough time to adapt itself to the ever changing state and will not or barely be distorted, such that an increase in dielectric constant does not appear. 

According to Debye and Falkenhagen\footnote{$\varrho$ = the average friction coefficient of the ions in the double layer.} the relaxation time of the double layer is\footnote{\BLUE{Something appears wrong with the dimensions of this equation, the dimension of $\Theta$ would be $s/m^3$ instead of $s$?}}
\begin{equation}
\nonumber
\Theta = \frac{\rho}{\kappa^2 kT}. 
\end{equation}
From their derivation, it can be seen, that this time is \textit{not} dependent on the size of the particle, if one ignores the Brownian motion of the particle. 

For ions with a normal mobility ($\Lambda = 70$) and $\kappa = 10^5$ to $10^6$, corresponding to a electrolyte concentration of $10^{-5}$ to $10^{-3}$, the relaxation time will be $\Theta =\frac{1}{2 \times 10^5}$ sec. to $\Theta =\frac{1}{2 \times 10^7}$ sec. which is in good agreement with the fact that the dispersion of the dielectric constant extends to the whole by Kunst measured frequency domain ($1.5 \times 10^5$ - $5 \times 10^7)$.  

\newpage
\section*{GLOSSARY OF SYMBOLS}
\noindent
$a$ radius of the sphere \\
$A$ integration constant from (55b)\\
$\alpha$ charge density on the sphere, $\alpha = \frac{n \varepsilon}{4\pi a^2}$\\
$B$ integration constant from (61)\\
$c$ electrolyte concentration\\
$c_1$ to $c_6$ integration constants of the hydrodynamical equations of chapter IV-A\\
$C$ integration constant from (66)\\
dielectric constant of the fluid \BLUE{ $=4\pi\epsilon_0 \epsilon_r$, in modern SI units, with $\epsilon_r$ the relative dielectric constant of the fluid}\\
$D_i$ dielectric constant of the sphere\\
$D_+, D_-$ ion diffusion constants \\
$E(x)$ = $\int_\infty^x \frac{e^{-t}}{t} dt$ (56)\\
$e$ charge of the proton \\
$\varepsilon_i$ charge of an ion "i"\\
$+Q$ charge of the sphere \\
$\epsilon_0$ permittivity of vacuum\\
$\epsilon_r$ relative permittivity of the fluid\\
$\epsilon_{ri}$ relative permittivity of the material ("i" for "internal")\\
$\zeta$ electrokinetic potential, $\zeta$-potential or "zeta"-potential\\
$\eta$ viscosity of the fluid\\
$\theta$ angle between the radius vector and field direction\\
$k$ Boltzmann constant \\
$n_i =  n_i^0 + \delta n_i$ number concentration of ion of species $i$ in presence of field $E$\\
$n_i^0$ equilibrium number concentration of ion of species $i$ in the double layer\\
$\delta n_i$ perturbation number concentration of ion of species $i$ \\
$n_i^\infty$ number concentration of ion of species $i$ in bulk \\
$\mu$ ratio of the conductivity constants of the sphere and the liquid \\
$p$ pressure in the liquid\\
$r$ distance form the center of the spherical particle\\
$R$ used in $\delta \psi=-ER \cos\theta$ (17)\\
$\rho_m$ mass density of the fluid\\
$\rho$ charge density\\
$\delta\rho$ additional charge density during electrophoresis\\
$\rho^0$ charge density at equilibrium (i.e. with no electric field applied)\\
$1/\lambda_+$, $1/\lambda_-$ friction coefficients of the + and - ions respectively\\
$T$ absolute temperature\\
$u$ velocity at a point in the fluid\\
$U$ electrophoretic velocity\\
$x=\kappa r$ (51)\\
$E$ strength of the (dc) electric field, whereby electrophoresis is taking place\\
$y=\frac{e \psi^0}{kT}$ (52) \\
$z_i$ unsigned valence of ion of species $i$\\
$\varphi$ angle between $r-\theta$ plane and the vertical plane\\
$\delta\psi$ additional potential during electrophoresis\\
$\psi^0$ potential of the double layer at rest i.e at equilibrium given say by the Poisson-Boltzmann equation

\newpage
\section*{Propositions (\RED{STELLINGEN})\footnote{\BLUE{According to a 'peculiar' Dutch custom, the PhD defence is accompagnied by 10 "Stellingen" (propositions). As in the case of Overbeek's, this is a loose leaflet addition to the actual thesis. The subject of these propositions does not necessarily have to be the same as that of the thesis (for example proposition IX on peneplains). The candidate can be challenged on the claims made in these propositions during the actual defence.} \RED{The original Dutch propositions are indicated in red color.}}}

\RED{I De manier waarop EYRING en medewerkers afleiden dat de smeltentropie ongeveer gelijk $R$ is, is niet bevredigend.}

I The way in which Eyring and coworkers deduce that the melting entropy is equal to $R$, is not satisfactory.\\
J.HIRSCHFELDER, D.STEVENSON and H.EYRING, J. Chem. Phys. \textbf{5}, 896 (1937).\\
R.W.GURNEY and N.F.MOTT, J. Chem. Phys. \textbf{6}, 222 (1938).\\
N.F.MOTT and R.W.GURNEY, Trans. Far. Soc. \textbf{35},364 (1939).\\

\RED{II De kinetische afleiding van de wetten van RAOULT en van VAN 'T HOFF voor verdunde oplossingen door FRAHM bevat eenige elkaar compenseerende denkfouten.}

II The kinetic derivation of the laws of Raoult and Van t'Hoff for dilute solutions by Frahm contain some mutually compensating logical errors.\\
H.FRAHM, Z. Physik. Chem. \textbf{A 184}, 399, (1939). Z. Physik . Chem. \textbf{B 48}, 119 (1941).\\

\RED{III SCHULZ heeft niet aannemelijk gemaakt, dat de vorming van stoffen met hoog moleculair gewicht in het levende organisme op reacties moeten berusten die in vitro niet realiseerbaar zijn.}

III Schulz has not shown plausibly that the creation of substances with high molecular weight in living organisms are based on reactions that cannot be realized in vitro. \\
G.V.SCHULZ, Z. physik. Chem. \textbf{A 182}, 127 (1938).\\

\RED{IV De uitspraak van JAEGER, dat een optisch actieve stof  $A$ gelijke affiniteit zou hebben tot de optische antipoden van een stof $B$ is onjuist.}

IV The claim from Jeager, that an optically active substance $A$ has the same affinity to the optical antipods of a substance $B$ is not correct.\\
F.M.JAEGER, Chem. Weekblad \textbf{33}, 522 (1936).
F.M.JAEGER, Optical Activity and High Temperature Measurements (1930), 205\\
M.M.JAMISON and E.E.TURNER, J. Chem. Soc. (1940), 264.\\

\RED{V Bij het bepalen van $\zeta$-potentialen heeft DE BRUYN de gemeten potentiaalverschillen ten onrechte gecorrigeerd voor de invloed van het toegevoegde indifferente electrolyt op de activiteits- co\"efficient der potentiaalbepalende ionen.}

V In the determination of the $\zeta$- potentials De Bruyn has erroneously corrected the measured potential difference for the influence of the added indifferent electrolyte on the activity coeffient of the potential determinining ions. \\
H.DE BRUYN, PhD thesis, Utrecht (1938) 47.\\

\RED{VI Uit het verloop van de conductometrische titratie van zure AgJ solen en S-solen met loog, blijkt dat de verhouding van Na$^+$- en H$^+$- ionen in de dubbellaag gelijk is aan die in de evenwichtsvloeistof.}

VI From the progress of the conductometric titration of acid AGJ sols and S sols with a base, it turns out that the ratio of NA$^+$ and H$^+$ ions in the double layer is equal to that in the equilibrium fluid. \\
H.DE BRUYN and J.TH.G.OVERBEEK, Kolloid Z. \textbf{84}, 186 (1938).\\
A.CHARIN, Acta Physicochimica, U.R.S.S. \textbf{12}, 703 (1940).\\

\RED{VII Bij metingen van de extinctie van geconcentreerde kolloide systemen speelt de herhaalde verstrooiing vaak een belangrijke rol. Men kan een eenvoudige methode aangeven om de hierdoor veroorzaakte complicaties te vermijden. }

VII In measuring the extinction of concentrated colloidal systems, the repeated scattering often plays an imprtant role. One can easily indicate a method to avoid the hereby caused complications.\\

\RED{VIII Het toepassen van het begrip "intermicellaire vloeistof" op geconcentreerde kolloide systemen voert tot tegenstrijdigheden.}

VIII The application of the concept of "intermicellar fluid" on concentrated colloid systems leads to contradictions.\\

\RED{IX Op morphologische gronden zijn de twee typen van schiervlakken die door W.PENCK "Prim\"arrumpf" en "Endrumpf" genoemd zijn, niet van elkaar the onderscheiden.}

IX On morphological grounds the two types of peneplains named "Prim\"arrumpf" and "Endrumpf" by W.Penck are indistinguishable.\\
W.PENCK, Die morphologische Analyse (1924).\\
J.C.OVERBEEK-EDIE and Th.RAVEN, Comptes rendus du Congr\`es International de G\'eographie, Amsterdam (1938) II, 169.\\

\RED{X WARBURG en CHRISTIAN hebben niet bewezen, dat de binding van eiwit en prosthetische groep in \textit{d}- aminozuuroxydase volgens het eenvoudige door hen aangegeven schema verloopt.}

X Warburg and Christian did not prove, that the binding of the prostethic group of protein in \textit{d}- amino-acid-oxydase is according to the simple by them indicated manner. \\
O.WARBURG and W.CHRISTIAN, Biochem. Z. \textbf{298}, 150, 369 (1938).\\
E.NEGELEIN and H.BR\"OMEL, Biochem. Z. \textbf{300}, 225 (1938).

\newpage

\section*{Translator's note}

\BLUE{Since the original PhD thesis of Overbeek is not readily available and is written in Dutch, we decided to provide a translation here as a service to the English speaking colloid community.}

\BLUE{
A copy of this PhD thesis made its way to the Australian colloid community, where Robin Arnold\footnote{\BLUE{See for example: Arnold, Robin and  Warren, Leonard J.. (1974), "Electrokinetic properties of scheelite", Journal of Colloid and Interface Science, 47, 134-144.}} gave it to Tom Healy who passed it on to Derek Chan. After a thorough office clean up, it finally ended up in the hands of a Dutch researcher in Singapore. A further reason to translate this work into English is that most other works by Overbeek are equally difficult to obtain (J.Th.G.Overbeek Kolloid-Beihefte, 1943, 54, 287) or do not contain as many details of the derivations as in his PhD thesis (Philips Research Reports, Vol. 1, No.4, 315-319, 1946: On Smoluchowski's equation for the electrophoresis of colloidal particles; available online).} 

\BLUE{In order to conform with modern notation and to enhance readibility, we have changed some of the original symbols to modern convention (for example $E$ is used to denote the electric field instead of $X$ as in the thesis) and use SI units throughout. To provide an authentic historical perspective, a version of this translation with the original symbols was also prepared. Our comments in the translation appear in blue color.}\footnote{\BLUE{More information on electrophoresis can also be found in the book: "Electrophoresis: Theory, Methods, and Applications, Volume II" edited by Milan Bier; Academic Press, New York,  1967, Chapter 1, "The interpretation of electrophoretic mobilities" by J.Th.G.Overbeek and P.H.Wiersema.}}

\end{document}